\documentclass{aa}
\usepackage[pdftex]{graphicx}
\usepackage{graphicx}
\usepackage{amsmath}
\usepackage{wasysym}

\usepackage{marvosym}
\usepackage{txfonts}
\pdfoutput=1
\usepackage[pdftex]{color,xcolor}
\usepackage[pdftex]{graphicx}
\usepackage{hyperref}               
\hypersetup{pdfauthor=Christoph Mordasini}
\hypersetup{backref=true, pagebackref=true, hyperindex=true, breaklinks=true,colorlinks=true,urlcolor=blue, linkcolor=blue,  citecolor=blue,pagecolor=red, bookmarks=true, bookmarksopen=true}
\usepackage{mathtools}

\def\mearth{M_\oplus}
\def\rearth{R_\oplus}

\def\mcore{M_{\rm core}}

\def\f1{f_{\rm I}}
\def\mj{M_{\textrm{\tiny \jupiter }}}
\newcommand{\lj}{L_{\textrm{\tiny \jupiter}}}
\newcommand{\rj}{R_{\textrm{\tiny \jupiter}}}

\def\mstar{M_*}

\def\tstart{t_{\rm start}}

\def\miso{M_{\rm iso}}

\def\beq{\begin{equation}}
\def\eeq{\end{equation}}

\def\fopa{f_{\rm opa}}
\def\t2{\tau_{\rm II}}

\def\sigmas0{\Sigma_{\rm s,0}}

\newcommand{\mz}{M_{\rm Z}}
\newcommand{\mzn}{M_{\rm Z,0}}
\newcommand{\mxy}{M_{\rm XY}}

\def\tcr{\tau_{\rm cross}}
\def\trun{\tau_{\rm run}}
\def\tg{\tau_{\rm g}}
\def\zp{Z_{\rm pl}}
\def\zstar{Z_{\rm star}}

\def\aj{AJ}                   
\def\araa{ARA\&A}             
\def\apj{ApJ}                 
\def\apjl{ApJ}                
\def\apjs{ApJS}               
\def\aap{A\&A}                
\def\mnras{MNRAS}             

\def\({\left(}
\def\){\right)}
\def\<{\left<}
\def\>{\right>}

\begin{document}

\title{Grain opacity and the bulk composition of extrasolar planets. I.}
\subtitle{Results from scaling the ISM opacity}

\author{C. Mordasini\thanks{Reimar-L\"ust Fellow of the MPG}\inst{1} \and H. Klahr\inst{1} \and  Y. Alibert\inst{2,3}  \and N. Miller\inst{4} \and T. Henning\inst{1} }

\institute{Max-Planck-Institut f\"ur Astronomie, K\"onigstuhl 17, D-69117 Heidelberg, Germany \and
Center for space and habitability, Physikalisches Institut, University of Bern, Sidlerstrasse 5, CH-3012 Bern, Switzerland \and Institut UTINAM, CNRS-UMR 6213, Observatoire de Besan\c{c}on, BP 1615, 25010 Besan\c{c}on Cedex, France \and Department of Astronomy and Astrophysics, University of California, Santa Cruz, USA }

\offprints{C. Mordasini, \email{mordasini@mpia.de}}

\date{Received 15.03.2013  / Accepted 10.03.2014}

\abstract
{{The opacity due to grains  in the envelope of a protoplanet $\kappa_{\rm gr}$ regulates the accretion rate of gas during formation, meaning that the final bulk composition of planets with a primordial H/He envelope is a function of  it. Observationally, for extrasolar planets with known mass and radius it is possible to estimate the bulk composition via internal structure models.} }  
{{We want to study the global effects of $\kappa_{\rm gr}$  {as a poorly known, but important quantity} on synthetic planetary populations.}}
{{We first determine the reduction factor of the ISM grain opacity $\fopa$ that leads to a gas accretion timescale consistent with  grain evolution models  {for specific cases}. In the second part we compare  the mass-radius relationship of low-mass planets and the heavy element content of giant planets for different values of the reduction factor with observational constraints.}}
{{For $\fopa$=1 (full ISM opacity) the synthetic super-Earth and Neptunian planets have too small radii (i.e., too low envelope masses) compared to observations,  {because at such high opacity, they can not efficiently accrete H/He during the formation phase.} At $\fopa$=0.003, the value calibrated with the grain evolution models, the synthetic and actual planets occupy a similar  mass-radius domain. Another observable consequence is  the metal enrichment of giant planets relative to the host star, $\zp/\zstar$. We find that the mean enrichment of giant planets as a function of mass $M$ can be approximated as $\zp/\zstar = \beta(M/\mj)^{\alpha}$ both for synthetic and actual planets. The decrease of $\zp/\zstar$ with mass follows $\alpha$$\approx$-0.7 independent of $\fopa$ in synthetic populations, in agreement with the value derived from observations (-0.71$\pm$0.10). The absolute enrichment level $\beta$ decreases from 8.5 at $\fopa$=1 to 3.5 at $\fopa$= 0. At $\fopa$=0.003, one finds $\beta$=7.2 which is similar to the  result derived from observations (6.3$\pm1$.0).}}
{{We find observational  {hints} that the opacity in protoplanetary atmospheres is much smaller than in the ISM  {even if the specific value of $\kappa_{\rm gr}$ can not be constrained in this first study as $\kappa_{\rm gr}$ is found by scaling the ISM opacity}. Our results for the enrichment of giant planets are also important  {to distinguish} core accretion and  gravitational instability. In the simplest picture of core accretion,  where first a critical core forms, and afterwards only gas is added,  $\alpha$$\approx$-1. If a core accretes all planetesimals inside the feeding zone also during runaway gas accretion $\alpha$$\approx$-2/3. The observational result  (-0.71$\pm$0.10) lies between these two values, pointing to core accretion as the likely formation mechanism.}}

\titlerunning{Grain opacity and planetary bulk composition}
\authorrunning{C. Mordasini et al.}

\maketitle


\section{Introduction}\label{sect:grainopa}
It is well known (e.g., Ikoma et al. \cite{ikomanakazawa2000}) that a reduction of the opacity $\kappa$ in the gaseous envelope of a forming giant planet leads to a reduction of the formation timescale. {In simulations of concurrent core and envelope accretion} (e.g., Pollack et al. \cite{pollackhubickyj1996}, hereafter P96) the reason is that a lower opacity  during the so-called phase II (where gas must be accreted in order to allow further core growth) leads to a more efficient transport of released potential energy out of the envelope. {Phase II is an intermediate phase that occurs for in situ formation between the moment the core has  reached the isolation mass, and the moment when rapid gas accretion starts. This happens when the core has reached  the crossover mass when envelope and core mass are equal.} 

For in situ calculations the overall formation timescale of the planet is dominated by the duration $\t2$ of  phase II, which means that at low $\kappa$ the overall formation timescale is reduced, too. For example, P96  find that an (arbitrary) reduction of the grain opacity $\kappa_{\rm gr}$ to 2\% of the interstellar medium (ISM) value leads to $\t2=2.2$ Myrs for Jupiter formation, while with the full ISM value, $\t2=6.97$ Myrs, longer than the mean disk lifetime. 

Obviously, one is therefore interested in knowing an {estimate} for the effective grain opacity in the envelope, instead of having to use arbitrary scalings like the 2\%. The grain opacity ({with opacity we mean in this work always the Rosseland mean}) can be calculated  from the microphysics of grain growth via coagulation, grain settling, and evaporation at high temperatures. Podolak (\cite{podolak2003}) presented such a numerical model, finding that grain growth leads to opacities up to three orders of magnitude smaller than in the ISM because grains grow efficiently. The limitation of this  work was a lack of self-consistency as the envelope structure in which the grain growth was studied was calculated with pre-specified different opacities.  

Only recently Movshovitz et al. (\cite{mbpl2010})  (hereafter MBPL10), presented the first self-consistently coupled calculations of grain evolution and giant planet growth. In their work, the envelope structure is used to {calculate} at each radius the evolution of the grains at each time step. From the grain size distribution the Rosseland mean opacity is calculated, which is then  {fed} back into the envelope calculation. The main result of these calculations is that the grain opacity is much reduced. For Jupiter formation, this leads to a duration of phase II of only 0.52 Myrs. This corresponds to a reduction by a factor 13.4 relative to the P96 full opacity case, and still a factor 4.2 to the P96 ``low opacity case'' with  a 2\% ISM opacity. As forming a giant planet within the typical lifetime is a timing issue (at least if migration is not taken into account cf. Alibert et al. \cite{alibertmordasini2004}), these factors matter. 

The complex {(and computationally heavy)} calculations of grain evolution made by MBPL10 are {not} the scope of this first paper.  Instead, we here follow a practical approach, and determine in {the first part of} this work the reduction factor $\fopa$ by which interstellar opacities must be reduced in order to obtain the same duration of  phase II as found by MBPL10. As will be shown, a grain opacity of only about 0.3\% the ISM value leads to the best reproduction of the MBPL10 results. {Using one uniform reduction factor of the ISM opacity can of course not reproduce the complex structure of the opacity  {which depends on planetary properties like the core or envelope mass} as found in grain evolution models (Movshovitz \& Podolak \cite{movshovitzpodolak2008}).  {The simple ISM scaling approach therefore has some important limitations (see the discussion in Sect. \ref{sect:generality})}. Our interest in still deriving a global, uniform $\fopa$ is  to  have an  {intermediate value for the opacity between the two extremes (full ISM opacity vs. grain free)} at a low computational cost for population synthesis simulations.} {The  goal of this work is rather to study the global effects of $\kappa$ on  planetary populations. For this, we compare in the second part of the paper important statistical properties of  synthetic planets found with different values of $\fopa$ with observational constraints from extrasolar planets.} 

{The opacity controls the rate at which a core of given mass accretes gas during the formation phase, therefore different $\fopa$ lead to different bulk compositions that manifest in the mass-radius relationship that we can observe after a few Gyrs of evolution. Population synthesis calculations that are built on self-consistently coupled planet formation (Alibert et al. \cite{alibertmordasini2005}) and evolution models (Mordasini et al. \cite{mordasinialibert2011})  can predict the synthetic mass-radius relationship. While there are other observational constraints on $\fopa$ that can be derived from the mass distribution of exoplanets alone (in particular the frequency of giant planets), transiting exoplanets are therefore of special interest in this study. What is needed for the comparison are transiting extrasolar planets with a well-defined mass and radius, and a rather large semimajor axis $\gtrsim0.1$ AU to minimize the impact of stellar irradiation. The extrasolar planets of this class must also have a mass-radius relationship that implies that they contain significant amounts of primordial H/He, since this is the envelope type considered here. It also means that we avoid the part of the mass-radius relationship that is most degenerate (e.g., Valencia et al. \cite{valenciaikoma2010}, Rogers \& Seager \cite{rogersseager2010}). Instead, for such planets it is possible to infer, at least in a rough way, the bulk composition (global heavy element content) as has been demonstrated by, e.g., Guillot et al. (\cite{guillotsantos2006}), Burrows et al. (\cite{burrowshubeny2007}), Guillot (\cite{guillot2008}), or  Miller \& Fortney (\cite{millerfortney2011}). The results from the latter study are used in the second part of this study.  Our simulation therefore also  {aims} at establishing a bridge between physical processes like grain evolution that govern gas accretion during formation, and observable quantities. This information can be used in the ideal case to feed back into the microphysical models for the grains (or into  specialized models in general).

The structure of the paper is as follows: in Section \ref{updatedopacity} we show the modifications of our giant planet formation model to take into account reduced grain opacities. The formation model itself was first described in Alibert et al. (\cite{alibertmordasini2005}), while the version used here is described in Mordasini et al. (\cite{mordasinialibert2011}). In Section \ref{sect:determinationfopa}, we determine $\fopa$ by comparison with MBPL10. {We then turn to the population synthesis calculations (Sect. \ref{sect:obsconstr}) and use them to study the envelope mass as  a function of core mass (Sect. \ref{sect:menveofmcore}). The associated mass-radius relationship mainly of low-mass planets is addressed in Sect. \ref{sect:MRR}.   In Section \ref{sect:zpzstar} we compare the enrichment of giant planets relative to the host star for different $\fopa$ with the results derived by Miller \& Fortney (\cite{millerfortney2011}) for actual extrasolar planets.} Finally, in Sect. \ref{sect:conclusion} we give a summary and present our conclusions.  {In Appendix \ref{sect:semianalyticalsolution} we derive a semi-analytical model for the core and envelope mass in phase II and determine its parameters in Appendix \ref{sect:paramstkh}. }

{In the second paper of this series (Mordasini \cite{mordasini2014}), hereafter Paper II, we present a simple analytical model for the opacity due to grains in protoplanetary atmospheres. It calculates the grain opacity based on the comparison of the timescales of the governing microphysical processes like grain settling, coagulation and evaporation (Rossow \cite{rossow1978}). It therefore takes into account that the grain opacity is a dynamically changing quantity that depends on planetary properties like core and envelope mass, or the accretion rates. In a subsequent work, we will couple this model with our upgraded population synthesis code that can now simulate the concurrent formation of many embryos in one disk during the formation phase (Alibert et al. \cite{alibertcarron2013}) and includes atmospheric escape during the long term evolution (Sheng et al. \cite{shengmordasini2014}).}

\section{Updated opacity model}\label{updatedopacity}
The opacity of the gas in the envelope is calculated using two different tables: first, combined ISM grain opacity (multiplied by a factor $\fopa$) and high-temperature gas opacities from the Bell \& Lin (\cite{belllin1994}) analytical opacity laws. Second, low-temperature molecular and atomic opacities of a grain-free, solar composition ([M/H]=0) gas from Freedman et al. (\cite{freedmanmarley2008}).

For the scaled grain and high-temperature gas opacities we use the expressions of Bell \& Lin (\cite{belllin1994}) that  analytically give the opacity as a function of density and temperature for eight different opacity regions, where the transitions between the different  regions are smoothed.  The  opacity regions are in increasing temperature order dominated by: ice grains, ice grain evaporation, metal grains, metal grain evaporation, molecules, H-scattering, Kramer's law and finally electron scattering. This means that at low temperatures, the opacity is assumed to be caused by grains only. While for a full ISM grain opacity this is justified, this is no more the case if the grain opacity is reduced by a large factor.  The code was modified so that the opacity in the regions where grains dominate can be scaled freely with a grain opacity reduction factor $\fopa\leq 1$ in a way  that the opacities remain continuous also with such a factor. The high-temperature gas opacities remain unchanged. 

When the grain opacity is very low, gas opacities begin to dominate also at low temperatures. This effect is seen in the calculations of MBPL10. Therefore it is necessary to include opacity tables for a grain-free gas also at low temperatures.  Freedman  et al. (\cite{freedmanmarley2008}) provide such tables of molecular and atomic opacities of a grain-free solar composition ([M/H]=0) gas. These tables were also used by MBPL10. For the present study, the results of Freedman  et al. (\cite{freedmanmarley2008}) were converted  into tables giving $\kappa(T,\rho)$, using the {SCvH EOS} (Saumon et al. \cite{saumonchabrier1995}) for the conversion of pressure $P$ and temperature $T$ into density $\rho$.   

Finally, the (scaled) opacities from Bell \& Lin (\cite{belllin1994}) and Freedman al. (\cite{freedmanmarley2008}) were combined, taking for simplicity always the maximum of the two values. This underestimates the opacities in regimes where both gas and grains contribute in a similar way, but avoids problems in the overlapping regime where both Bell and Freedman opacities are due to gas. This means that even when the grain opacity is very low, the combined opacity will not fall below a lower limit given by the gas. 

\section{Determination of $\fopa$ by comparison with Movshovitz et al. (\cite{mbpl2010})}\label{sect:determinationfopa} 
\begin{figure*}
\includegraphics[width=0.475\textwidth]{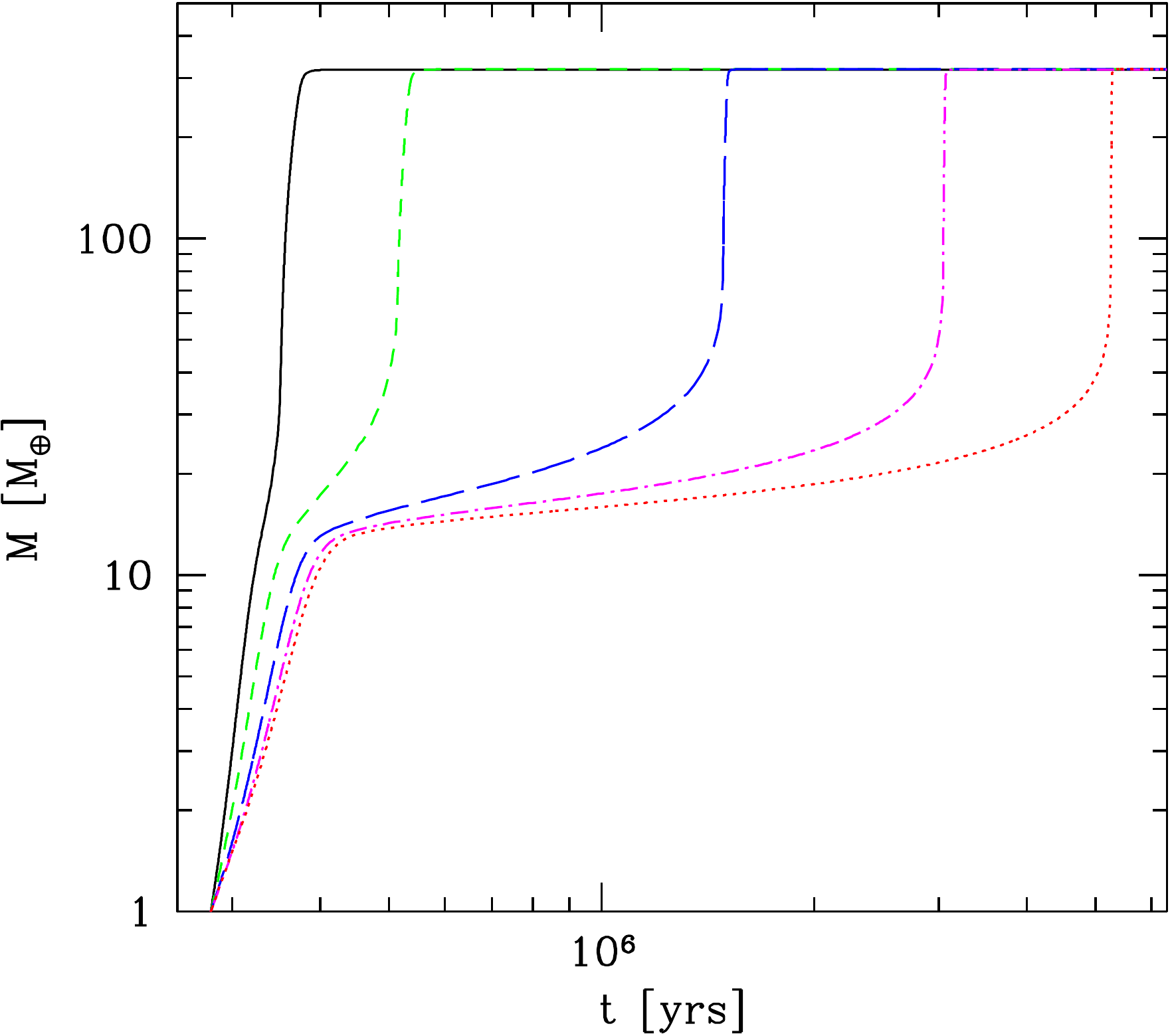}
\includegraphics[width=0.5\textwidth]{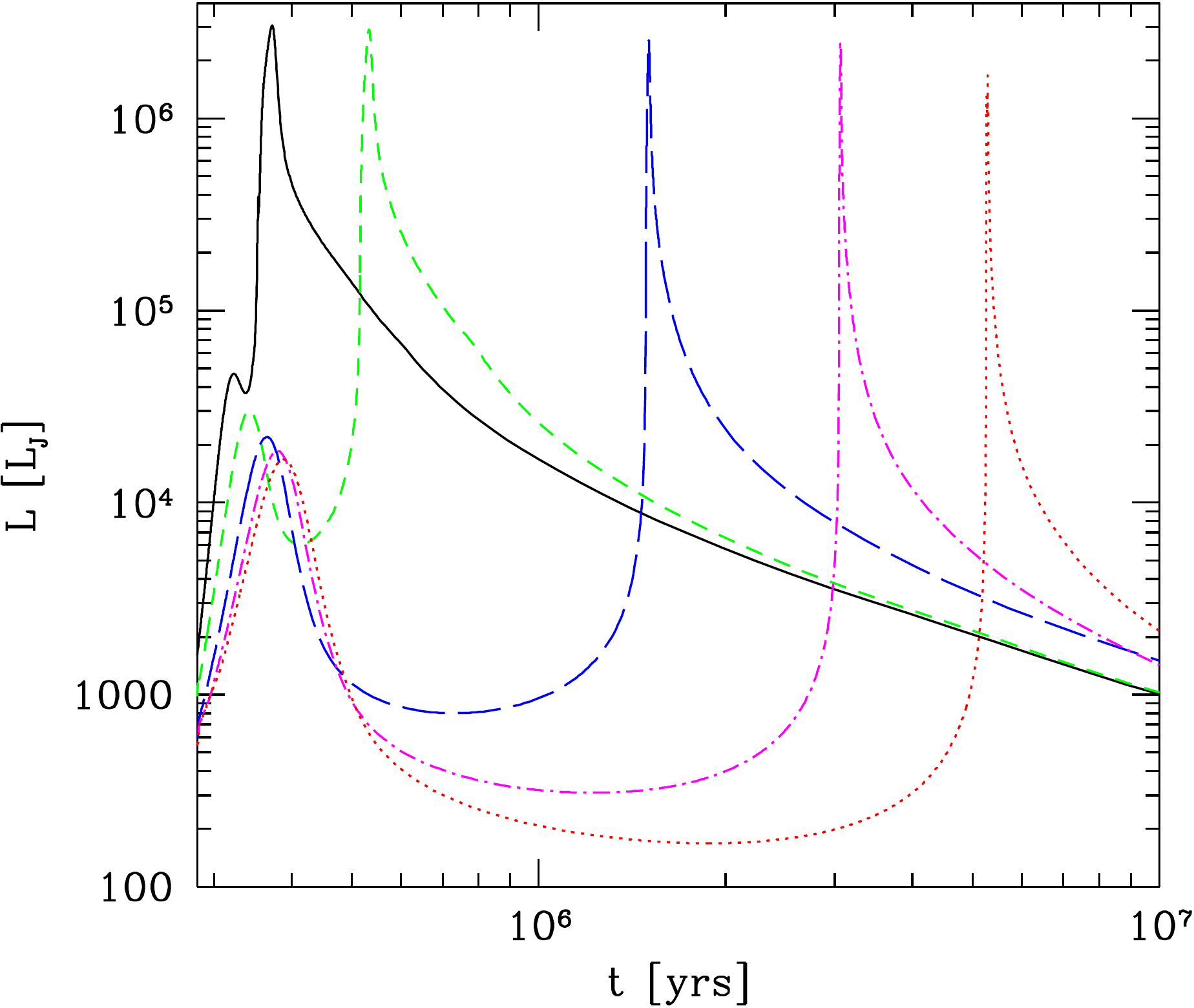}
\caption{Simulations of the in situ formation of Jupiter for $\sigmas0$ = 10 g/cm$^{2}$ as a function of the grain opacity reduction factor. The left plot shows the total mass as a function of time. The right plot is the total luminosity (including the intrinsic and the shock luminosity) in units of  the intrinsic luminosity of Jupiter today. The curves correspond, from left to right, to $\fopa$= $10^{-4}$ (black solid), 0.001 (green short-dashed), 0.01 (blue long-dashed), 0.1 (magenta dash-dotted) and 1 (red dotted line). The influence of $\fopa$ on the duration of the plateau phase II is obvious.}\label{fig:sigma10mlt}  
 \end{figure*}
 
The question we want to answer in this section is simple: What is the ISM grain opacity reduction factor that leads to  the same duration of  phase II for Jupiter in-situ formation as in the coupled grain evolution and envelope accretion calculations of  MBPL10? Only the duration of the phase II is considered because the opacity is the dominant factor for its length\footnote{{It is   far from clear that phase II actually occurs at all during giant planet formation: low planetesimal accretion rates (Fortier et al. \cite{fortierbenvenuto2007}), gravitational shepherding (Zhou \& Lin \cite{zhoulin2007}), gap formation in the planetesimal disk (Shiraishi \& Ida \cite{shiraishiida2008}), and orbital migration (Alibert et al. \cite{alibertmordasini2005}) all tend to suppress it. In any case, in this work we are interested in it only to calibrate the magnitude of $\kappa_{\rm gr}$. }}. 

\subsection{MBPL10 comparison mode}
For the MBPL10 comparison calculations, we made a number of modifications to our model in order to emulate as closely as possible the formation code in MBPL10. In particular we assume, in contrast to normal operation, that no migration or disk evolution occurs. The other setting, which are if possible identical as in MBPL10, are listed in Table \ref{tab:mov}. It is clear that the two codes still differ in some aspects.  Examples are the planetesimal-envelope interaction calculation, or the fact that we include for the outer boundary conditions the effect that the planet radiates into the nebula (Papaloizou \& Terquem \cite{papaloizouterquem1999}).  This is however known not to make a large difference ({Bodenheimer \& Pollack \cite{bodenheimerpollack1986}}).

\begin{table}
\caption{Settings for the Movshovitz et al. (\cite{mbpl2010}) comparison.}\label{tab:mov}
\begin{center}
\begin{tabular}{lccc}
\hline\hline
 $\sigmas0$ [g/cm$^{2}$]                         & 10         &  6    & 4 \\ \hline                                             
$P_{\rm neb}$ [dyn/cm$^{2}$]           &   0.75          & 0.45  &  0.3 \\ 
$\tstart$ [Myrs]    & 0.28              &  0.585            &  0.728 \\ \hline
$T_{\rm neb}$ [K]           &\multicolumn{3}{c}{115}\\
a [AU] & \multicolumn{3}{c}{5.2 AU}\\
Initial embryo mass [$\mearth$] & \multicolumn{3}{c}{1 }\\
Dust to gas ratio  &\multicolumn{3}{c}{1/70}\\
$\dot{M}_{\rm XY,max}$ [$\mearth$/yr]& \multicolumn{3}{c}{0.01}\\
Migration & \multicolumn{3}{c}{not included}\\
Disk evolution & \multicolumn{3}{c}{not included}\\
Planetesimal ejection & \multicolumn{3}{c}{not included}\\
Core density & \multicolumn{3}{c}{constant, 3.2 g/cm$^{2}$}\\ \hline
\end{tabular}
\end{center}
\end{table}

\subsection{Simulation setup}
As in MBPL10, three initial planetesimal surface densities $\sigmas0$ are studied: 10, 6, and 4 g/cm$^{2}$. The corresponding durations of  phase II $\t2$ found by MBPL10 are 0.52, 0.79 and 2.43 Myr, respectively.  For each of the three $\sigmas0$, the best fitting $\fopa$ was determined individually. {In the ideal case} the three values  are identical or at least similar. It is however clear that the grain growth process  depends  on $\sigmas0$. Different $\sigmas0$ directly lead to different core masses. This in turn leads to different dust settling velocities  (MBPL10) meaning that the timescales of the various processes of grain evolution will differ depending on $\sigmas0$. Therefore it is expected that the best fitting $\fopa$ will not be exactly the same for all three cases.  {The limitations of one general reduction factor are further discussed in Sect. \ref{sect:generality}.}
                          
\subsubsection{Initial planetesimal surface density $\sigmas0$ = 10 g/cm$^{2}$ }
To illustrate the impact of $\kappa_{\rm gr}$, we show in Fig. \ref{fig:sigma10mlt} the result for the calculations for   $\sigmas0$ = 10 g/cm$^{2}$. The left panel shows the total mass  as a function of time for  $\fopa$= $10^{-4}$, 0.001, 0.01, 0.1 and 1, while the right panels shows the total luminosity including internal and shock luminosity. The decrease of  $\t2$ with $\fopa$ is obvious. For $\fopa=10^{-4}$ no real phase II exists any more when looking at the mass evolution. The planet passes almost directly from phase I into phase III, even though that still two local maxima can be distinguished. For all cases the final core mass is  close to 36 Earth masses (36.1-36.7 $\mearth$), and the final total mass is 317.5-319.4 $\mearth$ (we artificially ramp down the gas accretion rate on a short timescale once a total mass of 1 $\mj$ is approached). The core mass at crossover is similar in all cases, covering a range from 15.6 to 17.9 $\mearth$.  MBPL10 found a crossover mass of 16.09 Earth masses.

\subsection{Duration of phase II as a function of $\fopa$}\label{durationphase2fopa}
Figure \ref{fig:t2m10} shows the duration of the phase II  as a function of $\fopa$ for the three surface densities. For  $\sigmas0$ = 10 g/cm$^{2}$ one sees that the length varies from just about 12\,000 years to about 5 Myrs. Beginning at the smallest opacities with $\fopa\lesssim 5\times 10^{-5}$, $\t2$ is first approximately independent of $\fopa$. This is because here, the grain opacity is so small that only the gas opacities matter that are independent of $\fopa$. These cases thus correspond to a grain-free regime that was  studied by Hori \& Ikoma (\cite{horiikoma2010}).  For $10^{-4}\leq\fopa\leq10^{-2}$, $\t2$ increases with the grain opacity reduction factor, following the same slope.  For $\fopa\geq 10^{-2}$, $\t2$ still increases, but following  another less steep slope. 

\begin{figure}
\begin{center}
\includegraphics[width=\columnwidth]{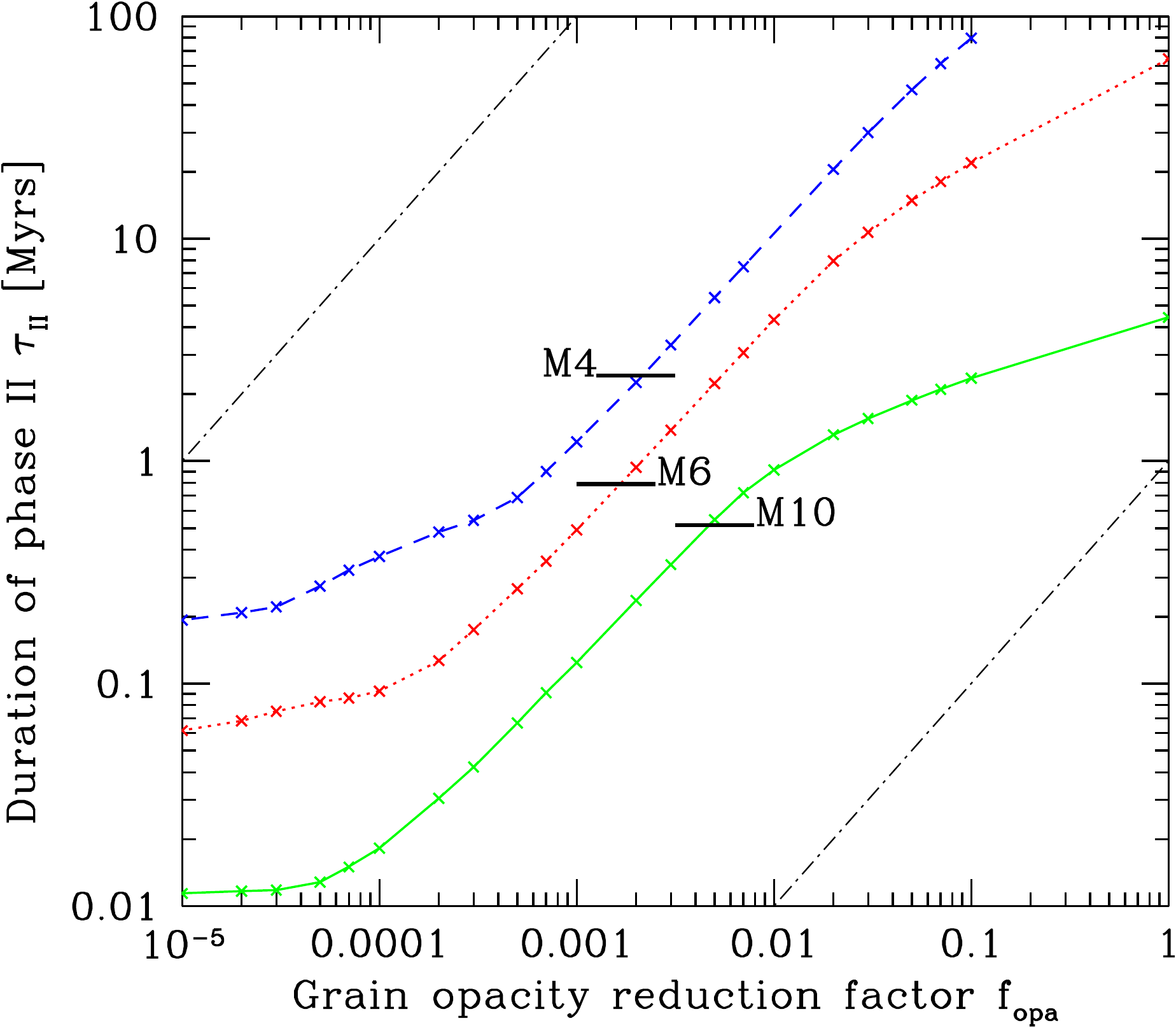}
\caption{Green solid curve: Duration of the phase II as a function of $\fopa$ for an initial planetesimal surface density of 10 g/cm$^{2}$. The short black horizontal line labeled M10 is the result of MBPL10, 0.517 Myrs. Red dotted curve: 6 g/cm$^{2}$ case. The short black horizontal line labeled M6 is the result of MBPL10, 0.787 Myrs. Blue dashed curve: 4 g/cm$^{2}$ case. The short black horizontal line labeled M4 is the result of MBPL10, 2.428 Myrs. {The black dashed-dotted lines in the top left and bottom right corner represent a linear slope to guide to eye.}}\label{fig:t2m10}
\end{center}
\end{figure}

In Figure \ref{fig:t2m10}, the short black horizontal curve labeled M10 marks the duration of phase II found by MBPL10, 0.517 Myrs. The two curves cross slightly below $\fopa=0.005$, which is thus the best fitting value. The value is remarkably low. Previous works (P96, Hubickyj et al. \cite{hubickyjbodenheimer2005}, Lissauer et al. \cite{lissauerhubickyj2009}) used as (arbitrary) ``low opacity case'' $\fopa=0.02$. The fitting value found here is a factor four smaller. This difference has a substantial influence on the formation time, as $\t2(\fopa=0.02)=1.33$ Myrs, i.e., almost a factor three longer.

The red dotted curve in  Fig. \ref{fig:t2m10} again shows the duration of phase II, but now for $\sigmas0$ = 6 g/cm$^{2}$. For this case, MBPL10 found $\t2=0.787$ Myrs.  As expected (e.g., P96)  the  formation  timescales are longer because of the smaller core mass at a lower planetesimal surface density. The duration of phase II we find is now between 62\,000 years and 64.3 Myrs. The dependence of $\t2$ on $\fopa$ follows a similar pattern as before. The curve meets with the value of MBPL10 at $\fopa=0.002$, which is thus the best fitting value. This is again a very low value. The value is smaller than the one for 10 g/cm$^{2}$ by a factor 2.5, but  still in the order of a few times $10^{-3}$. As discussed earlier, we cannot expect to find an identical value for the three simulations. 

The blue dashed line in  Figure \ref{fig:t2m10} finally shows the dependence of $\t2$ on $\fopa$ for the lowest surface density case of $\sigmas0=4$  g/cm$^{2}$. Here, the formation time with full ISM opacity would be much in excess of 100 Myrs.  Comparison with the result of MBPL10 (2.428 Myrs) leads to a best fitting value for this case of 0.002, i.e., again a few times $10^{-3}$.

{The dashed-dotted black lines in the top left and bottom right corner represent lines with linear slope. The comparison of these lines with $\t2(\fopa)$ shows that over an intermediate domain of $\fopa$ which itself depends on $\sigmas0$ (i.e., $\mcore$), there is roughly a linear correlation between  the opacity and the duration of phase II. Such a behavior is theoretically expected, see Appendix \ref{sect:paramstkh}. }

\subsection{Final result for $\fopa$}\label{sect:finalresfopa}
Table  \ref{tab:bestfitfopa} summarizes the results for the best fitting values $f_{\rm opa, best fit}$. In the table, we also give the mean of the three values. It seems reasonable for a fitting parameter to take simply the mean. This is especially the case as the three best fitting values do not depend in a strictly monotonic way on $\sigmas0$. The mean value $f_{\rm opa, best fit}=0.003$ is thus the {nominal} value  for {the population synthesis} calculations.  {The fact that the results for the three $\sigmas0$ differ by a few 0.001 indicates that the mean  $f_{\rm opa, best fit}=0.003$ is accurate only to factors of a few for a specific surface density. This is an additional limitation besides the fact that using a constant reduction factor throughout the envelope is itself an   {oversimplification} relative to actual grain evolution models, as discussed in the next section.

\begin{table}
\caption{Grain opacity reduction factors fitting MBPL10. }\label{tab:bestfitfopa}
\begin{center}
\begin{tabular}{l|rrr|r}
$\sigmas0$  [g/cm$^{2}$]&10 & 6 & 4 & mean\\ \hline
$f_{\rm opa, best fit}$ & 0.005 & 0.002 & 0.002& 0.003\\
\end{tabular}
\end{center}
\label{fopafit}
\end{table}%

\subsection{ {Limitations of a general scaling factor} }\label{sect:generality}
 {The numerical grain evolution models of Movshovitz \& Podolak (\cite{movshovitzpodolak2008}) and MBPL10 find a characteristic radial structure of $\kappa_{\rm gr}$ in the atmosphere (outer radiative zone) of protoplanets: near the outer boundary of the planet, the opacity is relatively high and comparable to the ISM opacity ($\kappa_{\rm gr} $$\sim$1 cm$^{2}$/g) but it becomes very low ($\kappa_{\rm gr} $$\sim$0.001 cm$^{2}$/g) in the deeper parts near the boundary to the convective zone. This is clearly a consequence of the growth of the grains. Such a radial structure of $\kappa_{\rm gr}$ does not result when simply scaling the ISM opacity, which rather leads to a roughly speaking radially constant $\kappa_{\rm gr}$. This is shown in the comparisons in Paper II. On the other hand, it is also found that for $\fopa=0.003$, the total optical depth of the outer radiative zone is similar as predicted by the grain evolution models, at least for the cases that were studied and, importantly, that were used to calibrate $\fopa$. The comparisons with MBPL10 presented here indicate that this is sufficient to lead to a similar gas accretion rate. While getting a physically motivated radial structure of $\kappa_{\rm gr}$ is an important task in its own right (and therefore addressed in Paper II), in the context of this first paper, we are only interested in the gas accretion rate, as this eventually determines the planetary bulk composition.}

{However, even if the optical depth is similar for the cases for which $\fopa$ was calibrated, there is no a priori reason to assume that for planets which {differ substantially}  (e.g., in core or envelope mass) from the calibration cases, the same $\fopa$ is applicable. This is because the microphysical processes that control $\kappa_{\rm gr}$ depend on, e.g., the gravitational acceleration, the density and temperature  in the envelope, or the grain accretion rate. This is a major limitation of the ISM scaling approach used in this paper, and means that no definitive conclusions about the value of $\kappa_{\rm gr}$ can be derived from the population syntheses presented further down, but only hints that the opacity might be much smaller than in the ISM. It is also a motivation to derive in Paper II an analytical model that dynamically calculates $\kappa_{\rm gr}$ based on the planet's properties. This is because a numerical model as in MBPL10 can currently not be used in population syntheses due to computational time limitations.}

 {In this context it is also important to clarify that the calibrated $\fopa$ is unlikely to apply to an, e.g., 1 $\mj$ planet due to its much higher mass than the planets used for the calibration. On the other hand, for the subject of this work, the final bulk composition, the value of $\fopa$ is relevant mainly for $\sim$one order of magnitude in core mass, namely from $\sim$1 to $\sim$15 $\mearth$ because of the following: for cores with an even lower mass, the envelope is not influencing much the planet's formation (it is given by the accretion of solids only) due to the tiny envelope mass. Additionally, only planets with $M\geq2\mearth$ are considered in the syntheses. Cores more massive than $\sim$15 $\mearth$ in contrast trigger gas runaway accretion on a timescale that is much shorter than typical disk lifetimes for all reasonable $\kappa_{\rm gr}$ (see, e.g., Ikoma et al. \cite{ikomanakazawa2000}). Once their total mass has reached $\sim$40 to 80  $\mearth$ (e.g., MBPL10), they pass into the disk limited gas accretion regime. In this phase, the gas accretion rate is no more dependent on the envelope opacity, but  controlled by external factors only (gap formation, accretion rate in the disk, etc). Interestingly, the opacity in the disk limited accretion phase does, however, affect the post-formation entropy and thus luminosity at young ages, see Mordasini (\cite{mordasini2013}). But this is not discussed here. }

{We have derived $f_{\rm opa, best fit}$ based on three simulations of MBPL10 that cover only a small part of the possible parameter space in which planets form in terms of core mass, luminosity, semimajor axis, and boundary conditions. Using the same $f_{\rm opa, best fit}$ for all parameters is therefore  {an oversimplification}. Therefore, we use this reduction value in the population synthesis calculation below as the nominal case, but consider it simply as  {an intermediate value between the extreme cases with $\fopa=0$ and 1.}

The simulations of MBPL10 cover a range in core mass of a factor 4 (about 4 to 16 $\mearth$ at crossover). The fact that  $f_{\rm opa, best fit}$ varies  {over this mass range by a factor of a few indicates that there is as expected a dependency, but potentially not a very strong one. A hint towards such a behavior is that the radial opacity structures in MBPL10 for the different core masses are all, at least to order of magnitude, similar. To investigate this quantitatively, we will couple in a forthcoming work the analytical model of Paper II to our updated population synthesis model. This will also allow to investigate the impact of different outer boundary conditions. }

\begin{figure}
\begin{center}
\includegraphics[width=\columnwidth]{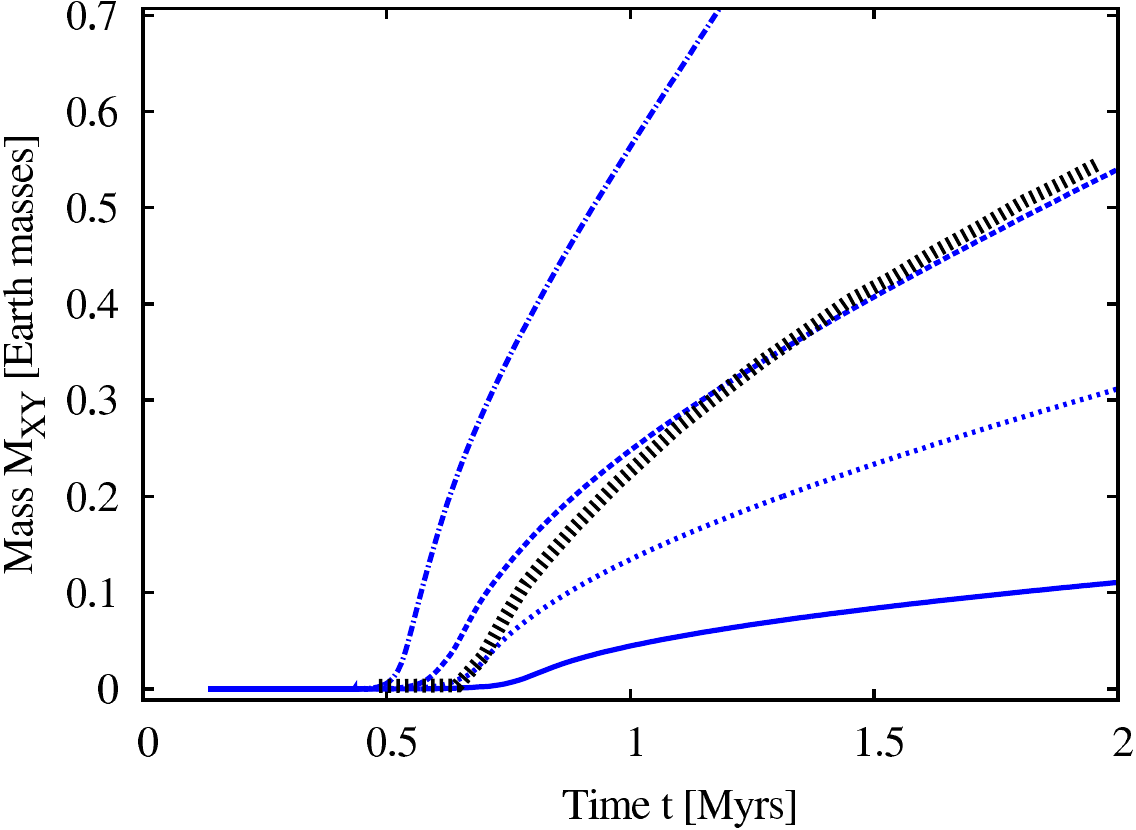}
\caption{{Envelope mass as a function of time for initial conditions as in run IIa in Rogers et al. (\cite{rogersbodenheimer2011}). The thick  dotted line is the result of Rogers et al. (\cite{rogersbodenheimer2011}) obtained in a combined grain evolution and core accretion simulation. The blue lines are our results for different values of the grain opacity reduction factor. The lines are $\fopa$=0.02 (solid), 0.003 (dotted), 0.001 (dashed) and 0 (dot-dashed).} }\label{fig:rogersR2a}
\end{center}
\end{figure}

{To investigate this further, we have compared the envelope mass as a function of time for different $\fopa$ in our model with the another  simulation of  combined grain evolution and core accretion  (run IIa in Rogers et al. \cite{rogersbodenheimer2011}). This study uses the same core accretion and grain evolution model as MBPL10, but additionally simulates the formation of a very low-mass planet at an orbital distance of 4 AU instead of 5.2 AU as in the three cases of MBPL10. The initial surface density is 6 g/cm$^{2}$ leading to a low isolation mass of about 2.4 $\mearth$. The simulation is continued to a time $t$=2 Myrs when Rogers et al. (\cite{rogersbodenheimer2011}) find that the core has grown to $2.65\mearth$, while the envelope mass is 0.54 $\mearth$.}  

{Figure \ref{fig:rogersR2a} shows the envelope mass as a function of time as found by Rogers et al. (\cite{rogersbodenheimer2011}) and in our simulations for $\fopa$=0.02, 0.003, 0.001, and 0 (grain free). The plot shows that for these initial conditions, $\fopa$$\approx$0.001 would lead to the best reproduction of the results of Rogers et al. (\cite{rogersbodenheimer2011}), while the nominal value of 0.003 calibrated with MBPL10 underestimates the actual envelope mass by a factor 1.7. The $\fopa=0.02$ simulation leads to an $\mxy$ that is too low by a factor 5, while the $\fopa=0$ case results in an envelope that is too massive by a factor 2.6. At a full ISM grain opacity, the envelope is 25 times less massive than in Rogers et al. (\cite{rogersbodenheimer2011}). This indicates that the mean calibrated value of  $\fopa=0.003$ is as expected only accurate to factors of a few, but that it still leads to differences that are significantly smaller than with arbitrary (or no) reduction factors. We must however  {recognize that  even this} fourth independent comparison case lies  relatively close to the simulations of MBPL10 in the parameter space of initial conditions.}

\subsection{Effect on long term evolution}\label{sect:longterm}
We have recently extended our  formation model to a combined formation and evolution model (Mordasini et al. \cite{mordasinialibert2011}). We can therefore study the effect of different grain opacities also on the long term evolution.  {To do so we followed the evolution} of a 1 $\mj$ planet at 5.2 AU for different $\fopa$  up to an age of 4.6 Gyr. We assume that $\fopa$ is constant in time, also after the gas accretion has finished. In reality  the grain opacity would probably quite rapidly tend to vanish after the planet has reached its final mass. The reason is that grains quickly rain out and get vaporized once the accretion of gas ceases (Rogers et al. \cite{rogersbodenheimer2011}).

{One way to quantify the impact of $\fopa$ on the long term evolution} is to study the planet's radius and luminosity at 4.6 Gyr. We  find that  the radius $R$ is constant for $\fopa\lesssim 0.004$, and equal to about 0.99 $\rj$. This means that for $\fopa\lesssim 0.004$, the contribution of the residual grain opacity is so small that is does not significantly affect the evolution, but that it is determined by the gas opacities only. For $\fopa$ larger than 0.004, we see as expected a gradual increase of the radius with $\fopa$ due to the less efficient cooling, reaching about 1.15 $\rj$ at the ISM opacity. For the luminosity we see an identical pattern with $L\approx 1.1 \lj$ for  $\fopa\lesssim 0.004$, rising to  about 1.8 $\lj$ at $\fopa=1$.  {This is a consequence of the fact that at a higher opacity, the planet's interior remains longer at a higher entropy due to the stronger insulation. At early times, the luminosity of the high opacity case is in contrast lower, but the more efficient cooling of the low opacity case makes that its luminosity eventually falls below the one of the high opacity case (the post-formation entropies are nearly identical).} {This result indicates that using one constant $\fopa=0.003$ during both the formation and evolution phase is an acceptable simplification. {This is confirmed in Sect. \ref{sect:MRR} also in population synthesis calculations.}
 
\subsection{Relationship of $\mz$ and $\mxy$ in phase II}\label{sect:relationmzmxyphaseII}
The simulations with different $\fopa$ show that the envelope mass as function of time during  phase II depends on the opacity. It is however also interesting to investigate the envelope mass  as a function of core mass for different opacities during phase II, because in this phase, there is a special relationship between core, envelope  and total mass: If a planet of mass $M$ at a semimajor axis $a$ from a star of mass $\mstar$ has accreted all planetesimals (surface density  $\Sigma_{\rm P}$) within its feeding zone of half width $B_{L}$ times the Hill sphere radius, it must hold 
\beq
\mz=4 \pi B_{L} a^{2} \Sigma_{\rm P} \left(\frac{M}{3 \mstar}\right)^{1/3}.
\eeq
If the total mass of the planet is equal to the core mass ($M=\mz$), this leads to the well-known isolation mass (Lissauer \cite{lissauer1993})
\beq\label{eq:miso2}
\miso^{2}=\frac{(4 \pi B_{L} a^{2} \Sigma_{\rm P})^{3}}{(3 \mstar)}. 
\eeq
In a more general case, $M=\mz+\mxy$. From the above equations we can then write
\beq\label{eq:mmisomz}
\mz=\left(M \miso^{2}\right)^{1/3},
\eeq
which leads to 
\beq\label{eq:mzmisop2}
\mxy=\frac{\mz^{3}}{\miso^{2}}-\mz.
\eeq
This equation shows that for a given $\miso$, there is only one $\mxy$ for some core mass during phase II.  The equation in particular  means that in phase II, the envelope mass for a given core and isolation mass is independent of $\kappa$. At crossover for example, $\mz=\mxy=\sqrt{2}\miso$ independently of opacity, as already noted by P96, again if there was sufficient time to accrete all planetesimals in the feeding zone\footnote{ {The timescale of gas accretion and the associated expansion of the feeding zone becomes very short at very low $\fopa$, so that some planetesimals in the feeding zone are not accreted. This is the explanation why in the $\sigmas0=10$ g/cm$^{2}$ simulation, the crossover mass at $\fopa=10^{-5}$ is about 2 $\mearth$ smaller  than for $\fopa=1$.}}. 

\begin{figure}
\begin{center}
\includegraphics[width=0.9\columnwidth]{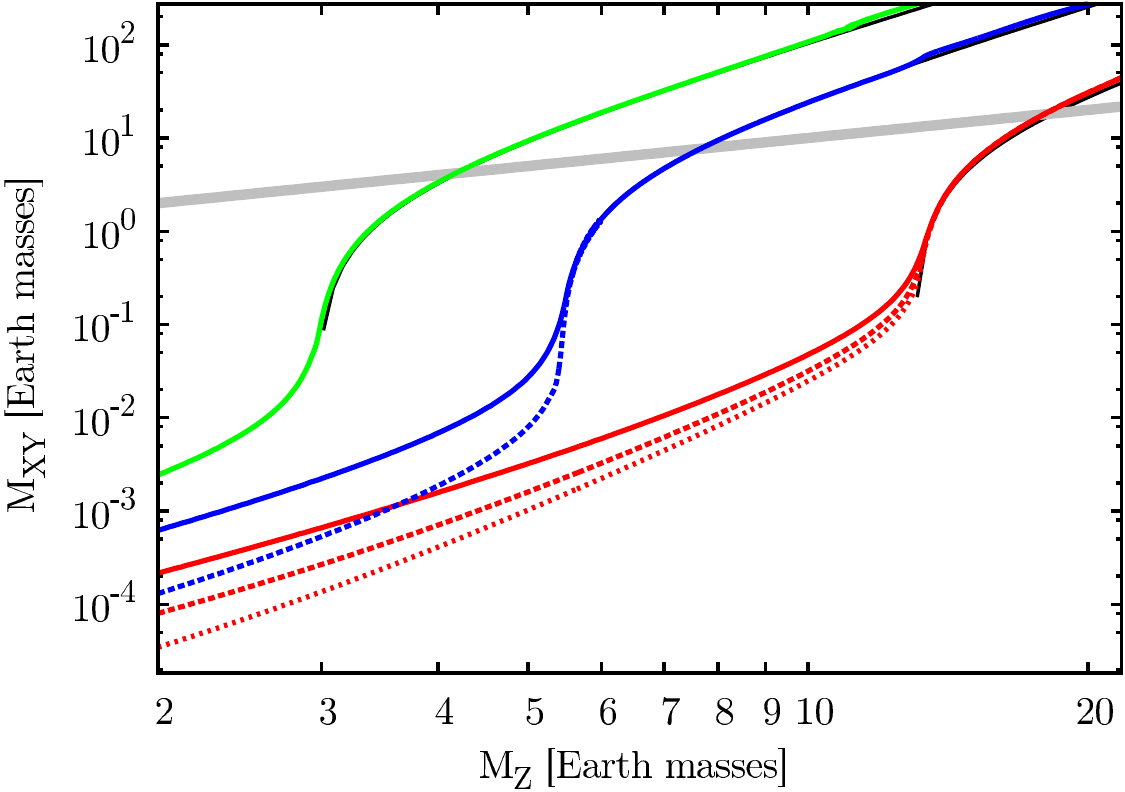}
\caption{Envelope mass $\mxy$ as a function of core mass $\mz$ for in situ simulations. The green, blue, and red lines correspond to isolation masses of about 3, 5.5, and 13 $\mearth$, respectively. For the latter two, different $\fopa$ are shown, with smaller values of $\fopa$ leading to larger $\mxy$ in phase I. The two blue lines show $\fopa$=0.003 and 0.1, while the three red lines shown $\fopa$=1, 0.1, and 0.01. In phase II, the $\fopa$ lines converge. Thin black lines (partially covered) show  Eq. \ref{eq:mzmisop2}. The thick gray line shows the crossover point where $\mxy=\mz$.}\label{fig:mxymxsigma}
\end{center}
\end{figure}

To illustrate this, Figure \ref{fig:mxymxsigma} shows $\mxy$ as function of $\mz$ for in situ simulations as in Sect. \ref{sect:determinationfopa}. Planets evolve in time along the lines, starting on the left. During phase I, the envelope mass is small and depends on $\fopa$. But once planets have reached the isolation mass and phase II begins, the envelope mass  becomes independent of $\fopa$, so that the lines merge. The relationship of core and envelope mass is now given by  Eq. \ref{eq:mzmisop2} which is also shown in the plot using the corresponding $\miso$. In these simulations without ejection, it  remains valid also in the gas runaway accretion phase, which sets in after the crossover point.  

But how does the temporal evolution come in this picture, i.e., what determines how quickly the planet goes through these different  stages if $\mz(\mxy)$ is independent of $\kappa$? We know that the gas accretion timescale depends on the opacity. The mechanism can be understood when considering the mass of the envelope as a function of luminosity $L$ and opacity. Both the fully radiative envelope of Stevenson (\cite{stevenson1982}), but also the generalization of Rafikov (\cite{rafikov2006}) to envelopes that are convective in the interior and radiative at the surface (which is the kind of envelope we are dealing with here) have the property that $\mxy\propto 1/(L  \kappa)$.

This means that  {if  $L$ is larger, then $\kappa$ must be smaller} to get the same $\mxy$, as required for a given isolation and core mass in phase II. But the luminosity $L$ (in phase II $\approx L_{\rm Z}$, the core luminosity) is proportional to the accretion rate of planetesimals $\dot{M}_{\rm Z}$ (which is in turn proportional also to  $\dot{M}_{\rm XY}$, see P96 and Appendix \ref{sect:semianalyticalsolution}), so that a smaller $\kappa$  {goes along with} a larger gas and solid accretion rate, and this means a shorter growth timescale.

The special relation of core and envelope mass  together with a  parametrization of the envelope growth timescale makes it possible to derive semi-analytical solutions for the core and envelope mass in phase II. This is presented in Appendix \ref{sect:semianalyticalsolution} while in Appendix \ref{sect:paramstkh} these semi-analytical solutions are used to derive the parameters $k$ and $p$ by comparison with numerical results.
 
\section{Observational  {consequences} of different $\fopa$}\label{sect:obsconstr}
Up to this point, we have determined in this paper  {a nominal grain opacity reduction factor. We now turn to the question if different $\fopa$ lead to potentially observable consequences.}

Observational evidence for low envelope opacities could come from the bulk composition of planets with primordial H/He, and in particular from transiting low-mass low-density planets like the planets around Kepler-11  (Lissauer et al. \cite{lissauerfabrycky2011}). The amount of primordial H/He these low-mass planets can accrete {during the finite lifetime of a protoplanetary disk} is a  function of the grain opacity. A complication is that a significant spread of  primordial H/He envelope mass for a given planet mass is expected, as the envelope masses depend besides the opacity also on, e.g., the core luminosity, and therefore the individual formation history of a planet. This means that  a statistically large enough sample of exoplanets is necessary for the comparisons.

In order to establish possible statistical consequences of grain opacity on observable quantities (e.g., the mass-radius relationship) we have conducted planetary population synthesis calculations  with different values of $\fopa$.  {Regarding the conclusions that can be drawn from the comparisons, one should keep in mind the limitations discussed in Sect. \ref{sect:generality}, in particular the likely lack of generality of one general $\fopa$.} 

\subsection{Population synthesis calculations}\label{populationsynthesis}
We calculated four populations of synthetic planets around solar-like stars varying the value of $\fopa$.  The main assumptions for  the populations are listed in Table \ref{tab:fourpops}.  Otherwise, the model and probability distributions are identical as in Mordasini et al. (\cite{mordasinialibert2012}). The reader is referred to this paper for details. The first three populations were calculated with the updated prescription for non-isothermal migration introduced in Mordasini et al. (\cite{mordasinialibert2011}). These three populations only differ by $\fopa$ which was set to 0 (grain-free gas, only molecular and atomic opacities from Freedman et al. \cite{freedmanmarley2008}), 1 (full ISM grain opacity from Bell \& Lin \cite{belllin1994}), and 0.003, the value best fitting the MBPL10 results as derived in Sect. \ref{sect:finalresfopa}. The $\fopa=0.003$ synthesis is the nominal case, while the other two syntheses represent limiting cases.  An additional fourth population was calculated again with $\fopa=0.003$, but this population differs in two other aspects: First, orbital migration was completely neglected, so that planets form in situ. This allows to see the differential impact of a migration as a mechanism that is currently not fully understood  {(e.g., Dittkrist et al. \cite{dittkristmordasini2014})}. It also allows to study a situation that is closer to the semi-analytical solutions of  Appendix \ref{sect:semiwcore}. Second, the accretion of planetesimals by the protoplanet in the  disk limited gas accretion phase was set to zero ($\dot{M}_{\rm Z,run}=0$). In the nominal case where $\dot{M}_{\rm Z,run}\ne0$ the core continues to accrete planetesimals also after the gas accretion rate due to the contraction of the envelope has exceeded the limiting rate set by the disk (see Mordasini et al. \cite{mordasinialibert2011}). The amount of planetesimals that get accreted in this detached phase could be significant due to the rapid expansion of the feeding zone. Hubickyj et al. (\cite{hubickyjbodenheimer2005}) or Lissauer et al. (\cite{lissauerhubickyj2009}) in contrast assume that the core accretion rate falls to zero once the planet is detached. With this synthesis, we study the global effects of this assumption.

\begin{table}
\caption{Settings for the four synthetic  populations.}\label{tab:fourpops}
\begin{center}
\begin{tabular}{lccc}
\hline\hline
Population     & $\fopa$         & Migration &  $\dot{M}_{\rm Z,run}$ \\\hline 
Grain free      & 0     &  yes & $\ne0$\\                                             
Nominal       & 0.003     &  yes & $\ne0$\\  
Full ISM   & 1     &  yes & $\ne0$\\  
In situ       & 0.003 & no & $=0$ \\  \hline
\end{tabular}
\end{center}
\end{table}

\section{Envelope mass as a function of core mass}\label{sect:menveofmcore}
In Appendix \ref{sect:mxymzsemi}, we discuss the envelope mass as a function of core mass as predicted by simple semi-analytical models. Figure \ref{fig:mcoremenvepop}  shows the envelope mass as a function of core mass as found numerically  in the  four populations. Planets with a semimajor axis of $0.11\leq a/$AU$\leq 5 $ are included. For comparison, we plot for the subcritical planets a dotted line showing $\mxy=10^{-b} \mz^{p+1} t$, corresponding to the simple approximation of Eq. \ref{eq:mxymcorenos} which is strictly speaking not applicable as it assumes $\dot{M}_{\rm Z}=0$ (and $\mxy\ll\mz$).  The exponent was set to $p=1.70$ which corresponds to the mean for a core mass of roughly 6 $\mearth$ as given in Table \ref{tab:paramstkh}, while  $b$ was set to 7.87. This corresponds to the approximate result for $\fopa=0.003$, again for $\mz\sim6\mearth$. The time $t$ was set to 2 Myrs because  the mean lifetime of the synthetic disks is 2.2 Myrs. The dashed lines simply indicate scalings of $\mxy$ as $\mz^{q}$ to guide the eye, where $q$ is 2, 3, and 4. The absolute value was adjusted to fit the $\mz$-$\mxy$ relation of low mass planets in the in situ population.   

\begin{figure*} 
\begin{minipage}{0.47\textwidth}
\includegraphics[width=1\textwidth]{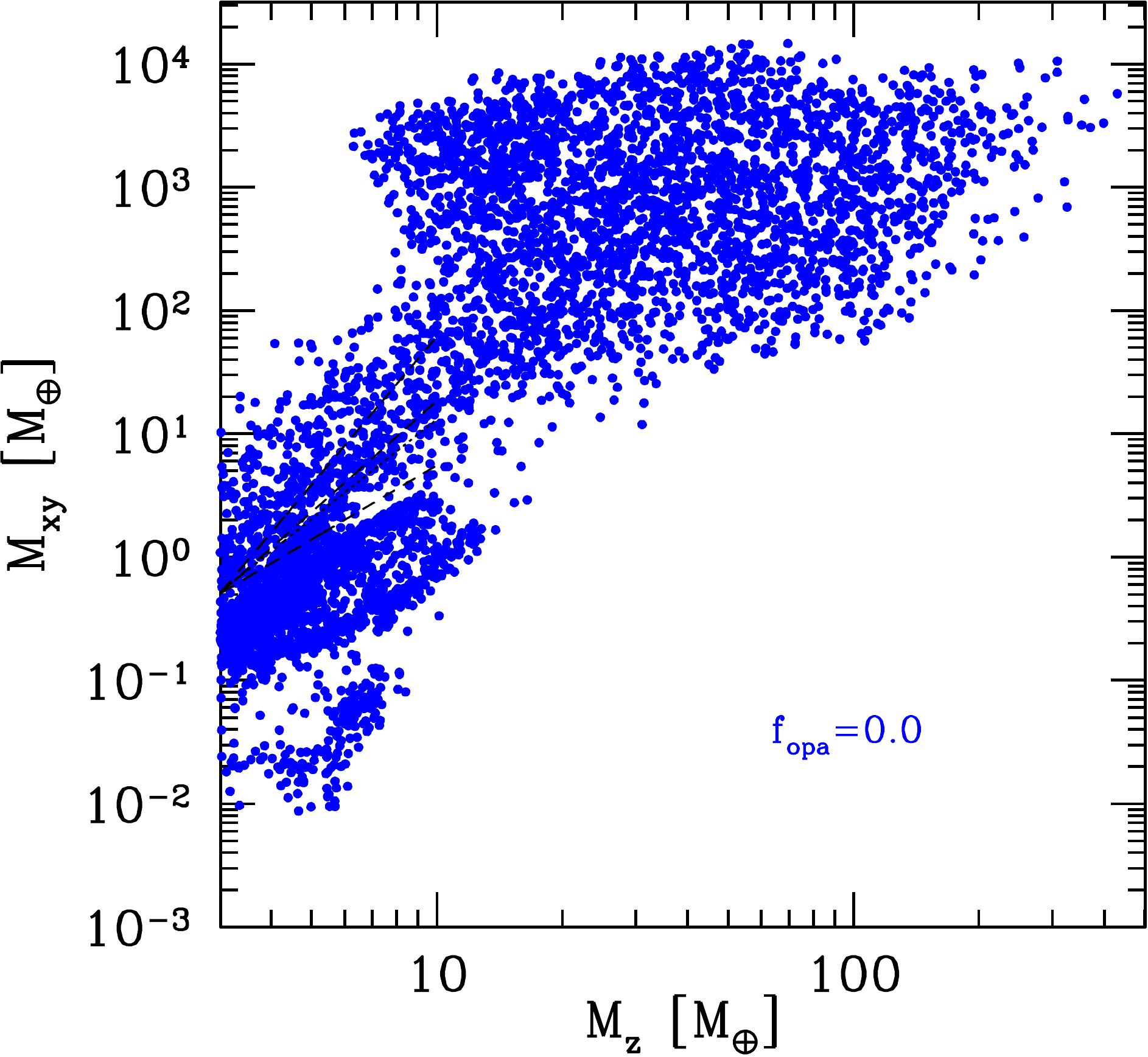}
\includegraphics[width=1\textwidth]{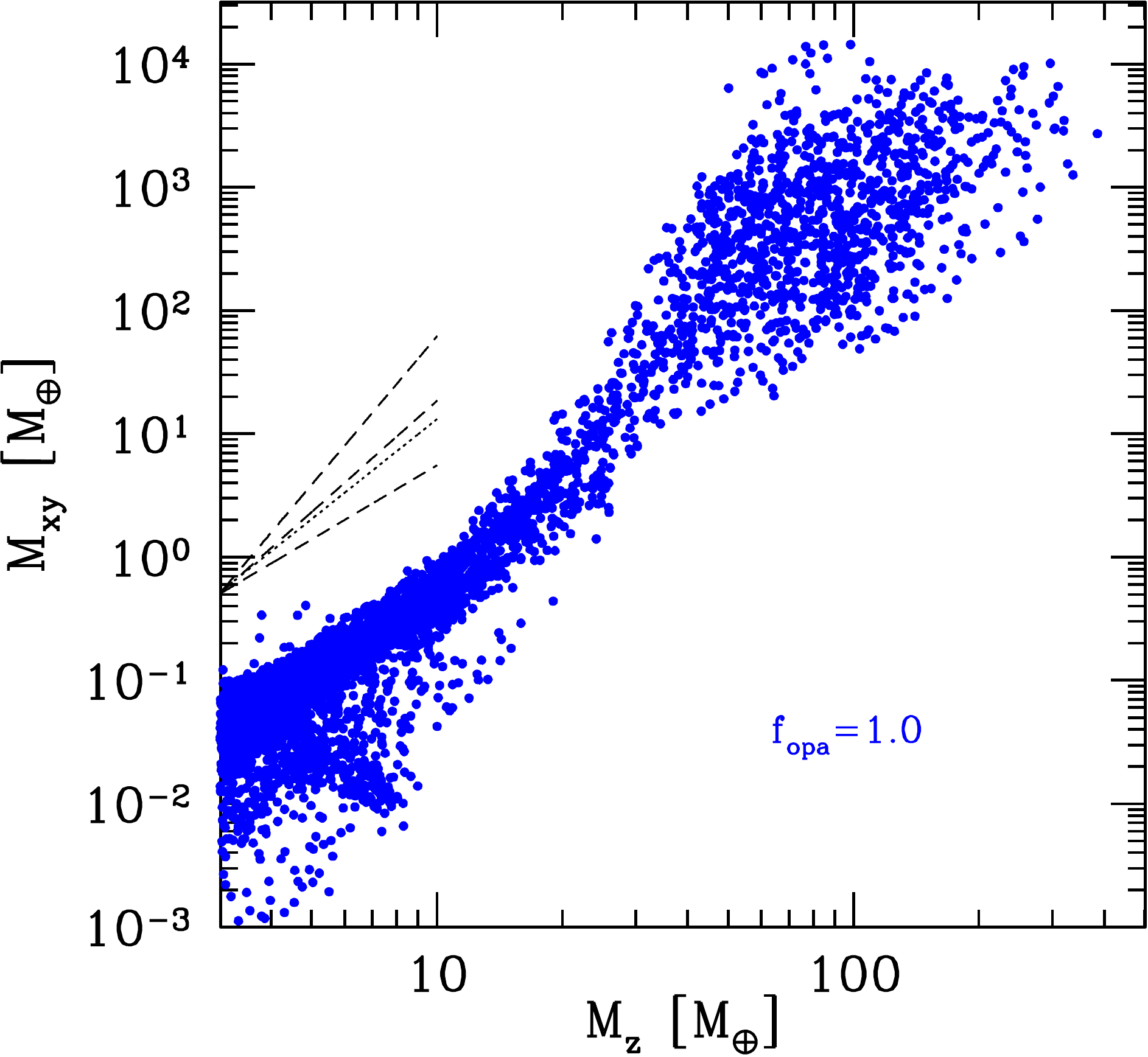}
\end{minipage}
          \hspace{0.5cm}
\begin{minipage}{0.47\textwidth}
\includegraphics[width=1\textwidth]{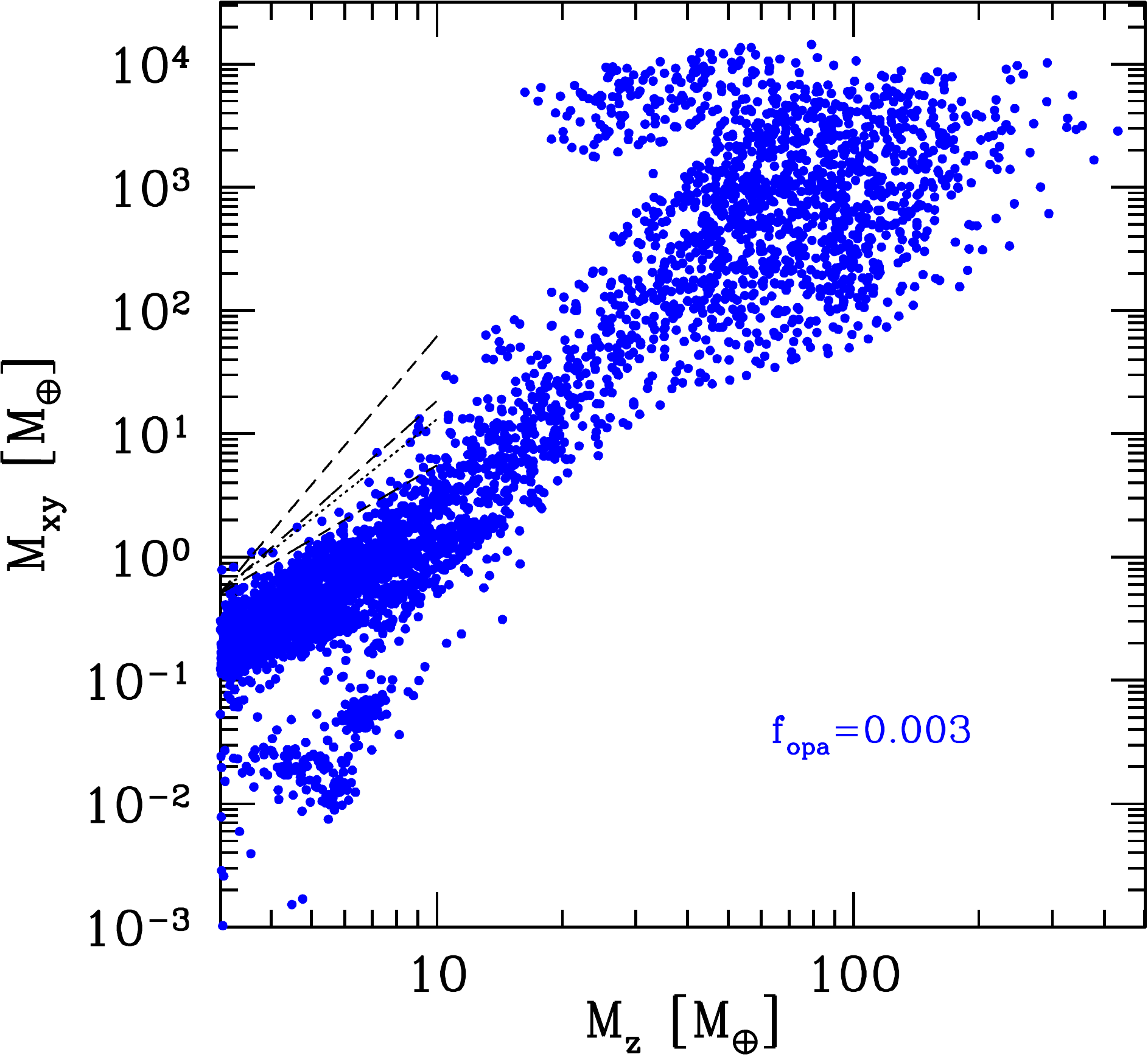}
\includegraphics[width=1\textwidth]{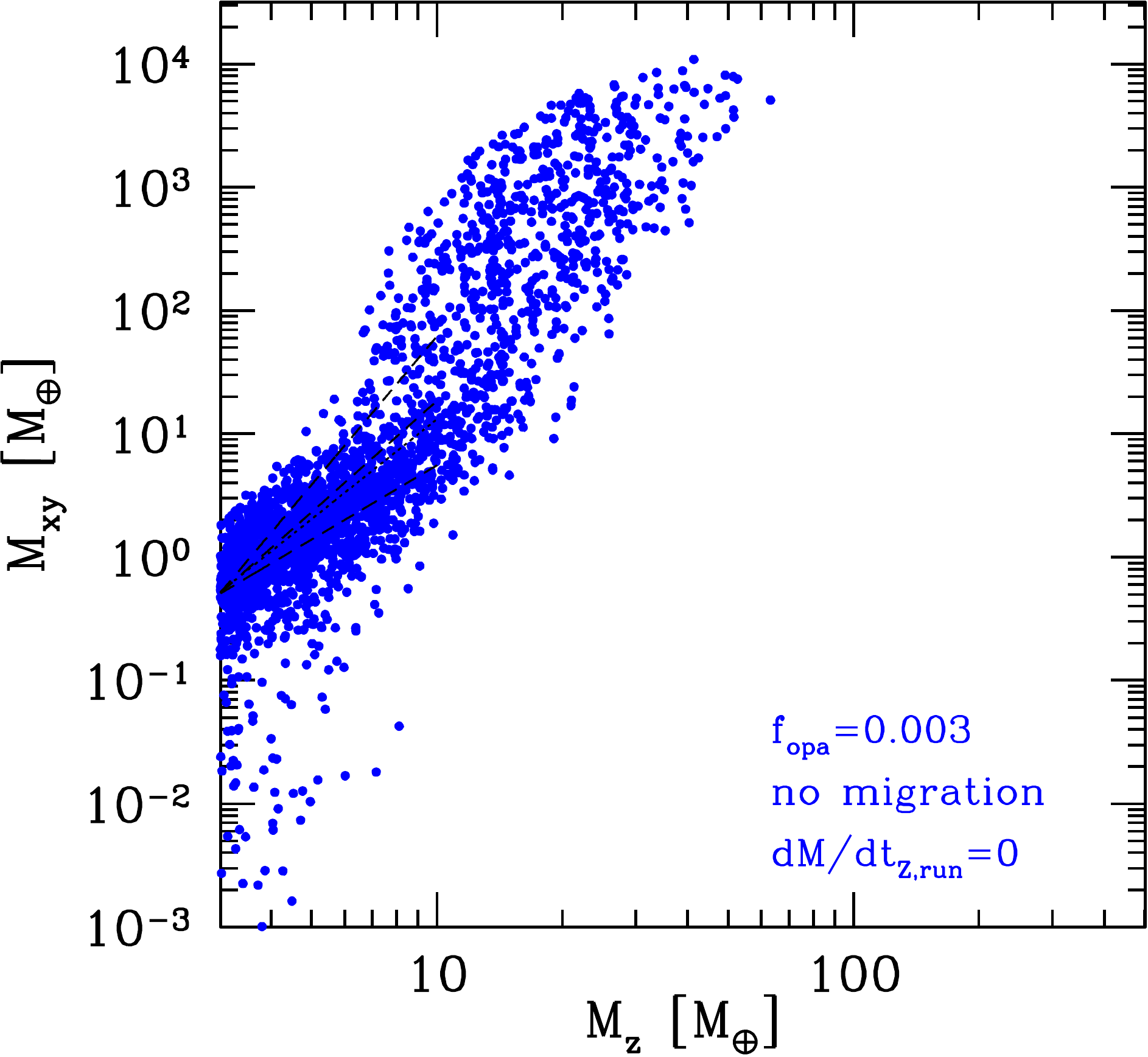}
\end{minipage}

\caption{H/He envelope mass $\mxy$ as a function of heavy element mass $\mz$ for synthetic planets with semimajor axes 0.11$\leq a/AU\leq$5. The populations in the  top panels and the bottom left panel only differ by the value of $\fopa$ as indicated in the plots. The populations were calculated assuming full non-isothermal type I migration rates. The population in the bottom right was calculated with  $\fopa$=0.003, but without orbital migration and assuming $\dot{M}_{\rm Z,run}$=0. The  lines show simple analytical estimates and scalings  (see text).}\label{fig:mcoremenvepop}  
 \end{figure*}
 
\subsection{Low-mass planets} 
We see that $\fopa$ leaves a clear imprint in the $\mz$-$\mxy$ relation. For low-mass planets, we see that for a fixed core mass, the typical envelope mass increases with decreasing opacity, as expected. When  going from $\fopa=1$ to $\fopa=0.003$, the envelope masses approximately increase by a factor 4. For a 10 $\mearth$ core for example, the typical envelope mass  in the syntheses including orbital migration is roughly 0.5, 2, and $>10$ $\mearth$ for $\fopa$=1, 0.003, and 0, respectively. In the $\fopa=0$ population, at a core mass of  10 $\mearth$ there are both subcritical planets and giants. There is a large spread in possible envelope masses for a given core mass that covers one to two and sometimes even more orders of magnitudes. These are imprints of different formation histories, which lead in particular to different core luminosities and therefore envelope masses. Also the semimajor axes and  the associated boundary conditions for the envelope given by the nebula affect the envelope mass for some planets (Rafikov \cite{rafikov2006}).

In the syntheses with orbital migration, there are more sub-structures in the  $\mz$-$\mxy$ relation than in the in situ simulation. This is due to the more complex formation tracks with orbital migration.  The tracks depend in particular on special convergence zones in the disk into which planets can get captured (e.g., Lyra et al. \cite{lyrapaardekooper2010}, Dittkrist et al. \cite{dittkristmordasini2014}). A low-mass planet captured in a convergence zone migrates at a lower rate, leading to a lower planetesimal accretion rate and core luminosity relative to a migrating planet that is not in a convergence zone.  But in general, in simulations with orbital migration, the core luminosity is higher than in the in situ case for the prescription of planetesimal accretion used in the model (taken from P96) as it leads to high $\dot{M}_{\rm Z}$. This higher core luminosity results in lower envelope masses. We indeed see that for the synthesis with $\fopa=0.003$ and migration, the envelope masses are roughly a factor five lower than in the population neglecting orbital migration that is also calculated with  $\fopa=0.003$. A number of low-mass planets in this population have clearly lower envelope masses $\lesssim 0.1 \mearth$. This are planets in phase I at the moment when the disk disappears that therefore have high core luminosities. 

\subsection{Giant planets and critical core mass}\label{giantsandmcrit}
It is known  from hydrostatic calculations (e.g., Mizuno et al. \cite{mizunoetal1978}) that the critical core mass decreases with decreasing opacity (but not in P96 type evolutionary calculations for a given isolation mass\footnote{This is, however, not contradictory: in the evolutionary calculations, the gas accretion timescale becomes shorter with decreasing opacity, so that again an initial conditions with a smaller core mass can trigger gas runaway gas accretion during the finite lifetime of a disk.}, cf. Sect. \ref{sect:relationmzmxyphaseII}) . The syntheses make it possible to study for a large number of evolutionary calculations how the minimal core mass $M_{\rm Z,min}$ required to form a giant planet depends on $\fopa$. From Hori \& Ikoma (\cite{horiikoma2010})  {it is known} that for a grain-free envelope ($\fopa=0$) and $\dot{M}_{\rm Z}$=0, a core of just 1.7 $\mearth$ can capture gas on a timescale that allows it to become a giant planet during the typical lifetime of a protoplanetary nebula. 

Adopting as criterion for classification as giant planet an envelope mass of at least 100 $\mearth$, we find the following $M_{\rm Z,min}$ in the syntheses with migration: $M_{\rm Z,min}$=6, 16 and 29 $\mearth$ for $\fopa$=0, 0.003, and 1, respectively. These values are the total final amount of heavy elements,  including the planetesimals accreted during the gas runaway accretion phase  {which likely will end up mixed into the gaseous envelope}. The actual critical core mass when gas runaway accretion is triggered is therefore smaller. In the synthesis without migration and $\fopa=0.003$,  $M_{\rm Z,min}$ is about 7$\mearth$.   An alternative definition of a gaseous planet is that the crossover point was reached during the presence of the disk, i.e., that $\mxy=\mz$. The minimal core masses that fulfill this definition are 2, 9 and 25 $\mearth$ for $\fopa$=0, 0.003, and 1, respectively, for the simulations with migration, and 4 $\mearth$ for the simulation without. This shows that at a low opacity, planets of a very small total mass can be gas dominated.

The value of $M_{\rm Z,min}$=6 $\mearth$ for the $\fopa$=0 case is higher than the result of Hori \& Ikoma (\cite{horiikoma2010}). This is expected because they considered isolated cores that do not accrete planetesimals, so that the luminosity is minimal. In the syntheses, the core luminosities are higher than in Hori \& Ikoma (\cite{horiikoma2010}). Due to the lower typical core luminosity without migration, the $M_{\rm Z,min}=7 \mearth$ found in the in situ population that assumes $\fopa=0.003$ approaches  the result for $\fopa=0$, but including migration (6 $\mearth$).  We thus see that there is a degeneracy between luminosity and opacity in controlling the envelope gas masses as mentioned by Hori \& Ikoma (\cite{horiikoma2010}).

\subsection{Maximal core mass}\label{sect:maximalcoremass}
Figure \ref{fig:mcoremenvepop} can be used to study the maximal core mass $M_{\rm Z,max}$. In the case where the core does not grow in the detached phase (bottom right panel), $M_{\rm Z,max}\approx$60$\mearth$. This value depends on the largest semimajor axis that is considered. If synthetic planets at all semimajor axis are considered instead of only those inside of 5 AU, $M_{\rm Z,max}$$\approx$100$\mearth$. This increase comes from the increase of the isolation mass with semimajor axis. In another synthesis with $\dot{M}_{\rm Z,run}$=0, but  normal orbital migration, $M_{\rm Z,max}$$\approx$100$\mearth$. With migration, planets reaching such a heavy element mass are also found inside of 5 AU, showing that orbital migration moves planets with very massive cores closer to the star.

For the three simulations where the core continues to accrete in the detached phase, the maximal mass is higher. Here, core masses can exceed 200$\mearth$, and there are planets with extreme cores  ($M_{\rm Z,max}$$\approx$400$\mearth$). This value does not increase if we include all semimajor axes instead of $a\leq5$ AU, an effect due to orbital migration. The measurement of mass and radius for some transiting exoplanets  indicates that they contain several 100 $\mearth$ of solids, like for example $\sim$300-600 (200-500) $\mearth$ of water (rock) in HAT-P-2b (Baraffe et al. \cite{baraffechabrier2008}). These extreme cores can only form in solid rich disk, i.e., when both the disk gas mass and the dust-to-gas ratio (i.e., [Fe/H]) come from the upper end of the probability distributions. The product of the dust-to-gas ratio times the gas surface density (which is proportional to the disk gas mass) sets the surface density of the planetesimals, and this controls together with the initial position of the embryo the maximal mass to which a core can grow. 

Rapid core growth the outer parts of the disk requires planetesimal random velocities to stay low as in our simulations that follow the eccentricity and inclination prescription of P96. Models of planetesimal dynamics (like Ida \& Makino \cite{idamakino1993}; Ormel \& Kobayashi \cite{ormelkobayashi2012};  Fortier et al. \cite{fortieralibert2013}) show that this  is difficult to achieve. In this sense, our model has a tendency to predict too heavy element masses.

\subsubsection{Factors affecting $\dot{M}_{\rm Z,run}$: the capture radius}\label{sect:factorsmdotzrun}
Figure \ref{fig:mcoremenvepop}  makes clear that the maximal core mass of giant planets depends in a significant way on the amount of solids that can be accreted in the detached phase with possibly observable consequences (Sect. \ref{sect:MRR}). Physically, the magnitude of $\dot{M}_{\rm Z,run}$ depends on the growth timescale of the planet, the timescale of gap opening in the planetesimals disk (and therefore the planetesimal size), as well as the capture radius of the planet and therefore its contraction timescale.  In a detailed study, Zhou \& Lin (\cite{zhoulin2007}) directly integrated the trajectories of  a planetesimal swarm under the action of gas and tidal drag due to the protoplanetary disk, and the gravity of the star and a growing protoplanet (see also Shiraishi \& Ida \cite{shiraishiida2008}). They studied what fraction of the planetesimals in the feeding zone get accreted during the gas runaway phase of the protoplanet (in the synthesis, we consider the limiting cases of 0 and 100\%). The mass growth of the protoplanet was parametrized, and the actual capture radius for planetesimals was not calculated. Instead, they simply used mean planetary densities of 1 g/cm$^{3}$ similar to Jupiter nowadays, and 0.125  g/cm$^{3}$ as typical for a young Jovian planet that has a radius twice as large as Jupiter.  For the  1 and 0.125 g/cm$^{3}$ cases, about 20\% and 28\% of the planetesimals got accreted. 

While this indicates a rather weak dependence on the capture radius (Shiraishi \& Ida \cite{shiraishiida2008} derive  a dependency $\propto R_{\rm capt}^{1/2}$), one should keep in mind that the capture radius can be very large during the formation phase. This is illustrated in Fig. \ref{fig:rcaptmpla}. The plot shows the capture radius for 100 km planetesimal as a function of mass for a growing giant planet in a simulation with the same initial conditions a J1 in P96. The planet grows in situ at 5.2 AU, and its final mass is 1 $\mj$.  The capture radius is found by directly integrating the trajectories of planetesimals in the envelope (Podolak et al. \cite{podolakpollack1988}; Mordasini et al. \cite{mordasinialibert2005}).  

\begin{figure}
\begin{center}
\includegraphics[width=0.99\columnwidth]{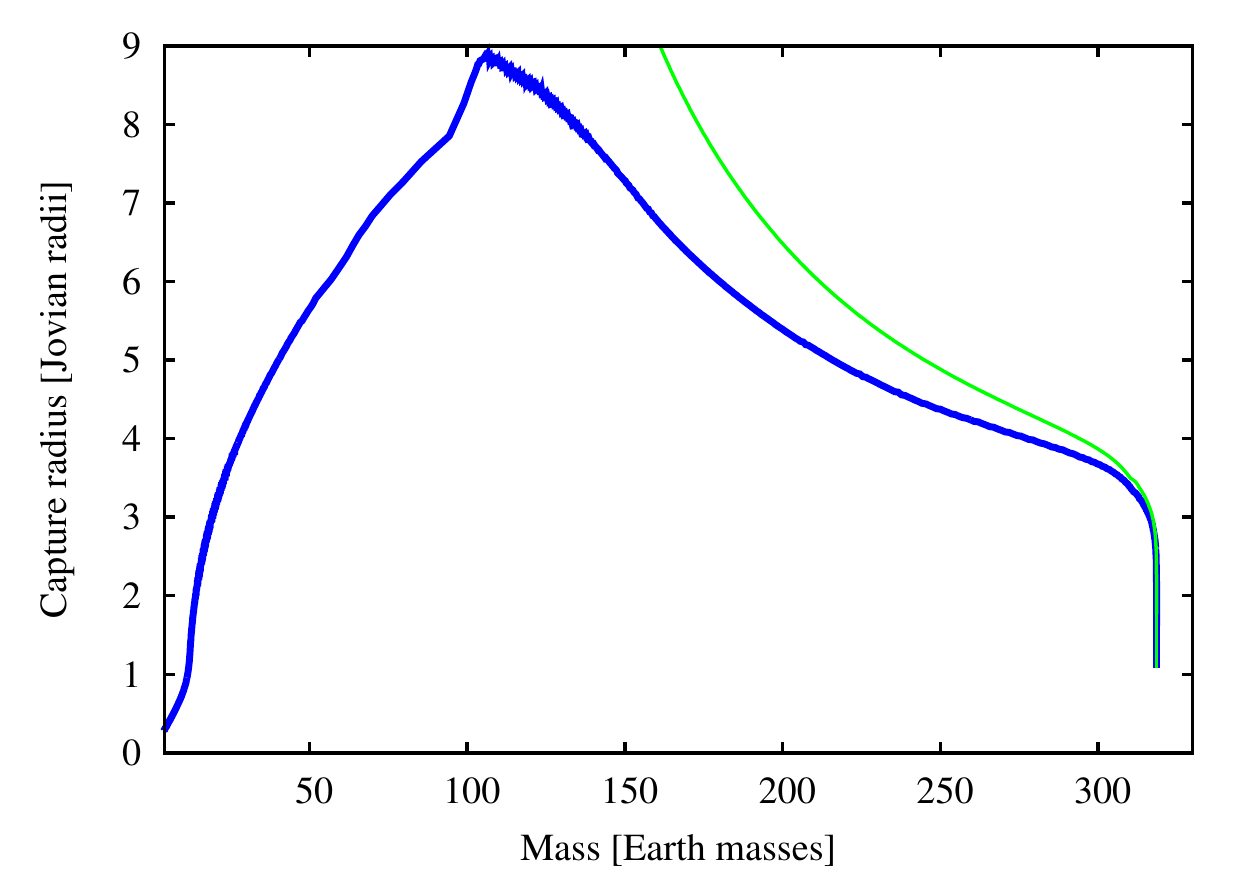}
\caption{Capture radius for 100 km planetesimals as a function of mass for a growing giant planet. The simulation uses the same initial conditions as the case J1 in P96. The capture radius is the thick blue line, while the green line is the total radius.}\label{fig:rcaptmpla}
\end{center}
\end{figure}

At a low mass, the capture radius is small and equal to the core radius. It then increases as the mass of the envelope grows, reaching a maximum of about 9 $\rj$ at the moment when the planet detaches from the nebula. At this moment, the mass of the protoplanet is about 110 $\mearth$. During the remaining runaway growth phase, it gradually decreases to about 4 $\rj$. We thus see that the capture radius might be considerably larger than 2 $\rj$. Using  the scaling with $R_{\rm capt}^{1/2}$, we crudely estimate with a mean capture radius of $\sim$6 $\rj$ that $\sim$50\% percent of the planetesimals contained in the feeding zone might get accreted. Then, the maximal core mass  $M_{\rm Z,max}$ would come to lie about halfway between the two limiting cases of the syntheses. Clearly, these are only  rough estimates, and dedicated simulations are necessary that concurrently model the planetesimal swarm and the evolution of the internal structure of the planet. 

We see that in principle, the maximal core mass could also constrains the contraction timescale of giant planets formed by core accretion. Helled \& Bodenheimer (\cite{helledbodenheimer2011}) made  similar considerations for the gravitational instability mechanism. They find that if grain growth is included, the timescale until collapse of the first core is much reduced. The amount of planetesimals that can be captured is much reduced, too, as the planet has a large capture radius only while it is extended. One finds an equivalent effect for core accretion: at low $\fopa$, the accretion of planetesimals in the detached phase is slightly reduced compared to the full opacity case. The reason is the faster decrease of the capture radius. For a 10 $\mj$ planet we  found for example that the timescale on which the capture radius decreases from 7 $\rj$ to 2 $\rj$ is roughly 84\ 000 and 37\ 000 years at $\fopa$=1 and 0.003, respectively. The impact on the final core mass is however very small (difference of a few 0.1 $\mearth$). This is because the difference in $R_{\rm capt}$ becomes only important at a moment when most of the planetesimal accretion has already ended.  
 
\section{Mass-radius relationship as a function of $\fopa$}\label{sect:MRR}
\begin{figure*}
\begin{minipage}{0.48\textwidth}
\includegraphics[width=1\textwidth]{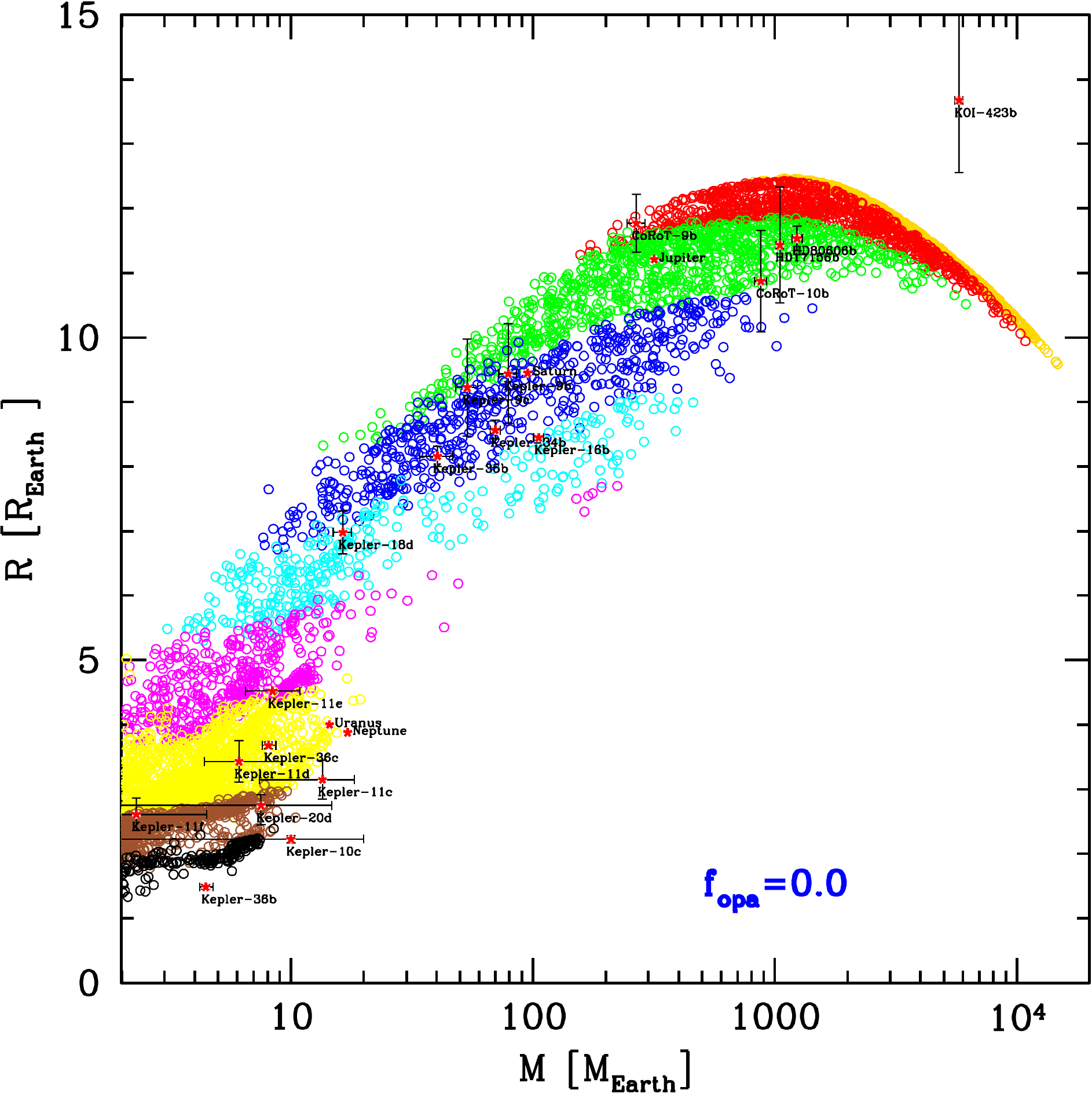}
\includegraphics[width=1\textwidth]{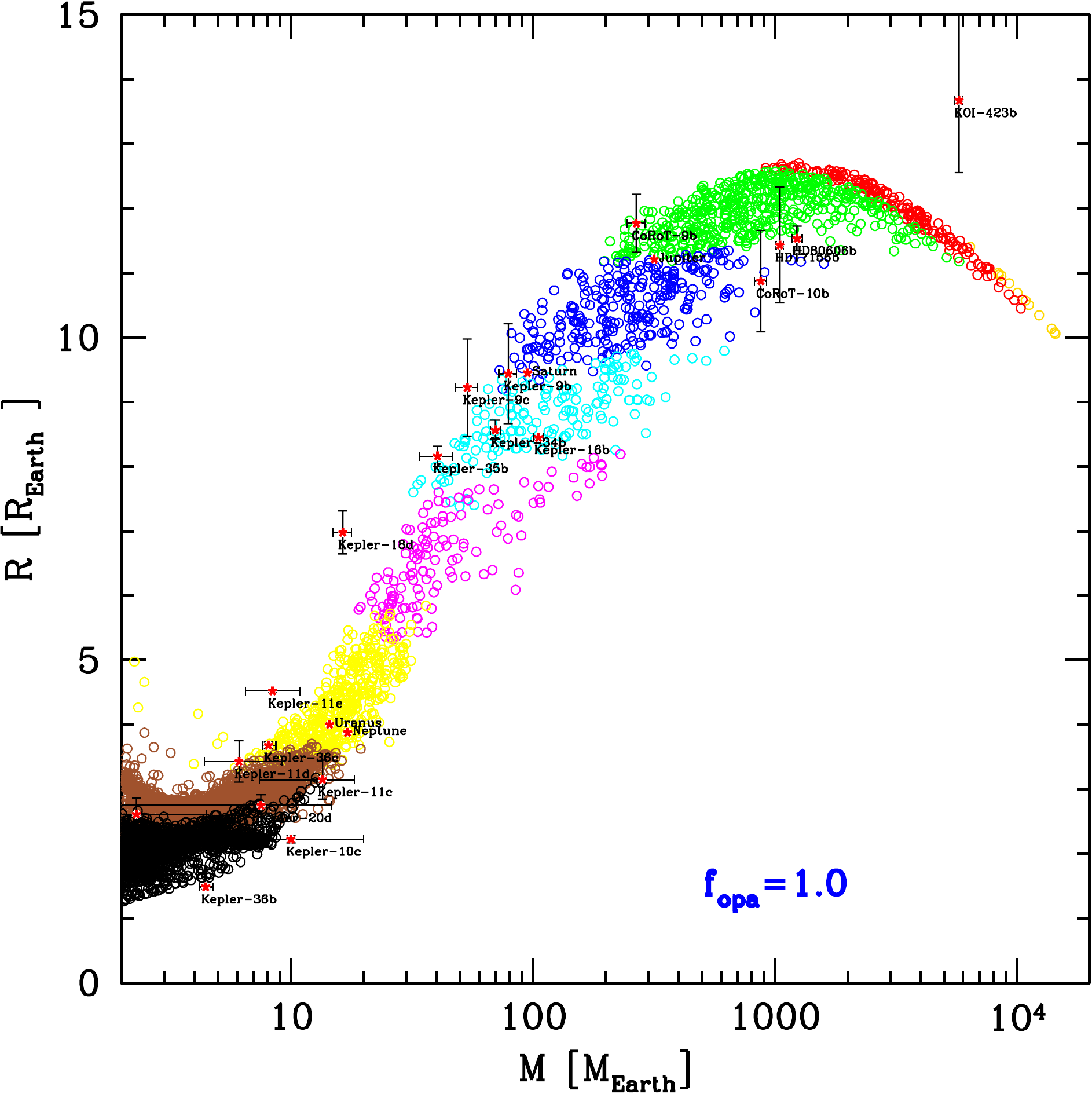}
\end{minipage}
\hfill
\begin{minipage}{0.48\textwidth}
\includegraphics[width=1\textwidth]{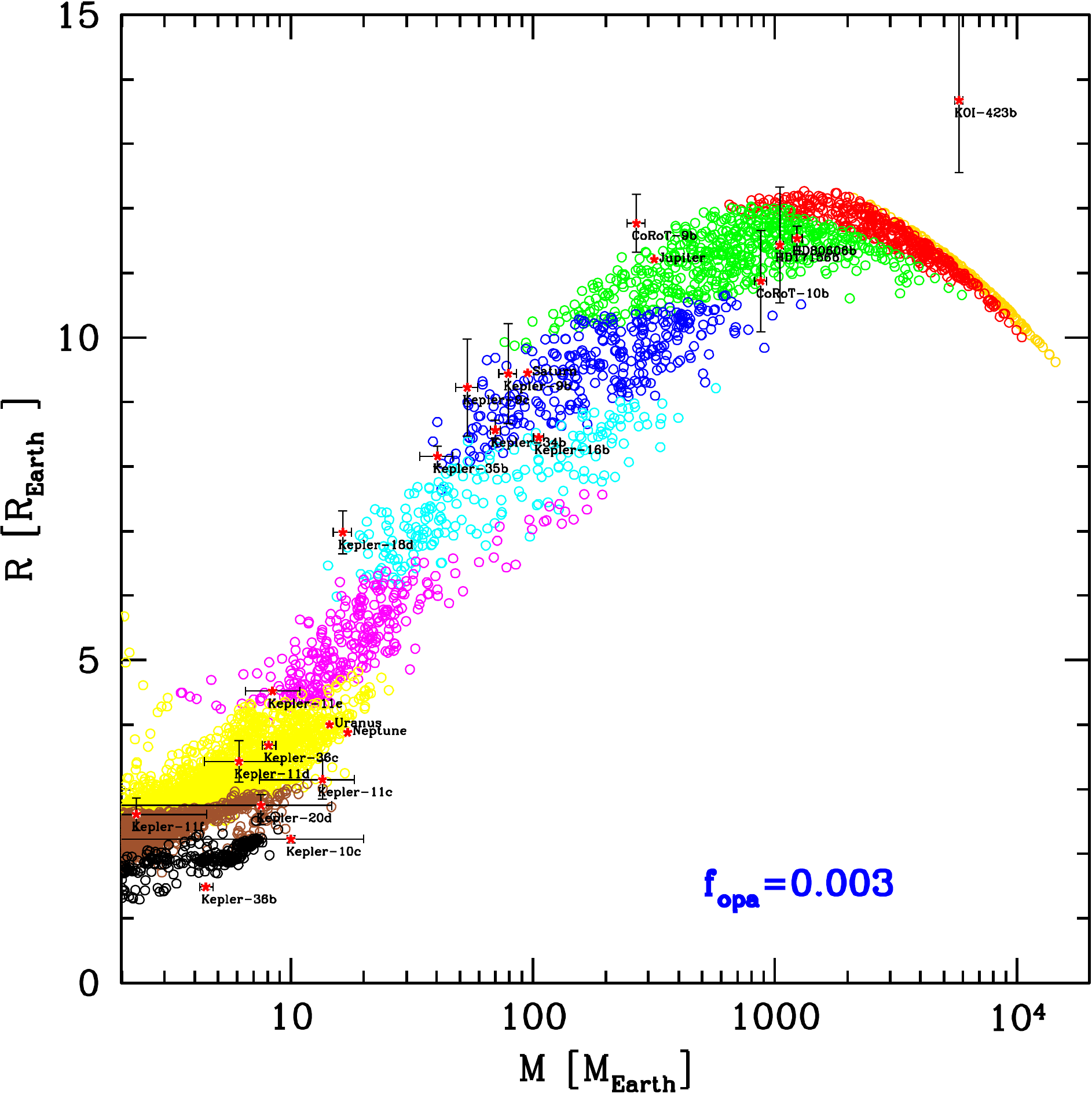}
\includegraphics[width=1\textwidth]{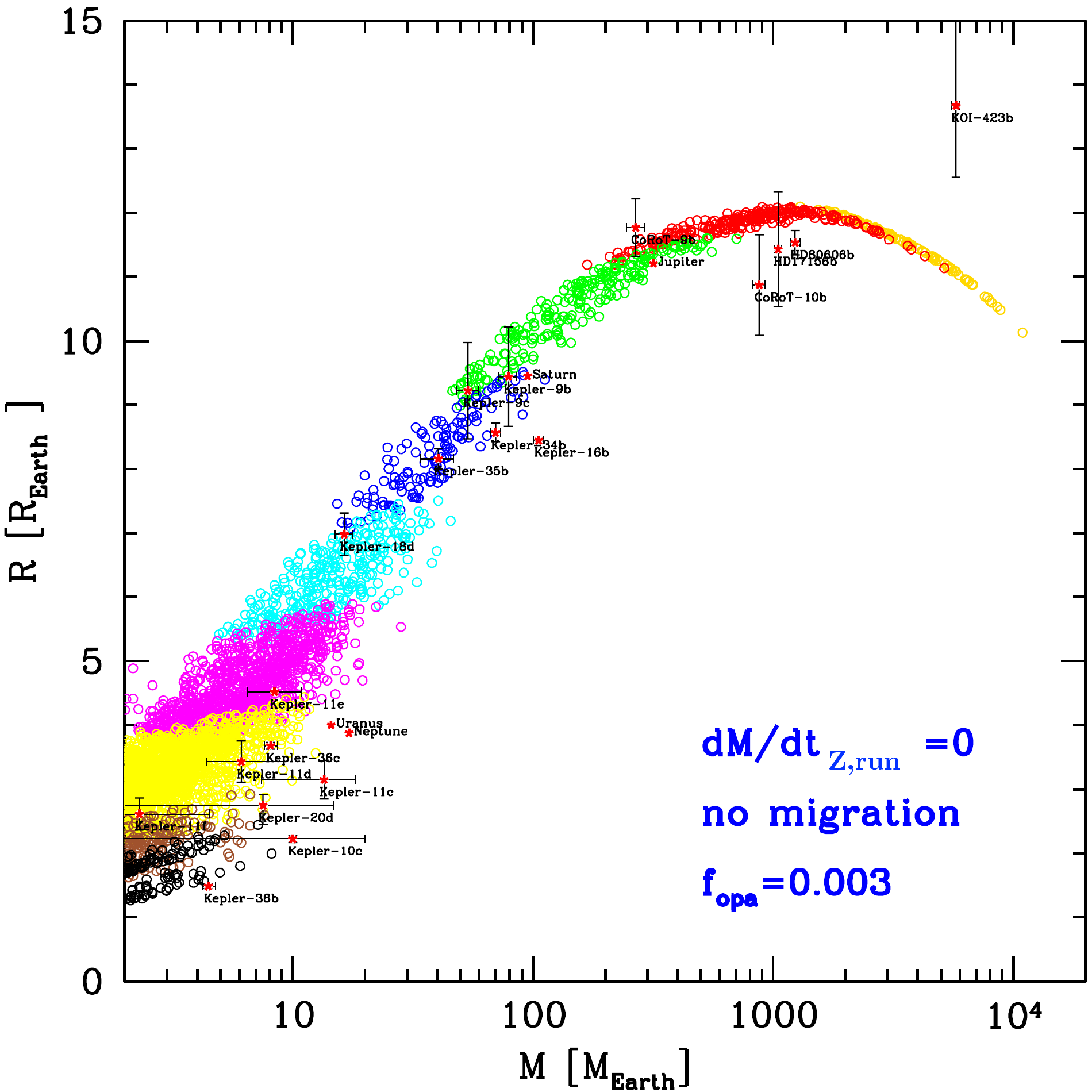}
\end{minipage}
\caption{Comparison of the mass-radius relationship of synthetic planets at an age of 5 Gyrs, the Solar System planets, and  extrasolar planets with $a>0.1$ AU and well constrained mass and radius. Synthetic planets have $0.1<a$/AU$<$5.  The colors show the fraction of heavy elements $Z=\mz/M$ in a synthetic planet. Orange: $Z\leq1\%$. Red: $1<Z\leq5\%$. Green: $5<Z\leq20$\%. Blue: $20<Z\leq40$\%. Cyan: $40<Z\leq60$\%. Magenta: $60<Z\leq80$\%.  Yellow: $80<Z\leq95$\%.  Brown: $95<Z\leq99$\%. Black: $Z>99$\%.}\label{fig:mrr}  
 \end{figure*}
 
The last section shows that $\fopa$ has important consequences for the bulk composition of planets. The envelope mass of low-mass planets  in particular increases by (very roughly speaking) one order of magnitude if $\kappa$ decreases from the full ISM to grain-free. Since the radius of solid low-mass planets increases rapidly even if only a tenuous H/He envelope is added  (e.g., Adams et al. \cite{adamsseager2008}, Mordasini et al. \cite{mordasinialibert2011}), different values of  $\fopa$ should leave imprints in the planetary mass-radius relationship.

To test that, we calculated the cooling and contraction of the synthetic planets after the formation phase has finished (i.e., after the dispersal of the protoplanetary disk), using the simple, but self-consistently coupled formation and evolution model of Mordasini et al. (\cite{mordasinialibert2011}).  Deuterium burning in massive companions is  included (Molli\`ere \& Mordasini \cite{mollieremordasini2012}), but not important at the age of interest here (5 Gyrs). Fig. \ref{fig:mrr} shows the resulting synthetic mass-radius relationships for the same  four populations as before. Planets with $0.11<a$/AU$<5$ are shown. Colors indicate the fraction of heavy elements in the planet, $Z=\mz/M$.
 
Compared to models of, e.g., Burrows et al. (\cite{burrowsmarley1997}), Fortney et al. (\cite{fortneyikoma2011}), or Baraffe et al. (\cite{baraffechabrier2008}), our evolutionary model is simplified in a number of aspects: Simple gray boundary conditions are used, special inflation effects are neglected, and all solids are put into the core, so that $\mz=\mcore$.  This might not be the case in reality since planetesimals can dissolve during impact. Our results for $\mz$ should therefore be associated with the total mass of heavy elements in a planet and not its core mass. Putting all solids into the core instead of homogeneous mixing with the envelope leads to radii which are somewhat bigger (Baraffe et al. \cite{baraffechabrier2008}).  Another simplification is that the opacity during evolution is identical for all planets in one population, while it is likely that different atmospheric compositions lead to different cooling curves (e.g.,  Burrows et al. \cite{burrowsetal2011}). 

In the combined formation and evolutionary calculations, $\fopa$ was for simplicity hold constant at the same value during both phases. In reality, a full ISM grain opacity over Gyrs are not expected, because grains quickly rain out after the accretion of gas has ended (Rogers et al. \cite{rogersbodenheimer2011}).  We also calculated a synthesis where $\fopa=0.003$ during the formation phase, but $\fopa$=0 during the evolution. For this population we found radii at 5 Gyrs which are virtually identical to the case where $\fopa$ is kept at 0.003 also during evolution. This shows that  during the evolution,  gas opacities control the cooling for such a strongly reduced value of  grain opacity as already found in Sect. \ref{sect:longterm}.

For the $\fopa=1$ case, the contraction is in contrast visibly slowed down for giant planets. This weakens the effect that for a given core mass, $\mxy$ is usually lower at $\fopa=1$ (Fig. \ref{fig:mcoremenvepop}).  At the maximum of the mass-radius relationship at about 4 $\mj$, we find a maximal radius of about 1.13$\rj$ for $\fopa=1$, while for $\fopa=0$ the maximal radius is about 1.09$\rj$. While this is not a large difference, it means that in the giant planet regime ($M\gtrsim1\mj$), the delayed cooling with $\fopa=1$ dominates over the fact that the $Z$ is higher, as indicated by the colors. 

For lower mass planets, and in particular for $M\lesssim 20 \mearth$, the protracted cooling at $\fopa=1$ is in contrast more than offset by the much smaller primordial $\mxy$ for a given total mass in this synthesis. The maximal radii of low-mass planets are much smaller in the $\fopa=1$ synthesis than in the nominal $\fopa=0.003$ case and even more so in the $\fopa=0$ case. At $M=10 \mearth$ for example, the maximal radius  in the syntheses with migration is approximately 4, 5.5 and 7.5 $\rearth$ for $\fopa$=1, 0.003, and 0, respectively.  The grain opacity thus leads to a clear  imprint in the planetary mass radius relationship via the presence or absence of low-mass, very large planets. This is interesting since otherwise, grain growth models are difficult to compare directly with observations.  In the synthesis without migration and $\fopa$= 0.003, the maximal radius is about 6.5 $\rearth$, i.e., between the value for $\fopa=0$ and 0.003 in the simulations including migration.
 
\subsection{Comparison with the observed mass-radius relationship}\label{sect:compmrrobs}
Figure \ref{fig:mrr} also shows the mass-radius relationship of  planets in the Solar System and extrasolar planets outside of 0.1 AU that have a well defined radius and mass. The observational data set  available for comparison with the synthetic mass-radius relationship is obviously still rather small, mainly because of $a>$0.1 AU. 

But already with the currently limited data set we note that for $\fopa=1$ (bottom left panel), there are two planets (Kepler-11e and Kepler-18d) that lie above the envelope covered with synthetic planets. For Kepler-11e, the distance from the synthetic population is not very large given the error bars, but Kepler-18d lies several $\sigma$ in mass and radius away.  Since both Kepler-11 and Kepler-18 are old stars with estimated ages of $8\pm2$ and $10\pm2.3$ Gyrs (Lissauer et al. \cite{lissauerfabrycky2011}; Cochran et al. \cite{cochranfabrycky2011}), the larger radii of the two exoplanets cannot be due to a younger age of the host star relative to the synthetic population at 5 Gyrs. 

We therefore find an interesting result: At $\fopa=1$, gas accretion is  so inefficient in the model that no synthetic planets can form containing a sufficiently high $\mxy$ to reproduce  the observations. The failure of the $\fopa=1$ population to reproduce actual low-mass exoplanets having a high gas mass fraction could therefore be an observational indication of grain growth in protoplanetary atmospheres. 

Alternative explanations are also possible, like a delay of the contraction due to an  opacity even higher than the full ISM grain opacity (but only during the evolutionary phase), or an additional energy source in the interior. Regarding the latter point it was found (Demory \& Seager \cite{demoryseager2011}) that at least for giant planets, no visible inflation seems to be active for orbital distances larger than $\sim$0.07 AU  around solar-type stars.   Given the small eccentricities also internal heating due to eccentricity damping seems unlikely, at least for Kepler-18d (Cochran et al. \cite{cochranfabrycky2011}). 

The population with $\fopa=0$ (top left) behaves clearly differently. Here, the planets  Kepler-11e and Kepler-18d are well within the envelope covered by synthetic planets. At their mass, there are many synthetic planets that are even significantly larger. The comparison with the observed planets shows that no such extremely large low-mass planets have been found to date. While it is  possible that such planets will be found in future, the  $\fopa=0$ population thus has a large sub-population of low-mass, very large (i.e., very gas rich) planets presently without observational counterparts. This could mean that with grain-free opacities, gas accretion in the model is too efficient, and that a small, but still non-zero contribution of grains  to the opacity is acting in protoplanetary atmospheres. An alternative explanation is that the planets  {lose} parts of their gaseous envelope after formation via evaporation (or giant impacts), an effect we do not include in the current model. Since atmospheric escape could be important (e.g., Lopez et al. \cite{lopezfortney2012}; Owen \& Wu \cite{owenwu2013}), the result that $\fopa=0$ leads to too large planets needs to be revised once this effect is taken into account (Sheng et al. \cite{shengmordasini2014}). 

The population with $\fopa=0.003$ (top right) lies between the two extreme cases. The domain in the $M-R$ plane covered by the synthetic planets seems similar to the observational result. Kepler-18d is at the upper boundary of the synthetic mass-radius relationship, but still in agreement when taking into account the error bars.  

In the figure, there is a group of actual extrasolar planets  with a mass similar to Saturn. In this mass domain, all three populations with migration  show a certain excess of planets with a radius smaller than found in the observational data. This could be an effect related to the semimajor axis and orbital migration: In the figure, synthetic planets out to 5 AU have been included, while the actual exoplanets (but of course not Saturn) typically have  smaller orbital distances. As discussed in Mordasini et al. (\cite{mordasinialibert2012}) we find that despite migration, there remains a positive correlation between the amount of heavy elements in a planet and the orbital distance, as expected from the isolation mass. The excess of small synthetic planets in this mass domain is therefore due to planets at larger semimajor axes and disappears if only planets with $a\lesssim 1$ AU are included. It is also absent in the population without migration, because the feeding zone is in general smaller for in situ formation, so that the core masses  are lower (also the setting that $\dot{M}_{\rm Z,run}=0$ in this population plays a role).

\subsubsection{Population with in situ formation and $\dot{M}_{\rm Z,run}=0$}
The population with $\dot{M}_{\rm Z,run}=0$ and in situ formation is shown in the bottom right panel. The change of a fundamental assumption in the formation model (in situ formation instead of orbital migration) does not completely change the mass-radius relationship, but does leave an imprint. We see that for low-mass planets, this population covers in terms of the largest associated $R$ an envelope in the mass-radius plane that lies  between the one of the $\fopa$=0.003 and 0 population with migration, as expected. For giant planets the mass-radius relationship becomes vertically very thin, because with $\dot{M}_{\rm Z,run}=0$, there are no massive cores that can reduce the radius significantly. This becomes particularly obvious for massive giant planets with $M\gtrsim1000\mearth$, where the low maximal core masses of only $\sim$60 $\mearth$ in this population (cf. Sect. \ref{sect:maximalcoremass}) cannot reduce the total radius in a significant way. This very thin vertical spread is of course also due to the fact that the planets have exactly the same age, and that the envelope opacity is exactly identical for all planets in the population. In reality, this will not be the case, so that a larger vertical spread is expected even if no very massive cores exist. 

When comparing with the observational data for giant planets, we see that CoRoT-10b and HD80606b lie slightly below the envelope of points predicted by this synthetic population.  Neither CoRoT-10 nor HD80606 are much older than 5 Gyrs. For CoRoT-10, Bonomo et al. (\cite{bonomosanterne2010})  find weak constraints for the stellar age but even favor values smaller than 3 Gyrs. For HD80606, Pont et al. (\cite{ponthebrard2009}) estimate an age of  $5\pm3$ Gyrs. Therefore the smaller radii should not be an age effect, but an indication that these planets might have higher metal contents than found in the model under the assumption of in situ growth and $\dot{M}_{\rm Z,run}=0$. Indeed, Bonomo et al. (\cite{bonomosanterne2010}) estimate that CoRoT-10b contains between 120 and 240 $\mearth$ of rocks, which is more than the maximal mass found in the in situ population for planets at $a<5$ AU (Sect. \ref{sect:maximalcoremass}). We however also see that the vertical distance of the planets from the synthetic population is  small compared to the error bars, so that this is a weak constraint.
 
\subsubsection{Observational hints of a strongly reduced opacity}
The comparison  of  the mass-radius relationship of low-mass planets for $\fopa$=0, 1, and 0.003 (including migration) with observations   {seems} in summary  {to indicate} that grain growth occurs in protoplanetary atmospheres, leading to an opacity that is much reduced compares to the ISM value. It is currently not possible to constrain from our simulations the  {specify} value of $\kappa_{\rm gr}$, because  {first}, a range of $\fopa$ between the extreme cases would likely lead to a similar $M$-$R$ as in the observations,  {as long as $\fopa$ is clearly smaller than unity.} This is due to the currently low number of actual exoplanets and the intrinsic large spread.  {Second, and more importantly, a $\kappa_{\rm gr}$ found by scaling the ISM by one general $\fopa$ can not reproduce the dependency of the microphysical processes that govern the grain growth on planet properties. This dependency eventually influences the resulting $M$-$R$. We will therefore use in a forthcoming work the analytical grain growth model of Paper II for improved population syntheses.} 

Clearly, our results are tentative as both the observational comparison sample is rather small and the theoretical results are obtained within a framework of a formation and evolution model that is much simplified compared to reality. The theoretical results will thus partially depend on other assumptions in the model like, e.g., the core accretion rate or the migration model.

\section{Relative heavy element content of giant planets}\label{sect:zpzstar}
 {We now turn to the second observable quantity which potentially contains an imprint of $\kappa_{\rm gr}$ besides the $M$-$R$ relation of low-mass planet. It is the heavy element content of giant planets.} In Figure \ref{fig:mrr} we compared the position of observed extrasolar planets relative to the envelope of points covered in the syntheses. The color code in the figure shows that not only  the envelope of $M$-$R$ points changes, but also the bulk composition of planets within the covered envelope of $M$-$R$ loci. In this section we use this result to compare the metal enrichment of giant synthetic planets with constraints derived from observations. The heavy element content of giant planets was studied in several works. Guillot et al. (\cite{guillotsantos2006}), Burrows et al. (\cite{burrowshubeny2007}), and Guillot (\cite{guillot2008}) have in particular found that there is a positive correlation between the stellar [Fe/H] and the planetary heavy element content, a correlation that is reproduced by population synthesis models based on the core accretion paradigm (Mordasini et al. \cite{mordasinialibert2009b}). This correlation will be re-addressed in future work.

\subsection{Relative metal enrichment in observed exoplanets}\label{sect:enrichmenrobs}
In this work, we focus on anther aspect. We compare the heavy element enrichment of planets relative to the host star $\zp/\zstar$ in synthetic and actual transiting extrasolar planets as derived by Miller \& Fortney (\cite{millerfortney2011}). The planet's metal fraction is $\zp=\mz/M$, while the stellar metal fraction is  $\zstar=Z_{\odot} 10^{\rm [Fe/H]}.$ 
The heavy element fraction of the Sun is taken as $Z_{\odot}$=0.0142 (Asplund et al. \cite{asplundgrevesse2009}) while [Fe/H] is measured spectroscopically. We thus assume that iron is a good indicator of the global metal content of a star which is generally the case as long as we are dealing with thin-disk stars (Adibekyan et al. \cite{adibekyansantos2012}).

The sample of transiting extrasolar planets analyzed by Miller \& Fortney (\cite{millerfortney2011}) consisting of 14 planets that receive a stellar irradiation flux of less than $2\times10^{8}$ erg/(s cm$^{2}$). Below this threshold, no significantly inflated planets are found. This is in agreement with the analysis of Demory \& Seager (\cite{demoryseager2011}) of the Kepler sample. The 14 planets  have masses greater than 20 $\mearth$ and have, with one exception, roughly solar-like host stars with 0.7$<M_{\star}/M_{\odot}<1.3$. For each planet,  Miller \& Fortney (\cite{millerfortney2011}) determine the heavy element fraction $\zp$ by internal structure modeling (see Fortney et al. \cite{fortneymarley2007}). In the model, no extra heating sources are included, and the atmospheres have solar composition.  The derived planetary heavy-element masses are therefore likely  minimum masses because higher opacities (e.g., Burrows et al. \cite{burrowshubeny2007}) or additional heating sources would require a higher heavy element fraction to obtain the observed radius. The results are obtained based on a number of general assumptions like a fully convective interior or a specific equation of state (Saumon et al. \cite{saumonchabrier1995}). Regarding the former point, one must keep in mind that there are also semi-convective interior models for Jupiter that agree with observations (Leconte \& Chabrier \cite{lecontechabrier2012}). These models can have a $\sim$60\% higher metal fraction. Concerning the latter point, uncertainties in the  EOS of  H/He alone  cause significant variations for the estimated bulk composition of Jupiter (e.g., Guillot \cite{guillot1999}, Militzer et al. \cite{militzerhubbard2008}, Nettelmann et al. \cite{nettelmannbecker2012}). Due to these reasons, $\zp$  is currently only poorly known even for Jupiter (and the other giant planets of the Solar System). This means that the results for the heavy element mass $\mz$ in extrasolar planets  come with large systematic uncertainties.   

The planetary heavy element fractions derived by Miller \& Fortney  (\cite{millerfortney2011}) are the average of the values for the limiting cases of either all metals in the core or homogeneous mixing. This leads strictly speaking to an inconsistency with the formation model where all solids are in the core. The difference in $\zp$ derived for the layered and  homogeneous case is however rather small with  typically 15\% higher $\zp$ in the layered case. Within the errors bars due to other uncertainties (e.g., mass and radius measurement) the $\zp$ for the two cases agree for all 14 planets.
 
\subsection{Relative enrichment for synthetic planets}
The relative enrichment $\zp/\zstar$ is straightforward to calculate for the synthetic planets, as one of the four Monte Carlo variables (i.e., initial conditions) in the syntheses is [Fe/H] (Mordasini et al. \cite{mordasinialibert2009a}).  Once a synthetic planet has formed, we thus have its $\zp$ and the $\zstar$ of the corresponding initial condition. $\zstar$ represents the bulk metal fraction of the star, and not the dust-to-gas ratio in the innermost $\sim$10 AU that we use to calculate the planetesimal surface density (Mordasini et al.  \cite{mordasinialibert2009a}). The latter value can be higher by a factor 2-4  (2.8 in the formation model) than $\zstar$ due to the advection of dust from the outer parts of the disk (Kornet et al. \cite{kornetetal2004}, Birnstiel et al. \cite{birnstielklahr2012}).

Two assumptions in the model are of relevance here: first, we assume that the entire metal content in the disk has been incorporated into planetesimals. Consequently, the  nebular gas accreted by the protoplanets is composed of pure H/He. Clearly, this is a limiting case. In Sect. \ref{sect:effectgascomp} we study effects of this assumption. Second, we assume that gap formation does not reduce the gas accretion rate of massive planets due to the eccentric instability (Kley \& Dirksen \cite{kleydirksen2006}). Again this is a limiting case, because in reality, the eccentric instability can only operate for certain configuration of the location of the Lindblad resonances and the width of the gap.  Without a reduction of $\dot{M}_{\rm XY}$ due to gap formation, very massive planets can form with masses up to $\sim$40$\mj$. The specific value depends on the largest initial disk mass. They are very rare though (Mordasini et al. \cite{mordasinialibert2009b}), and only appear numerous in the plots because we are dealing with large populations of $\sim$$30\ 000$ planets. These very massive planets were already visible in the plots above, but here they are more obvious because in the observational comparison sample of Miller \& Fortney (\cite{millerfortney2011}), the maximal mass is in contrast only about $4 \mj$. 

\begin{figure*}
\begin{minipage}{0.48\textwidth}
\includegraphics[width=\textwidth]{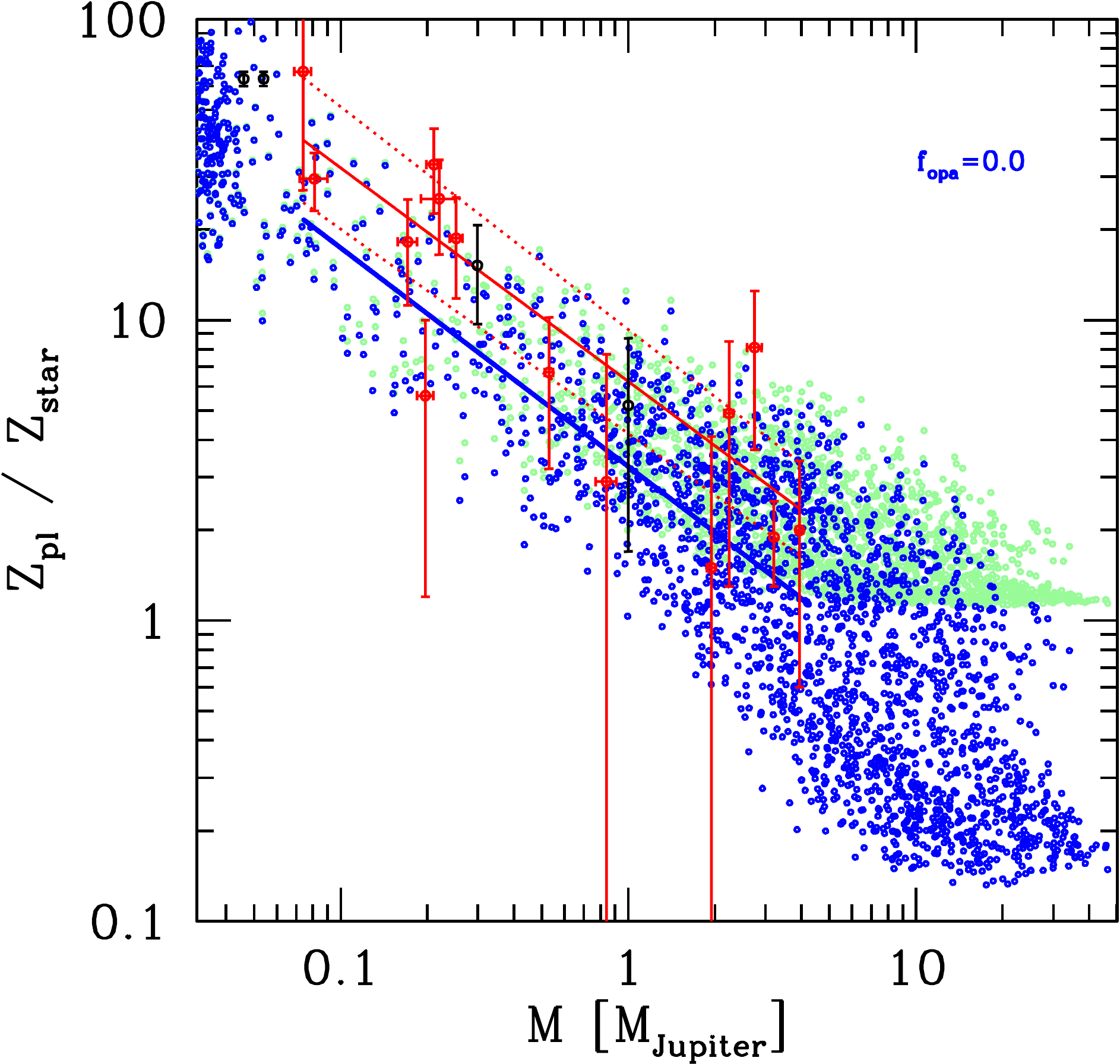}
\includegraphics[width=\textwidth]{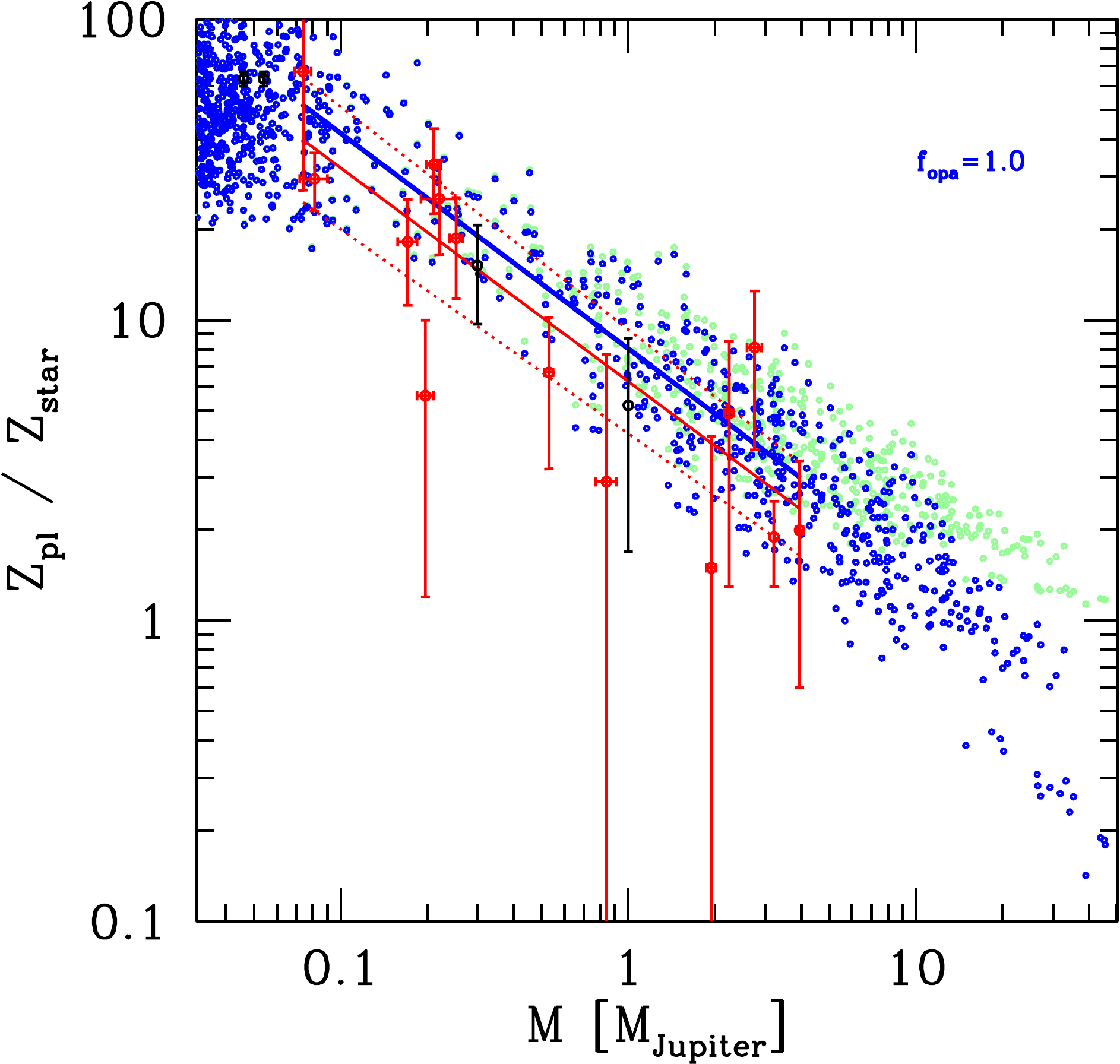}
\end{minipage}
\begin{minipage}{0.48\textwidth}
\includegraphics[width=\textwidth]{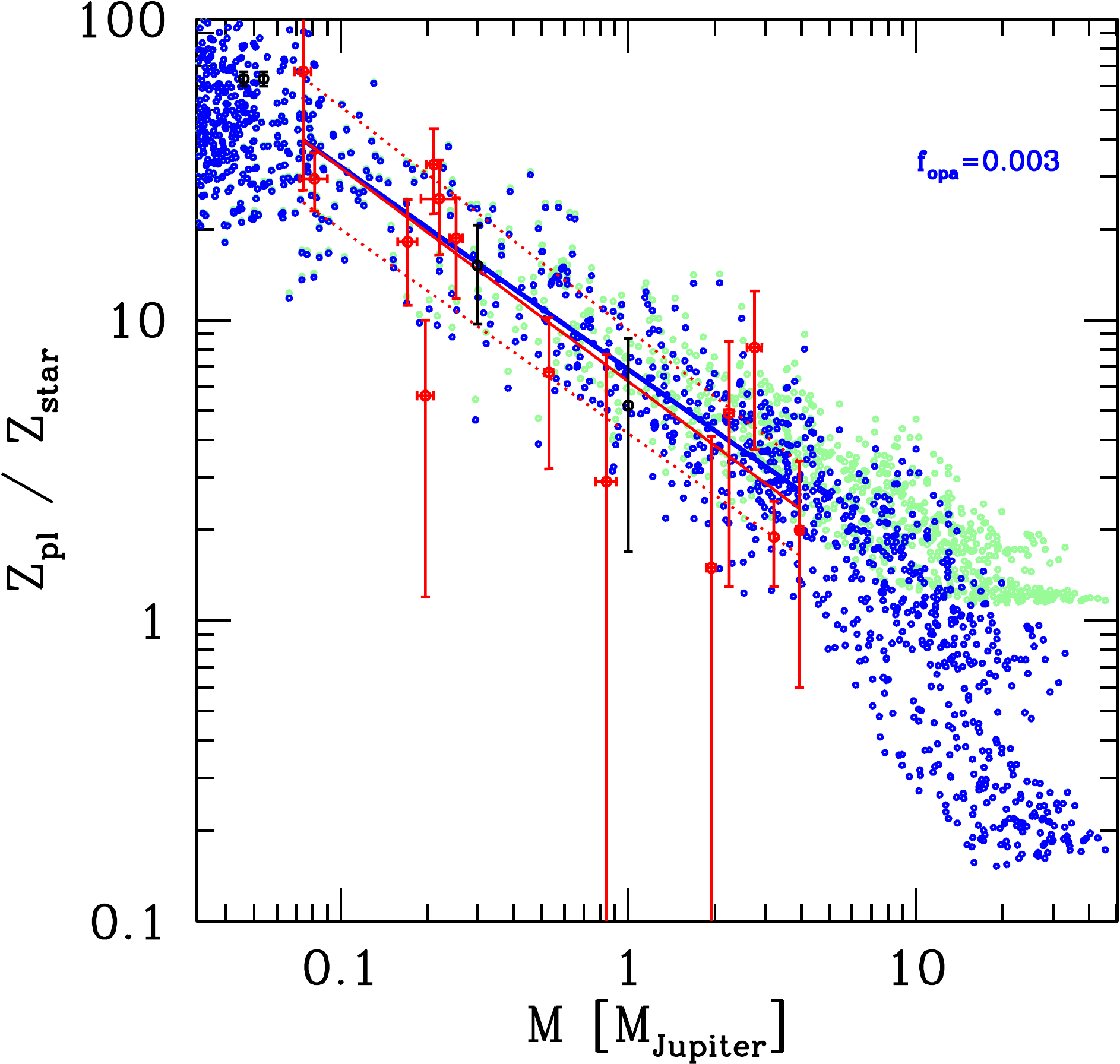}
\includegraphics[width=\textwidth]{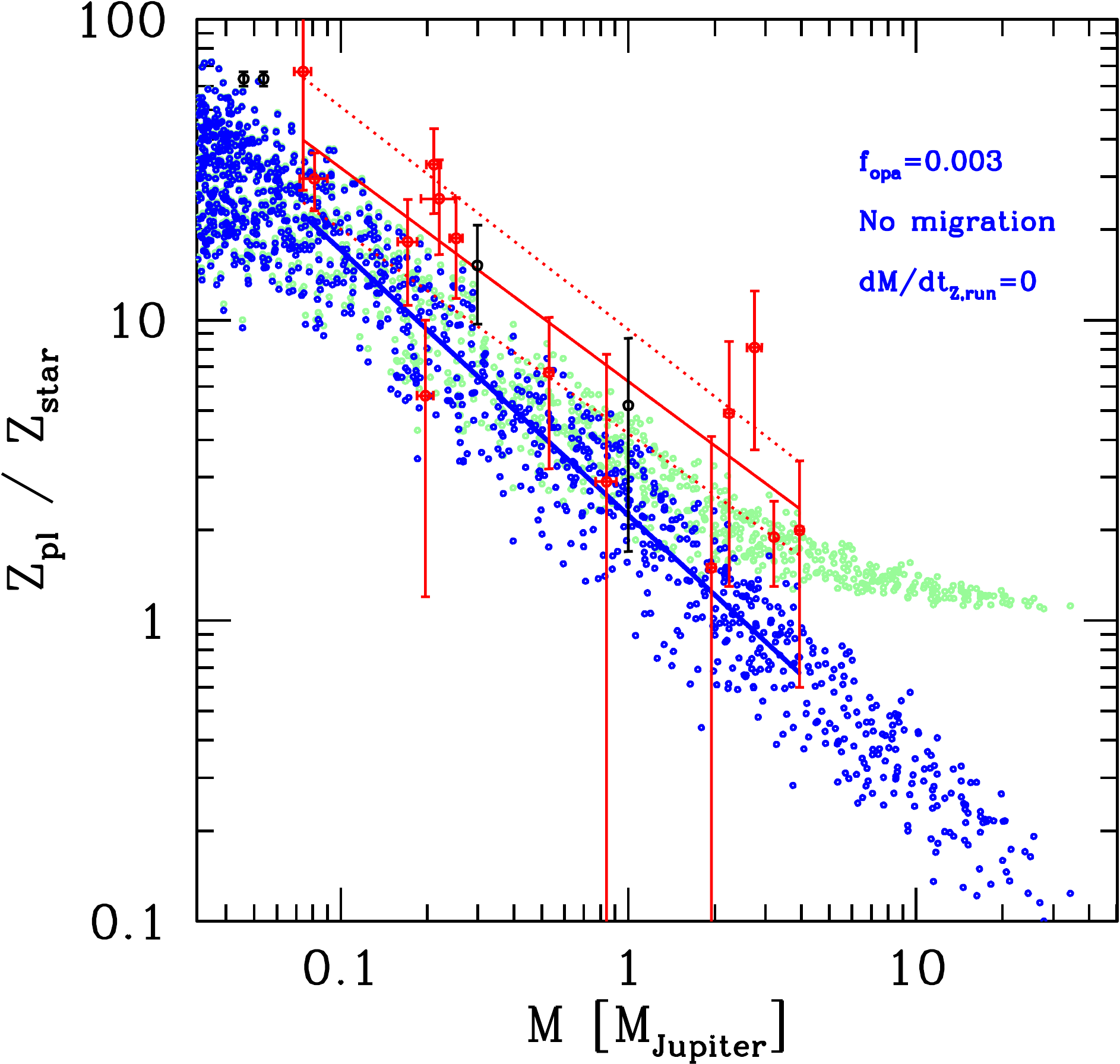}
\end{minipage}
\caption{Planetary heavy element fraction $\zp$ divided by the heavy element fraction of the host star $\zstar$. The relative enrichment is shown as a function of planetary mass $M$. Red dots with error bars are from Miller \& Fortney (\cite{millerfortney2011}) for transiting exoplanets.  The red solid line is a least square fit to the exoplanet data, while the dotted red lines give the 1-$\sigma$ errors of the fit. Giant planets of the Solar System are shown with black symbols. Blue dots are synthetic planets with $0.11\leq a$/AU$\leq 1$ (5  AU for the synthesis without migration). The blue line gives the fit to the synthetic data. The four panels show populations calculated with an $\fopa$ indicated in the plot. The bottom right panel assumes in situ formation and $\dot{M}_{\rm Z,run}$=0. The pale green dots are the same synthetic planets, but assuming that the accreted gas has the same metallicity as the host star instead of being metal free.}\label{fig:zpzstar}  
 \end{figure*}

\subsection{$\zp/\zstar$ as a function of planet mass}\label{zpzstarmass}
Figure \ref{fig:zpzstar}  shows $\zp/\zstar$ for the observational comparison sample of Miller \& Fortney (\cite{millerfortney2011})  and the synthetic populations. For the three populations with migration, planets inside of 1 AU are included (the actual exoplanets have $a<0.41$ AU). For the population without migration, planets inside of 5 AU are included because no giant planets form in situ inside of 1 AU.  The giant planets of the Solar System are also shown. For Jupiter and Saturn, we use the $\mz$ estimated by Saumon \& Guillot (\cite{saumonguillot2004}) for different EOS,  while for Uranus and Neptune the bulk composition given by Guillot \& Gautier (\cite{guillotgautier2009}) if used. The conversion into $\zp/Z_{\odot}$ is again made with the updated protosolar  $Z_{\odot}=0.0142$ of Asplund et al. (\cite{asplundgrevesse2009}). This value is substantially lower than earlier estimates, resulting in correspondingly higher enrichment of the planets.

While all populations (and the observed exoplanets) share  the property of a decreasing $\zp/\zstar$ with increasing mass, there are  substantial differences in the absolute level of the metal enrichment. At 1 $\mj$, the relative enrichment at $\fopa=1$ ranges for example between 4 and 20, while for $\fopa=0$, the enrichment is between 1 and 13.  The impact of $\fopa$ becomes particularly clear when one considers the least square fit to the observational data (red solid line with  red dotted lines indicating approximately the 1 $\sigma$ error bars). The fit is given as
\beq\label{eq:fitzpzstar}
\frac{\zp}{\zstar} =\beta \left(\frac{M}{\mj}\right)^{\alpha}.
\eeq
The parameters $\alpha$ and $\beta$ are shown in Table \ref{tab:zpzstar}. Corresponding fits to the synthetic data are also shown in the figure. These fits were made over the same mass domain as for the observations. The figure shows that the decrease of $\zp/\zstar$ with mass follows a similar slope $\alpha\approx-0.7$ independently of $\fopa$ for all three populations with migration. This slope is in good agreement with the observational value of $-0.71\pm0.10$. The absolute level given by $\beta$  decreases significantly from 8.5 at $\fopa=1$  to 3.5 at $\fopa=0$, corresponding to a decrease by a factor $\approx2.4$. When we compare these values with the observational result $\beta=6.3\pm1.0$ we see that  the full ISM  grain opacity leads to  too enriched planets, while the grain-free opacity leads to too low enrichments. At $\fopa=0.003$, $\beta=7.2$ which is similar to the observational result. We thus find for the bulk composition of giant planets a similar result as we had found for the mass-radius relation of low-mass planets: grain evolution in protoplanetary atmospheres seems to lead to an opacity that is much smaller than in the ISM, but potentially, there is still a non-zero contribution from the grains.

One can ask why the bulk composition also of giant planets depends on $\fopa$ despite the fact that in the disk limited gas accretion phase (when the planet is detached from the nebula), the gas accretion rate is independent of $\fopa$ (in contrast to phase II). The imprint of $\fopa$  on the bulk composition nevertheless remains also for giant planets because at small $\fopa$, the critical core mass is lower. Also the amount of solids accreted in runaway is lower due to the smaller surface density of planetesimals in the feeding zone.  In the population without migration and  $\dot{M}_{\rm Z,run}$=0, the decrease of the relative enrichment is as expected steeper ($\alpha$=-0.88), because in this population, the maximal cores masses are lower (Sect. \ref{sect:maximalcoremass}). Also the absolute level of  $\zp/\zstar$ is low ($\beta=2.4$). This is related to the fact that without migration, the feeding zones and luminosities are smaller. 

As stated earlier, it is clear that the results  for $\kappa_{\rm gr}$ in protoplanetary atmospheres have been obtained from models  that are strongly simplified compared to reality. We have for example not considered the effect of different chemical compositions of the envelope due to envelope pollution. Hori \& Ikoma (\cite{horiikoma2011}) have shown that a pollution with heavy elements reduces the critical core mass. The relation between composition and grain opacity on the other hand is complex: Movshovitz \& Podolak (\cite{movshovitzpodolak2008}) have found that an increase in the dust-to-gas ratio in the newly accreted can actually lead to a decrease of $\kappa_{\rm gr}$ in the deeper layer.  {This is extensively studied in Paper II.} We see that the results found here must be further investigated with models that combine self-consistently the structure of the envelope, the evolution of the grains, the deposition of mass by planetesimals, and  the chemical composition of the gas. 

\begin{table}
\caption{Fit to relative enrichment $\zp/\zstar$ (Eq. \ref{eq:fitzpzstar}).}\label{tab:zpzstar}
\begin{center}
\begin{tabular}{lcc}
\hline\hline
Data set                      &$\alpha$&  $\beta$\\\hline 
Miller \& Fortney (\cite{millerfortney2011})           &     -0.71$\pm$0.10     &  6.3$\pm$1.0 \\                                             
$\fopa$=0	 &  -0.73 & 3.5\\
$\fopa$=0.003	 &  -0.68 & 7.2\\
$\fopa$=1	 &  -0.72 & 8.5\\
$\fopa$=0.003, in situ, $\dot{M}_{\rm Z,run}=0$	 &  -0.88 & 2.4 \\ \hline
\end{tabular}
\end{center}
\end{table}

\subsection{Effect of gas composition}\label{sect:effectgascomp}
In the observational data set analyzed by Miller \& Fortney (\cite{millerfortney2011}) where $M\leq4\mj$,  all planets are consistent with being enriched relative to the host star, and several seem to be enriched even significantly. Figure \ref{fig:zpzstar} shows that in the synthetic populations, there are planets with $\zp/\zstar<1$ for sufficiently large masses. These cases are a direct consequence of the assumption that the accreted gas is pure H/He so that a massive gas envelope can deplete a planet relative to its host star (cf. Helled \& Bodenheimer \cite{helledbodenheimer2011}).  Extrapolating Eq. \ref{eq:fitzpzstar} with the parameters of Miller \& Fortney (\cite{millerfortney2011}) leads to a mass $M_{1}/\mj=\beta^{-1/\alpha}$  where  $\zp/\zstar=1$  of about 13.9 $\mj$. In  the nominal $\fopa=0.003$ population, the lowest planet mass where $\zp/\zstar=1$ is about 4 $\mj$, while the typical $M_{1}$ is  10-15 $\mj$.  These values are a function of $\fopa$: at $\fopa=0.0$, the lowest $M_{1}$ is about 1 $\mj$, while the typical value is about 5 $\mj$. At $\fopa=1$, the lowest $M_{1}$ is only about 5 $\mj$, while typically it is about 17 $\mj$. The relative enrichment  $\zp/\zstar$ decreases steeper with mass in the population without planetesimal accretion in the runaway accretion phase. The smallest $M_{1}$ is here about 1 $\mj$, and the typical value is about 2.5 $\mj$. These are clearly lower values. For the nominal populations,  $M_{1}$ is coincidentally similar to the deuterium burning limit at approximately 13 $\mj$ also for companions formed via core accretion (Molli\`ere \& Mordasini \cite{mollieremordasini2012}, Bodenheimer et al. \cite{bodenheimerdangelo2013}). We see that neither deuterium burning nor the absence of an enrichment relative to the host star necessarily excludes  core accretion as the formation mechanism.   

The observed infrared excess of many young stars shows that hot micrometer-sized dust remains in protoplanetary disks on timescales of  $10^{6-7}$ yrs (e.g., Haisch et al. \cite{haischlada2001}). This shows that in reality, not all dust grains get  incorporated into planetesimals leaving the gas essentially dust free as assumed in the formation model. Instead, the gas that is accreted by the protoplanets will itself contain a certain (unknown) fraction of metals. To investigate that, we also calculated $\zp/\zstar$ for the synthetic planets under the assumption that the gas has the same composition as the  star. This is not self-consistent with the way the solid surface density is derived from [Fe/H] in the model (it means that we artificially generate metals), but it gives an impression of the impact of different gas compositions. An identical metal fraction as in the star is, however, not necessarily the other limiting case, as the gas could also be enriched relative to the star due to, e.g., photoevaporation of gas only.  The enrichment of the synthetic planets under this assumption is also shown in Fig. \ref{fig:zpzstar}. 

The consequence is as  expected: At low masses, where the core dominates, $\zp/\zstar$ does not change appreciably, while for clearly envelope-dominated massive planets, $\zp/\zstar$ approaches  unity.  In the three populations with migration and accretion of planetesimals in the detached phase, $\zp/\zstar$ does not fall below  $\sim$1.1 because of the massive cores in the planets and the positive correlation of [Fe/H], heavy element mass (Guillot et al. \cite{guillotsantos2006}; Burrows et al. \cite{burrowshubeny2007}; Guillot \cite{guillot2008};  Mordasini et al. \cite{mordasinialibert2009b}) and total mass for the most massive planets (Mordasini et al. \cite{mordasinialibert2012pp}). In the population with $\dot{M}_{\rm Z,run}=0$ the convergence towards unity for the most massive planets is clearer. 

In the plot, we also include the planets of the Solar System. Even if the uncertainty of their heavy element content is large, we see that they follow the general trend seen for the exoplanets. The nominal values of Jupiter's and Saturn's enrichment fall quite close to the mean relation derived from the exoplanets. In the Neptunian mass domain, the enrichment becomes very high with $\zp/\zstar$ between 20 and 100 in the synthetic planets. This covers the enrichment estimated for Uranus and Neptune. For these planets, traditional three layer models predict a heavy element mass fraction of about 85 - 95 \%, corresponding to  $\zp/\zstar$ roughly equal to 60 (Guillot \& Gautier \cite{guillotgautier2009}).

\subsection{Comparison with gravitational instability}\label{compgi}
In the simplest picture of the core accretion mechanism, first a critical core forms, and afterwards only gas is added ``diluting'' the planet's enrichment. This would mean that for a given $\zstar$, the relative enrichment $\propto \mz/M$ would decrease with mass as $\alpha\approx$-1 for massive giant planets where $M$$\approx$$\mxy$. In the simplest picture of gravitational instability one would in contrast expect $\alpha\approx0$. The numerical (and observational) result that  $\alpha\approx$-0.7 indicates that for $\dot{M}_{\rm Z,run}\ne0$, the core mass itself increases weakly with total mass roughly $\propto M^{0.3}$. Analytically, without orbital migration and planetesimal ejection, and if the planet accretes all planetesimals in the feeding zone, Eq. \ref{eq:mmisomz} shows that $\mz/M$ falls as $\alpha$=-2/3. The observational result  ($\alpha\approx$-0.7) thus lies between the cases that the giant planet accretes all planetesimals in the feeding zone ($\alpha$=-2/3), and no accretion in runaway ($\alpha$=-1).  In the population where $\dot{M}_{\rm Z,run}$=0, $\alpha$ is indeed closer to -1 (-0.88).

Our results indicate that giant planets with masses less than $\sim$10 $\mj$ formed by core accretion are enriched due to the accretion of planetesimals roughly as $6 (M/\mj)^{-0.7}$. Figure \ref{fig:zpzstar}  shows that the spread in $\zp/\zstar$ around this relation  for a given mass is however large, typically about one order of magnitude. For more massive giant planets the composition of the gas becomes important in determining the bulk composition, and both enriched as well as depleted planets may form depending on the efficiency of dust and planetesimals growth.  

The composition of giant planets formed by the gravitational instability mechanism was studied by, e.g., Helled \& Schubert (\cite{helledschubert2009}), Nayakshin (\cite{nayakshin2010}), Boley et al. (\cite{boleyhelled2011}), and Helled \& Bodenheimer (\cite{helledbodenheimer2011}). The initial composition of a clump should be similar to the  composition of the disk and therefore  $\zp/\zstar\sim 1$ independent of mass.  Several  {mechanisms} can however lead to enrichment:  enrichment at birth, planetesimal capture (before and after the collapse of the first core), and differentiation plus tidal stripping.  These processes can lead to a large spread of final bulk compositions also for gravitational instability. 

Enrichment via planetesimal capture is efficient as long as the planet is extended  (i.e., before the second collapse) because of the large capture radius. This is similar to the behavior for core accretion (Sect. \ref{sect:factorsmdotzrun}). The duration of the pre-collapse phase depends on opacity. If grain growth is taken into account (Helled \& Bodenheimer \cite{helledbodenheimer2011}), the duration of the pre-collapse phase is very short ($\sim10^{3}$ years) so that enrichment by planetesimal capture is precluded over a wide domain of planetary masses and metallicities. This would result in roughly stellar compositions if this is the dominant enrichment mechanism. 

Due to the complexity and the multitude of interacting  {mechanisms} that determine the final composition of giant planets in both formation mechanisms (additional effects are, e.g., giant impacts), it seems likely that for many individual planets one cannot distinguish core accretion and gravitational instability based on the bulk composition only. More promising should be statistical correlations, namely the (mean) enrichment as function of mass (as studied here), stellar [Fe/H], and orbital distance. Regarding the orbital distance, Helled \& Bodenheimer (\cite{helledbodenheimer2011}) find that for giant planets formed by direct collapse, the metal enrichment strongly decreases with orbital distance. The impact of  distance and [Fe/H] on the bulk composition for core accretion will be addressed in future work. An important finding from internal structure modeling is here that the heavy element content of giant planets and the stellar [Fe/H] are correlated, as found by Guillot et al. (\cite{guillotsantos2006}) and  Burrows et al. (\cite{burrowshubeny2007}).

\section{Summary and conclusions}\label{sect:conclusion}
 {In this series of papers we investigate the impact of the opacity due to grains in protoplanetary atmospheres on the predicted bulk composition  (H/He fraction) of  extrasolar planets. In this first paper, we study the impacts found by scaling the ISM opacity. In Paper II, we present an analytical model to calculate $\kappa_{\rm gr}$. In future work, we will couple the analytical model to our updated population synthesis calculations.} The results of this first paper are summarized as follows: 
\begin{itemize}
\item In the context of the core accretion paradigm, we studied the duration $\t2$ of phase II  for in situ formation  of Jupiter as a function of the reduction factor of grain opacity $\fopa$ relative to  ISM grain opacity. We found that over an important range of $\fopa$, there is a linear relationship between $\fopa$ and $\t2$ as expected from theoretical considerations (Sect. \ref{durationphase2fopa}). 
\item We compared the duration of $\t2$ as function of $\fopa$ with the corresponding duration found by Movshovitz et al.  (\cite{mbpl2010}) who conducted simulations of combined giant planet formation and grain evolution.  We found that the ISM grain opacity must on average be reduced by  a factor $\fopa$$\approx$0.003 to reproduce their results (Sect. \ref{sect:finalresfopa}). As a caveat one must keep in mind that one uniform reduction factor cannot reproduce the complex radial opacity structure found in grain evolution calculations, and also that the calibration was only made in a small part of the parameters space of possible core masses, luminosities, and outer boundary conditions (Sect. \ref{sect:generality}). 
\item Without migration and planetesimal drift, there is a unique relation between isolation, core, and total mass during phase II if the planet accretes all planetesimals in the feeding zone. The reason is that the core mass $\mz$ is given by the size of the feeding zone that depends via the Hill sphere on the total mass $M$. This means that for a given core and isolation mass, the envelope mass $\mxy$ during this phase is independent of $\fopa$. The opacity merely determines how quickly a planet evolves through the different $\mz-\mxy$ states  (Sect. \ref{sect:relationmzmxyphaseII}).
\item We studied the global consequences of a scaled ISM opacity on planets forming via  core accretion  with population synthesis calculations (Sect. \ref{populationsynthesis}).  {In reality, the grain opacity in protoplanetary atmospheres can not be  obtained  as the ISM opacity simply reduced by one general $\fopa$. This is because the grain dynamics depend on a planet's properties (Movshovitz \& Podolak \cite{movshovitzpodolak2008}; Paper II). Therefore, no  constraints for the value of $\kappa_{\rm gr}$ for a specific planet can be derived from our calculations. Due to this limitation, we} considered besides the nominal value $\fopa$=0.003 also populations with full ISM grain opacity ($\fopa$=1) and grain-free opacity ($\fopa$=0)  {as limiting cases. This allows to see if the opacity leads at all to potentially observable imprints, which is the main goal of the paper}.
\item  Grain opacity leaves clear imprints in the planetary bulk composition (H/He envelope mass as a function of core mass). For sub-critical cores, the envelope mass for a given core mass increases with decreasing $\kappa_{\rm gr}$. In the synthesis with $\fopa=0.003$, the envelope masses of sub-critical cores ($\mz\lesssim10\mearth$) are about four times more massive than for full ISM opacity. There is  a large spread of at least one order of magnitude in $\mxy$ for a given $\mz$ (Sect. \ref{sect:menveofmcore}).  
\item The critical core mass for runaway gas accretion decreases with $\fopa$. The lowest core mass $M_{\rm Z,min}$ in giant planets (envelope mass $\mxy\geq 100 \mearth$) are  6, 16 and 29 $\mearth$ for $\fopa$=0, 0.003, and 1, respectively. These results are for populations where orbital migration is included. In a synthesis without orbital migration and $\fopa=0.003$,  $M_{\rm Z,min}=7 \mearth$. This difference is due to the lower mean luminosity without migration and the dependency of the critical mass on both the luminosity and  opacity (Sect. \ref{giantsandmcrit}). 
\item If planets continue to accrete planetesimals in the disk limited gas accretion phase,  {and if the planetesimal random velocities stay low}, then the maximal heavy element content in giant planets we find can be up to  $\sim$400 $\mearth$.  If in contrast no planetesimals are accreted in the disk limited phase, then the maximal heavy element content is about 50 to 100 $\mearth$  depending on the semimajor axis  (Sect. \ref{sect:maximalcoremass}). 
\item  Grain opacity leaves a clear imprint in the mass-radius relationship of low-mass planets.  At a low opacity, planets in the super-Earth and Neptunian mass domain have radii for a given mass that are much larger than for ISM opacity because of higher envelope mass fractions.  At $M=10 \mearth$, for example, the maximal radii at 5 Gyrs in the syntheses with migration are 4, 5.5 and 7.5 $\rearth$ for $\fopa$=1, 0.003, and 0, respectively (Sect. \ref{sect:MRR}). 
\item  We compared the envelope covered by actual and synthetic planets in the mass-radius plane assuming an age of 5 Gyr for the synthetic planets. One finds that for $\fopa=1$, the synthetic super-Earth and Neptunian planets have  too small radii (i.e., too low envelope masses) to cover the same domain as the observations. At $\fopa=0.003$,  the value  calibrated  with a grain evolution model,  the synthetic and actual planets occupy a similar loci in the mass-radius plane. This is a  {hint} that the opacity in protoplanetary atmospheres is much smaller than in the ISM (Sect. \ref{sect:compmrrobs}). 
\item A second consequence of grain opacity that can be compared with observational data is the bulk composition of giant planets. We compared the bulk composition of synthetic planets as expressed in the  enrichment of a planet relative to its host star $\zp/\zstar$ ($\zp=\mz/M$, $\zstar=Z_{\odot} 10^{\rm [Fe/H]}$) with the results derived from observations  (Miller \& Fortney \cite{millerfortney2011}). Miller \& Fortney (\cite{millerfortney2011}) derived the heavy element content of 14 transiting extrasolar planets obtaining low irradiation fluxes by internal structure modeling (Sect. \ref{sect:enrichmenrobs}). 
\item We found that there is a clear imprint of opacity on the bulk composition of giant planets. The mean relative enrichment  $\zp/\zstar$ is about  2.4 times higher for full ISM grain opacity compared to the grain-free case (\ref{zpzstarmass}). 
\item The mean relative enrichment of giant planets as a function of planet mass $M$ can be approximated as $\zp/\zstar =\beta (M/\mj)^{\alpha}$. The decrease of $\zp/\zstar$ with mass follows $\alpha\approx$-0.7 independently of $\fopa$ for all three populations with orbital migration. This slope is in good agreement with the value derived from observations ($-0.71\pm0.10$). In the simplest picture of core accretion, where first a critical core forms, and afterwards only gas is added, $\alpha\approx-1$, while in the simplest picture of gravitational instability $\alpha\approx0$. The  $\alpha\approx-0.7$ found in the simulations indicates a weak positive correlation of $\mz$ with total mass. A similar exponent ($\alpha\approx$-2/3) is expected if giant planets efficiently accrete all solids in their feeding zone (Sect. \ref{zpzstarmass}).  
\item The absolute level of enrichment  $\beta$ decreases significantly from 8.5 at $\fopa$=1  to 3.5 at $\fopa$=0. When we compare these values with the result derived from observations ($\beta=6.3\pm1.0$) we see that a full ISM grain opacity leads to too enriched planets, while a grain-free opacity leads to a too low enrichment. At $\fopa=$0.003, $\beta=7.2$, which is similar to the observational result.  The bulk composition of giant planets thus seems to  {hint} that the opacity in protoplanetary envelopes is much smaller than in the  ISM, but that there is possibly still a non-zero contribution from the grains  (Sect. \ref{zpzstarmass}). This result is similar as the one found for the mass-radius relationship of low-mass planets.
\item Giant planets with masses less than $\sim$10 $\mj$ formed by core accretion are enriched due to the accretion of solids, with the relative enrichment $\zp/\zstar$ following roughly  $6 (M/\mj)^{-0.7}$ in the  observational sample and similarly in the nominal synthetic population. The spread around this relation at a given mass is  large, about one order of magnitude. For more massive giant planets the composition of the gas becomes important in determining the enrichment, and both enriched and depleted planets are possible.  In the limit that the accreted gas is free of solids, planets more massive than $\sim$10-15 $\mj$ are depleted relative to the host star (Sect. \ref{sect:effectgascomp}). 
\item We derived a semi-analytical expression for the core and envelope mass as a function of time in phase II using a two parameter expression for the gas accretion timescale $\tg$ (Appendix \ref{sect:semianalyticalsolution}).  We determined the parameters of $\tg$ for different core masses and $\fopa$ by comparison with  numerical results (Appendix \ref{sect:paramstkh}). 
\end{itemize}

We have found in this work that the ISM grain opacity must be reduced by a factor $\fopa$$\approx$0.003 to reproduce the formation timescales of Movshovitz et al. (\cite{mbpl2010}). This  value is almost an order of magnitude lower than the previously considered ``low opacity case'' with 2\% ISM opacity (P96, Lissauer et al. \cite{lissauerhubickyj2009}). It is however clear that uniform reduction factors  {can not capture the physical} effects of grain evolution that in reality lead to complex opacity profiles in the envelope (Movshovitz \& Podolak \cite{movshovitzpodolak2008},  {Paper II})  {which differ from the ones found by scaling the ISM opacity}. The interest in a calibrated $\fopa$ is therefore merely to have an  {intermediate} case for the magnitude of $\kappa_{\rm gr}$ in simulations that study the global impact of opacity in a parametrized way  {between the limiting $\fopa$=0 and 1 cases}.

With this grain reduction factor, cores with a low mass can trigger gas runaway accretion during the typical lifetime of a protoplanetary nebula, the exact value depending on luminosity.  As discussed by Movshovitz et al. (\cite{mbpl2010}) and Hori \& Inaba (\cite{horiikoma2010}), the lower critical core masses found with more realistic opacities are an important result for the core accretion paradigm, as several studies  (e.g., Fortier et al. \cite{fortierbenvenuto2007}; Ormel \& Kobayashi \cite{ormelkobayashi2012}) indicate that building up a 10 $\mearth$ core at 5.2 AU within the typical lifetime of a disk is a critical timing issue.  These lower opacities contribute substantially in making this timing issue less stringent, even without taking into account additional mechanisms like migration or envelope pollution (Alibert et al. \cite{alibertmordasini2004}; Levison et al. \cite{levisonthommes2010}; Hori \& Ikoma \cite{horiikoma2011}).  {Also, at such low $\fopa$, cores of just a few $\mearth$ can accrete quite significant H/He envelopes, making them potential progenitors of the recently detected low-mass, low-density planets.}

The global consequences of grain opacity on planetary populations are strong and multiple. We have studied these imprints by running population synthesis calculations with a wide range of $\fopa$. We focused on two possibly observable consequences, namely the mass-radius relationship of low-mass, subcritical planets, and the bulk composition of giant planets. Additional imprints also exist in the planetary mass and radius distributions, and in particular in the frequency of giant planets. This frequency  increases in the synthetic populations approximately by a factor three when reducing  $\fopa$ from 1 to 0. 

The resulting comparison of synthetic and (statistical) observational results establishes a connection between the opacity in protoplanetary envelopes and observable quantities.  {In this work, we cannot directly derive constraints on the value of $\kappa_{\rm gr}$ because we find it by scaling the ISM opacity instead of calculating it based on planet properties.}  {However, in future, a similar approach should allow} to test microphysical models of grain evolution that are otherwise  difficult to test observationally. The result of this first paper that the observed mass-radius relationship can not be reproduced with a full ISM opacity is interesting. It could be an observational hint that the prediction of grain growth models  {(Podolak \cite{podolak2003}, Movshovitz \& Podolak \cite{movshovitzpodolak2008}, MBPL10, and Paper II)} that the opacity in protoplanetary atmospheres is much smaller than in the ISM, is  correct. 

 {However, it is clear that our results for the predicted bulk composition of extrasolar planets are preliminary.} Too complex and too numerous are the  actual physical  {mechanisms} occurring during formation and too simple is the model in comparison that uses, e.g., one uniform reduction factor, neglects the impact of the chemical composition of the gas, and only handles one embryo per disk.  {For a better understanding, it is necessary to couple a physically motivated grain opacity model like the analytical model of Paper II with planet formation codes, and to take into account additional important factors like the pollution of the envelope (Hori \& Ikoma \cite{horiikoma2011}), the concurrent formation of many planets in one disk (Alibert et al. \cite{alibertcarron2013}), or the loss of the H/He envelope during evolution due to atmospheric escape (Sheng et al. \cite{shengmordasini2014}).}  

On the observational side, a significant extension of the sample of relatively cold, transiting planets with well-constraint mass and radius would be very important, e.g., with CHEOPS (Broeg et al. \cite{broegfortier2013})  {and PLATO (Rauer et al. \cite{rauercatala2013})}. Then it should become possible to understand the processes that determine the opacity in protoplanetary atmospheres  much better than today.  The associated statistical results for the enrichment of the planets will allow to distinguish better different formation mechanisms like core accretion and gravitational instability, or to understand from a theoretical point of view the transition from solid to  gas dominated planets (Marcy et al. \cite{marcyetal2014}).

\acknowledgements{We thank K.-M. Dittkrist, P. Molli\`ere and especially Chris Ormel for enlightening discussions. This work was supported in part by the Swiss National Science Foundation and the European Research Council under grant 239605.  Computations were made on the BATCHELOR cluster at MPIA. We  thank {two anonymous referees for important comments that led to a substantial improvement} of the manuscript. CM thanks the Max-Planck-Gesellschaft for the Reimar L\"ust Fellowship.}

\appendix

\section{Semi-analytical solutions}\label{sect:semianalyticalsolution}
The left panel of Figure \ref{fig:sigma10mlt} in Sect. \ref{sect:determinationfopa} shows that  the curves giving the mass as a function of time for different opacities are self-similar in the sense that they can be approximately converted in each other by  stretching the x-axis. This indicates that there could possibly be (semi-)analytical solutions for the mass as a function of time in phase II. Since such solutions are interesting for  comparison with  numerical results, we derive a semi-analytical solution for the envelope and core mass as a function of time during phase II. 

During this phase, the envelope grows by the contraction of the outer layers (e.g., Ikoma et al. \cite{ikomanakazawa2000}), while the core grows due to the resulting expansion of the feeding zone (P96). This growth mode applies if the accretion rate of solids is limited directly by the availability of planetesimals in the feeding zone, rather than the probability that they get captured by the protoplanet. In other words, the timescale of accretion of planetesimals already within the feeding zone is much smaller than the timescale on which the feeding zone expands. This regimes can arise if the planetesimals are dynamically cold, so that the capture probability is high, as assumed in P96. 

The gas accretion rate is parametrized with the growth timescale $\tg$ of a planet of total mass $M=\mz+\mxy$ where $\mz$ is the total mass of heavy elements in the planet (equal to the core mass if all planetesimals sink to the core) while $\mxy$ is the envelope mass. It has become customary to approximate the growth timescale of the envelope in the form (e.g. Ikoma et al. \cite{ikomanakazawa2000}; Bryden et al. \cite{brydenlin2000}; Ida \& Lin \cite{idalin2004}; Miguel \& Brunini \cite{miguelbrunini2008})
\beq\label{eq:tkh}
\tg=k M^{-p}=10^{b}(\mz+\mxy)^{-p},
\eeq
where we adopted the same form as Ida \& Lin (\cite{idalin2004}) who use the total mass, while Ikoma et al. (\cite{ikomanakazawa2000}) use the core mass only. The core and  total mass are at least initially during phase II in any case similar.  The two parameters $k=10^{b}$ and $p$ ($b$ and $p$ are both parameters $>0$) are found by comparison with numerical simulations. Clearly, the parameters will depend on the core accretion rate and the associated luminosity, the core mass, and the opacity $\kappa$ in the envelope that is  represented in this study by $\fopa$. The explicit dependency of $k$ on $\fopa$ can be merged into the parameter, writing  $k=k_{0} \fopa^{q}$ (at least over a certain range of $\fopa$), where $k_{0}$ and $q$ are a priori unknown (in Sect. \ref{sect:paramstkh} we will see that $q\approx1$ over an important domain of $\fopa$). Table \ref{tab:littkh} gives an overview of values found in the literature. The units in the fit are years and Earth masses. Note that some fits were derived for planets accreting planetesimals, while others were derived for planets where the luminosity is coming from the accretion of the gas itself only. This naturally explains part of the differences. Still, as already discussed by Miguel \& Brunini (\cite{miguelbrunini2008}), the (remaining) significant differences and the high powers lead to highly uncertain results for the predicted gas envelope masses.  The comparison of the semi-analytical solution presented below and the numerical simulations  makes it possible to determine the parameters in a more systematic way.   

\begin{table}
\caption{Literature values of parameters for $\tg=10^{b} M^{-p}$. }\label{tab:littkh}
\begin{center}
\begin{tabular}{lcc}
\hline\hline
Reference                      &$b$&  $p$\\\hline 
Ikoma et al.  (\cite{ikomanakazawa2000})       &     8.3 &  2.47 \\                                             
Ikoma \& Genda  (\cite{ikomagenda2006})     &  10 & 3.5\\
Bryden et al. (\cite{brydenlin2000})	 &  10 & 3\\
Ida \& Lin (\cite{idalin2004}) 	 &  9 & 3\\
Miguel \& Brunini (\cite{miguelbrunini2008}) & 9.21 & 1.91\\ 
Miguel et al. (\cite{miguelguilera2011}) &10.92 & 4.89 \\ \hline
\end{tabular}
\end{center}
\end{table}

\subsection{Growing core mass: $\dot{\mz}\ne0$}\label{sect:semiwcore}
First, we consider the full problem where the core mass increases, assuming that the expansion of the feeding zone is the only limiting factor for core growth, and that the core instantaneously accretes all planetesimals within it. The feeding zone is proportional to the Hill sphere radius of the planet which in turn is proportional to $M^{1/3}$, therefore (Sect. \ref{sect:relationmzmxyphaseII})
\beq
M=\mz+\mxy\propto\mz^{3}.
\eeq
Differentiation with respect to the time gives
\beq
\dot{M}_{Z}\propto\frac{\dot{M}}{3 M^{2/3}}.
\eeq
This leads to the following system of coupled differential equation for the core mass $\mz(t)$ as a function of time $t$ (P96)
\beq\label{eq:dotmz}
\dot{M}_{Z}=\left(2+3\frac{\mxy}{\mz}\right)^{-1}\dot{M}_{XY}
\eeq
which holds very well during phase II in the numerical simulations above\footnote{The equations remain initially also valid in the runaway gas accretion phase as long as the extension of the feeding zone is the limiting factor for $\dot{M}_{Z}$. At some moment, however, the planet's envelope has collapsed so much that the capture radius for planetesimals becomes small. Concurrently, the ejection of planetesimals becomes important (depending on the semimajor axis of the planet). Then, $\dot{M}_{Z}$ diverges from Eq. \ref{eq:dotmz} and eventually falls to zero (see Mordasini et al. \cite{mordasinialibert2011}).} This equation is also found from differentiation of Eq. \ref{eq:mzmisop2}. For the envelope mass $\mxy(t)$ we have 
\beq\label{eq:mdotxysemi}
\dot{M}_{XY}=\frac{\mz+\mxy}{\tg}=\frac{\left(\mz+\mxy\right)^{p+1}}{k}.
\eeq
As boundary conditions we use $\mz(0)=\mzn$ which is the core mass at the beginning of phase II (typically the isolation mass), while we assume that the initial mass of the gaseous envelope $\mxy(0)=0$. This   approximation is valid if the antecedent core luminosity is high, and/or if the core mass is low. The time $t$ is counted relative to the beginning of phase II. 

While a closed form for $\mz(t)$ and $\mxy(t)$ solving the differential equations cannot be found, we still find an implicit equation which can be simply solved by determining numerically the root. One finds  
\beq\label{eq:mxywmcore}
\begin{multlined}
\shoveright[1cm]{\frac{1}{\mz(t)^{3+3p}}\left[\frac{\mz(t)}{2+3p}-\frac{3 \mz(t)^{3}}{3p\mzn^{2}}\right]=} \\ \frac{1}{\mzn^{2+3p}}\left[\frac{1}{2+3p}-\frac{1}{p}+\frac{\mzn^{p} t}{k}\right]
\end{multlined}
\eeq
and
\beq\label{eq:mxywmcore2}
\mxy(t)=\frac{\mz(t)^{3}}{\mzn^{2}}-\mz(t)
\eeq
The second equation is simply the solution to Eq. \ref{eq:dotmz} and  has as such no explicit temporal dependence. 

Two examples of the temporal evolution of the core, envelope and total mass obtained with these equations are shown in Fig. \ref{fig:semiwcore}. The initial mass of the core is $\mzn=5\mearth$, and the parameters of Ida \& Lin (\cite{idalin2004}) and Miguel et al. (\cite{miguelguilera2011}) are used to illustrate the impact of different parameters. As in numerical calculations, the evolution is characterized by a long plateau phase of slow growth, which accelerates rapidly once the core and the envelope mass become comparable. 

\begin{figure}
\begin{center}
\includegraphics[width=0.9\columnwidth]{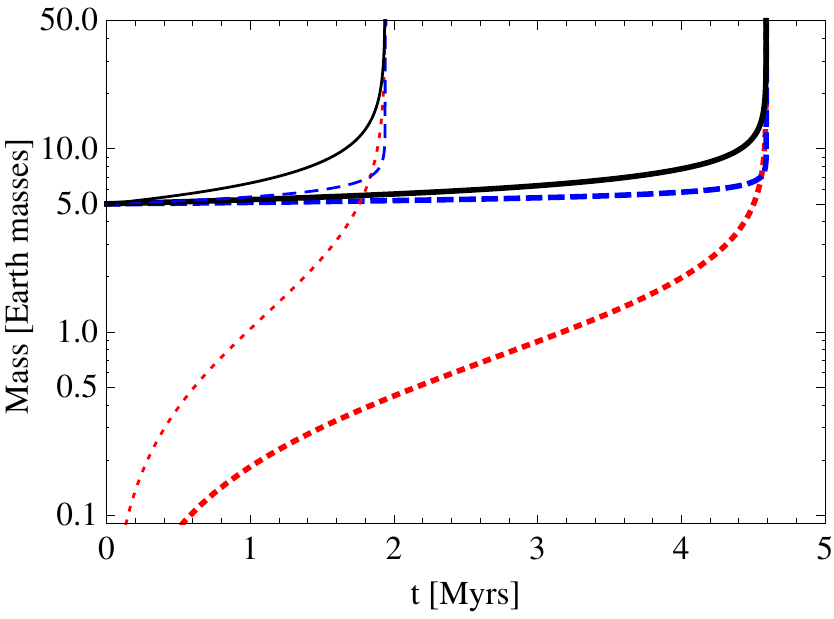}
\caption{Semi-analytical solution for the temporal evolution of the core mass (blue dashed), envelope mass (red dotted), and total mass (black solid lines). Thin lines use the parameters of Ida \& Lin (\cite{idalin2004}), while thick lines use those of Miguel et al. (\cite{miguelguilera2011}).}\label{fig:semiwcore}
\end{center}
\end{figure}

The semi-analytical solution can be used to calculate the time of crossover $\tcr$ when $\mz=\mxy$, and of runaway $\trun$ when $\mz$ and $\mxy$ approach infinity. In reality, the growth gets then limited due to the finite reservoirs of both gas and solids in the disk and/or gap formation. 

The core mass at crossover derived from the above equations is equal to $\sqrt{2}\mzn$ as already found by P96. From that we find  that the time when crossover occurs (relative to the beginning of phase II) is 
\beq
\tcr=\frac{k}{\mzn^{p}}\frac{2^{3 p/2 +2}(1+p) - 5p - 4}{2^{3 p/2 +1}(2 p+3 p^{2})}
\eeq
where $k/\mzn^{p}$ is the growth timescale of the initial core, while the rest of the equation corresponds to a numerical factor smaller than unity, as the growth timescale decreases with increasing mass. 

The time of runaway accretion when the masses diverge must occur when the right hand side of Eq. \ref{eq:mxywmcore} becomes equal to zero, therefore
\beq\label{eq:trunwcore}
\trun=\frac{k}{\mzn^{p}}\left(\frac{1}{p}-\frac{1}{2+3p}\right).
\eeq
This timescale is again proportional to the growth timescale of the initial core, while the positive numerical factor in the brackets decreases as expected with increasing $p$. 

A problem encountered when fitting numerical results to derive $b$ and $p$ is the degeneracy of the two parameters. It is therefore useful that the ratio of the two  characteristic times, $\trun/\tcr$ is a function of $p$ only. The ratio is given as  
\beq\label{eq:truntcrwcore}
\frac{\trun}{\tcr}=\left(1-\frac{5 p/4+1}{2^{3p/2}(1+p)}\right)^{-1}.
\eeq
While we cannot solve algebraically for $p$, we can invert the equation numerically if $\trun/\tcr$ is given from a simulation, and then use Eq. \ref{eq:trunwcore} to also calculate $k$. 

Figure \ref{fig:calibsolution} shows a comparison of a numerical simulation and the  corresponding calibrated semi-analytical solution for $\mz(t)$ and $\mxy(t)$ obtained in this way. In the numerical simulation, $\mzn=4.35\mearth$ and $\fopa=0.002$, leading to $\tcr=2.50$ Myr and $\trun$=3.57 Myr,  both relative to the beginning of phase II. For $\trun$, we use the moment when the limiting gas accretion rate of 0.01 $\mearth$/yr is reached, but the timescales are at that moment anyway very short, so that the specific choice is not important. These numerical results lead to fit parameters $p=1.269$ and $b=7.573$.    

\begin{figure}
\begin{center}
\includegraphics[width=0.9\columnwidth]{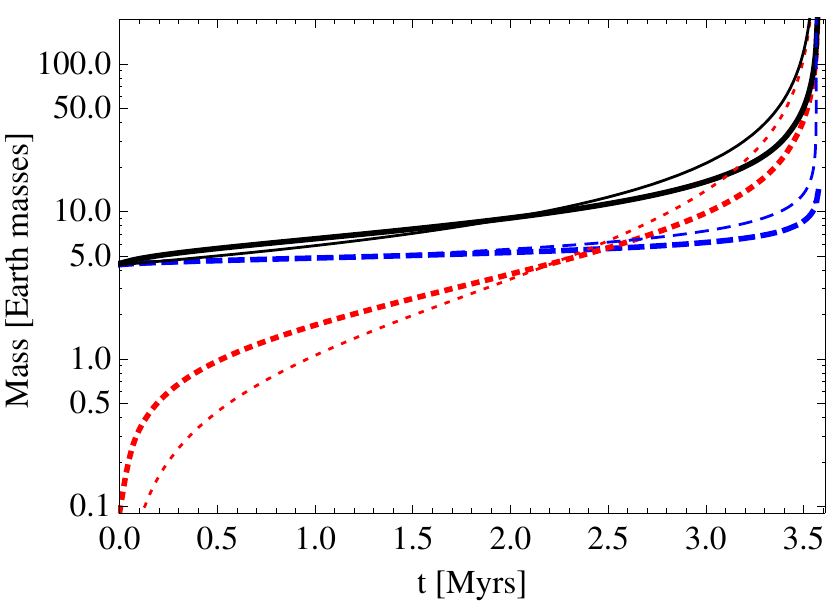}
\caption{Numerical result (thick lines) and corresponding calibrated semi-analytical solution (thin lines) for the temporal evolution of the core mass (blue dashed), envelope mass (red dotted), and total mass (black solid lines).}\label{fig:calibsolution}
\end{center}
\end{figure}

The comparison shows that the calibrated semi-analytical solution provides a fair general approximation to the numerical result. The times of crossover and runaway agree by construction. Before crossover, $\mxy$ is however underestimated, while it is overestimated afterwards. This seems to be a general property of the semi-analytical solution, and is a  sign that the fit parameters are in fact not constant during the entire evolution. The lower envelope mass initially predicted by the semi-analytical solution is partially also due to the initial condition that $\mxy(t)=0$. In the simulation, $\mxy$ is in contrast about 0.1 $\mearth$ at this moment.

\subsection{Constant core mass: $\dot{\mz}=0$}
Second, we consider the simpler case that the core does not grow, e.g., due to the presence of neighboring competing cores (Hubickyj  et al. \cite{hubickyjbodenheimer2005}) or the second passage through a region where the core has previously accreted the planetesimals due to outward migration followed by inward migration (Mordasini et al. \cite{mordasinialibert2011}).

In this case, we only have to consider Eq. \ref{eq:mdotxysemi}. With the boundary conditions that $\mxy(0)=0$, one finds
\beq\label{eq:mxymcoreno}
\mxy(t)=\left(\mzn^{-p}-\frac{p t}{k}\right)^{-1/p}-\mzn
\eeq

Figure \ref{fig:mcmxywocore} shows the temporal evolution for a core mass of 5 $\mearth$ and the same parameters $p$ and $k$ as in Fig. \ref{fig:semiwcore}.  The comparison with this figure shows that  while the growth timescales are for identical parameters $p$ and $k$ in principle longer  for $\mz$=cst. than for the case with a growing core, they are in reality shorter since e.g. $k$ is roughly one order of magnitude smaller when the core does not grow  due to the reduced luminosity without planetesimal accretion (Ikoma et al. \cite{ikomanakazawa2000}).

\begin{figure}
\begin{center}
\includegraphics[width=0.9\columnwidth]{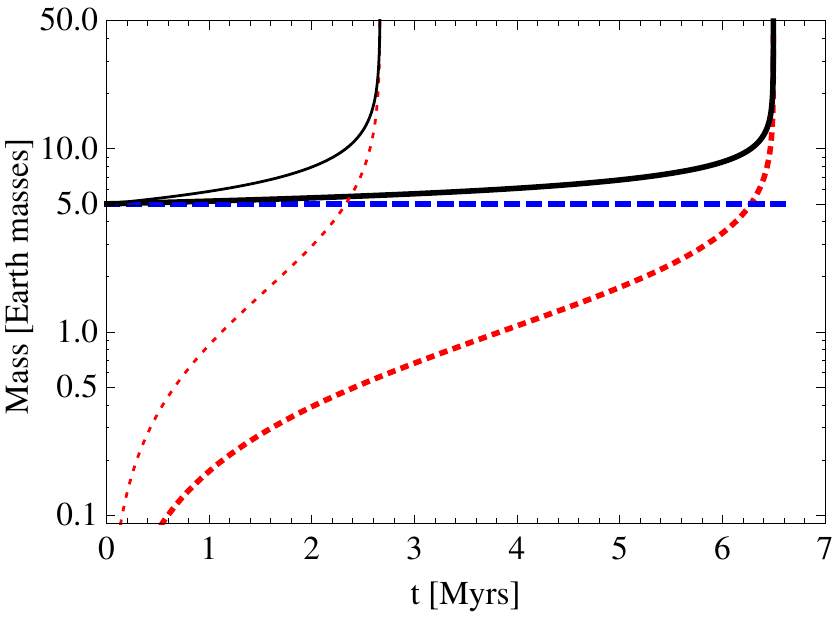}
\caption{Semi-analytical solution for the temporal evolution of the envelope mass (red dotted) for the case that the core mass is constant (blue dashed line). The total mass is shown with black solid lines. The thin lines use the parameters from Ida \& Lin (\cite{idalin2004}), while the thick lines use those of Miguel et al. (\cite{miguelguilera2011}).}\label{fig:mcmxywocore}
\end{center}
\end{figure}
From the solution we find that crossover occurs at 
\beq
\tcr=\frac{k}{\mzn^{p}}\frac{2^{p}-1}{2^{p}p}
\eeq
and runaway gas accretion at 
\beq
\trun=\frac{k}{\mzn^{p}}\frac{1}{p}.
\eeq
The ratio of the time of runaway to the time of crossover is here 
\beq\label{eq:truntcrwocore}
\frac{\trun}{\tcr}=\frac{2^{p}}{2^{p}-1}.
\eeq
For a fixed $p$ this ratio is larger than for the case with $\dot{\mz}\ne0$. In contrast to this case, we can here algebraically solve for $p$ if  $\trun/\tcr$ has been determined in a numerical simulation. 

\subsection{Constant core, small envelope: $\dot{\mz}=0$, $\mxy\ll\mzn$}\label{sect:mcorecstsmallmenve}
Finally we consider the simplest case where the core does not grow, and where the envelope mass is much smaller than the core mass, as it is the case during the early phase II long before crossover. Here, we immediately find (again with $\mxy(0)=0$) 
\beq\label{eq:mxymcorenos}
\mxy(t)=\frac{\mzn^{p+1}}{k}t
\eeq

\subsection{Estimates for $\mxy$ as function of $\mz$}\label{sect:mxymzsemi}
\begin{figure}
\begin{center}
\includegraphics[width=0.9\columnwidth]{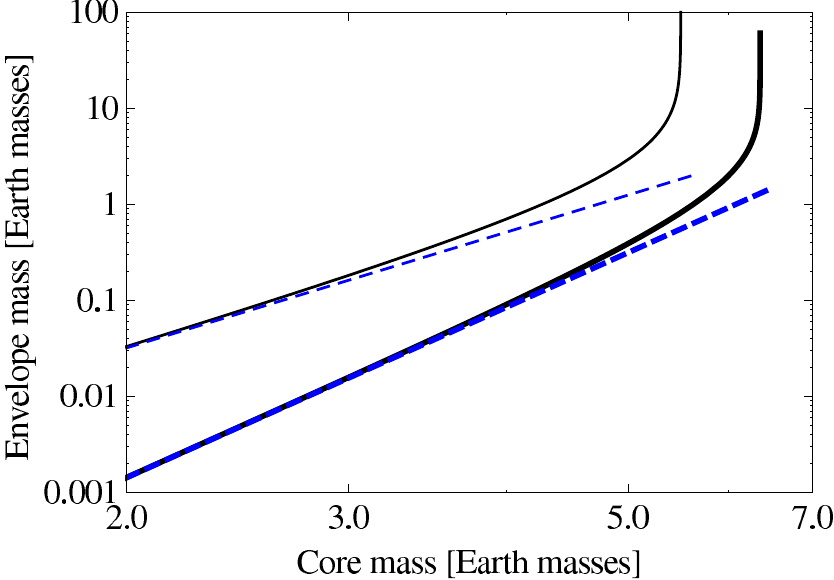}
\caption{Illustration of $\mxy(\mzn)$ at $t=2$ Myrs. The black solid lines show envelope masses as a function of core mass for the constant core mass case, while the blue dashed lines are the approximation for  $\mxy\ll\mzn$. The thin lines use the parameters from Ida \& Lin (\cite{idalin2004}), while the thick lines use those of Miguel et al. (\cite{miguelguilera2011}). }\label{fig:mcmemtana}
\end{center}
\end{figure}

In view of the numerical results for the envelope mass as a function of core mass for synthetic planetary populations, it is instructive to plot the envelope mass as a function of core mass for a given moment in time using the simple semi-analytical solutions. In Fig. \ref{fig:mcmemtana},  $\mxy$ calculated with Eqs. \ref{eq:mxymcoreno} and \ref{eq:mxymcorenos} is shown as function of $\mzn$ for $t=2$ Myrs, which is roughly the mean lifetime of  protoplanetary disks around T Tauri stars (e.g., Haisch et al. \cite{haischlada2001}; Fedele et al. \cite{fedelevandenancker2010}).  For $\mxy\ll\mz$, $\mxy$ is proportional to $\mz^{p+1}$, while it increases stronger and eventually diverges when $\mzn$ approaches the critical core mass for runaway $M_{\rm Z,0,crit}$ at $t=2$ Myrs.  In the plot, we again use the parameters from Ida \& Lin (\cite{idalin2004}) and Miguel et al. (\cite{miguelguilera2011}) for $\tg$. These parameters are chosen for  illustrative purposes only. But they make clear that one can gain in principle observational constraints on the parameters of $\tg$ by measuring the bulk composition of many low-mass, sub-critical planets with primordial envelopes, since this allows to determine the slope and absolute scaling of the $\mz-\mxy$  relation (see Sect. \ref{sect:menveofmcore} for a synthetic population).   

The curves showing $\mxy$ (Eq. \ref{eq:mxymcoreno}) in Fig. \ref{fig:mcmemtana} are functionally similar to the relationship between the core and the envelope mass in the classical hydrostatic calculations of Mizuno et al. (\cite{mizunoetal1978}) and the corresponding approximate analytical solution of Stevenson (\cite{stevenson1982}): at low core masses $\mxy\ll\mz$, then $\mxy$ increases rapidly, and beyond a certain critical core mass, there is no physical solution. It also shears the property that the smaller the opacity (and therefore $k$), the lower the critical mass. A significant difference is however the explicit temporal dependence in the case considered here. For a given time $t$, $M_{\rm Z,0,crit}$ is 
\beq
M_{\rm Z,0,crit}=\left(\frac{k}{t p}\right)^{1/p}.
\eeq
Figure \ref{fig:mz0crit} shows $M_{\rm Z,0,crit}$ as a function of time using the parameters of $\tg$ from Ikoma et al. (\cite{ikomanakazawa2000}) and Ikoma \& Genda (\cite{ikomagenda2006}). The latter work uses more realistic opacities. In both cases, the cores do not accreted planetesimals, and full a ISM grain opacity is assumed.

\begin{figure}
\begin{center}
\includegraphics[width=0.9\columnwidth]{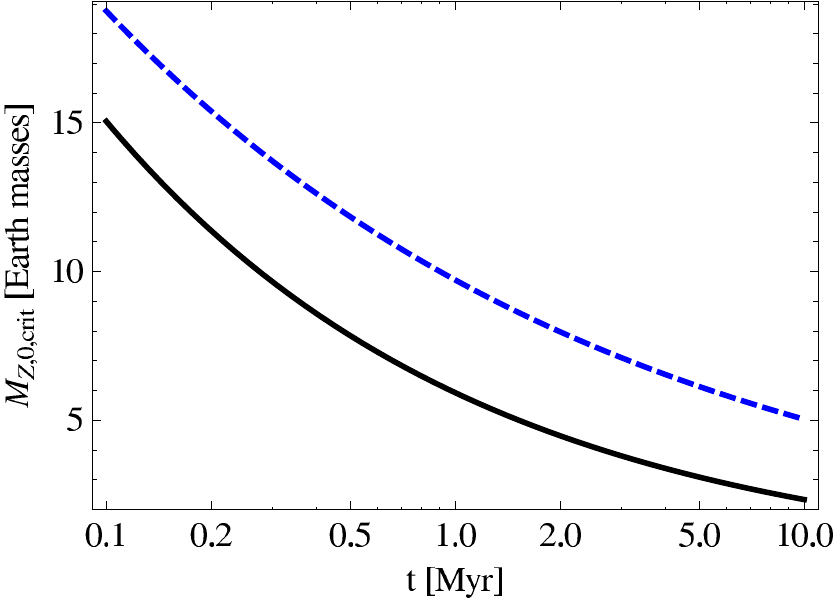}
\caption{Mass of a core $M_{\rm Z,0,crit}$ that undergoes gas runaway accretion after being embedded in the nebula without planetesimal accretion during a time $t$. The black solid line uses the parameters for $\tg$ from Ikoma et al. (\cite{ikomanakazawa2000}) while the blue dashed line is for the updated parameters of   Ikoma \& Genda (\cite{ikomagenda2006}).}\label{fig:mz0crit}
\end{center}
\end{figure}

The plot shows that without planetesimal accretion, gas runaway accretion sets in for massive cores ($\sim15\mearth$) relatively quickly after $1-2\times 10^{5}$ years even for full grain opacity. For lower mass cores ($\sim 5 \mearth$), the moment of runaway however approaches or exceeds the lifetime of the disk, so that lower opacities are needed for giant planet formation even for the case of minimal luminosity (Ikoma \& Genda \cite{ikomagenda2006}).

The explicit temporal dependence  of the critical core mass has been derived for a core that does not accrete planetesimals. In this case, the luminosity is generated by the contraction of the envelope itself, i.e., by the $-T \partial S/\partial t$ term ($T$ is the temperature, $S$ the entropy) in the energy equation which establishes a direct dependence on the earlier state of the envelope. This is different from the situation considered by  Stevenson (\cite{stevenson1982}) where the luminosity is generated by the accretion of planetesimals which is a free parameter. 

This is also different from phase II considered in the simulations in Sect. \ref{sect:determinationfopa} where the core accretes. There, the dominant part of the luminosity again comes from the accretion of planetesimals but the accretion rate of planetesimals in no more a free parameter, but depends via the extension of the feeding zone on the gas accretion timescale, which in turn depends on the opacity.  
 
\section{Parameters of the growth timescale}\label{sect:paramstkh}
With the semi-analytical solution of Sect. \ref{sect:semiwcore} and the numerical results for the time of crossover and runaway from Sect. \ref{sect:determinationfopa}, we can systematically study the parameters $p$ and $b$ (or $k$) of  the planetary growth timescale as a function of core mass and opacity for the MBPL10 comparison calculations.  We thus only consider the case where the core grows due to expansion of the feeding zone. Figure \ref{fig:tkhparams} shows the results as a function of $\fopa$ over the relevant domain. In both panels, the results for the three different planetesimals surface densities $\sigmas0=$10, 6, and 4  g/cm$^{2}$ are given. 

\begin{figure}
\begin{center}
\includegraphics[width=0.98\columnwidth]{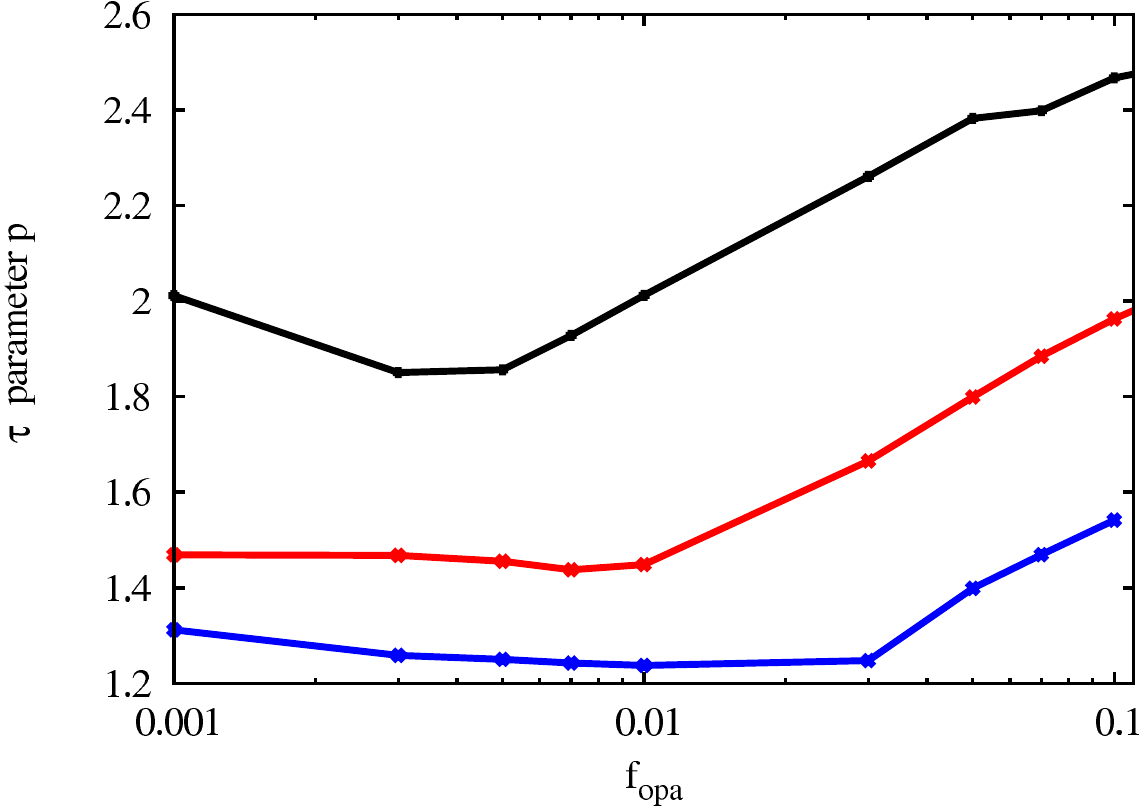}
\includegraphics[width=0.98\columnwidth]{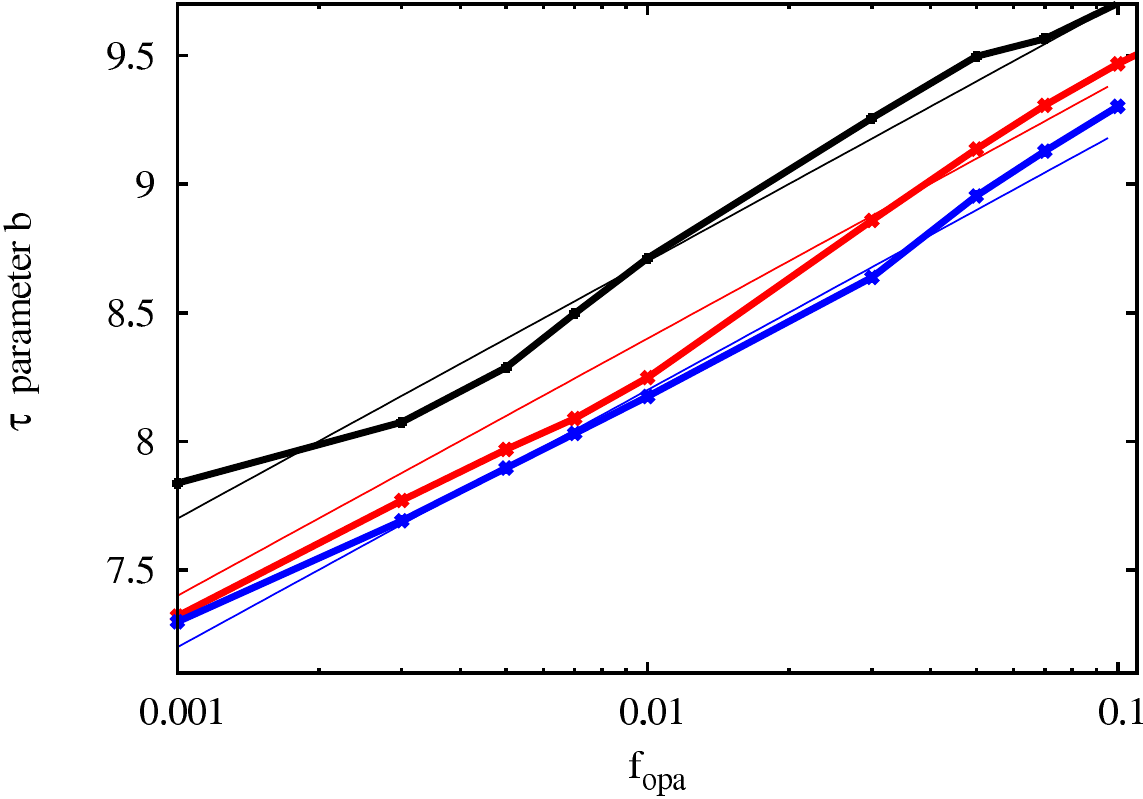}
\caption{Parameters for the planetary growth timescale (Eq. \ref{eq:tkh}) obtained from the comparison of numerical simulations (Sect. \ref{sect:determinationfopa})  with the semi-analytical solution of Sect. \ref{sect:semiwcore}.  The upper panel shows the exponent $p$ as a function of $\fopa$. The initial surface density of planetesimals is 10 (top, black), 6 (middle, red) and 4 g/cm$^{2}$ (bottom, blue line). The lower panel shows $b$, again as function of $\fopa$ and the three planetesimal surface densities. Thick lines are the numerical results, while thin lines give the approximative linear scaling with $\fopa$. }\label{fig:tkhparams}
\end{center}
\end{figure}

The upper panel shows the exponent $p$. One sees that the exponent varies by about 20\% for a variation of $\fopa$ over two orders of magnitude, so that there is only a weak, but still non-negligible dependency. There is a trend of an increase of $p$ with $\fopa$, bringing it to a value between 2.5 to 3.5 for full grain opacity and $\sigmas0=$10 g/cm$^{2}$. This agrees with previous results considering  a similar setup like e.g. Bryden et al. (\cite{brydenlin2000}) who fit the result of P96. The plot also shows that $p$ is an increasing function of the mass of the core, indicating that the functional dependence of $\tg$ on the total mass only (Eq. \ref{eq:tkh}) is insufficient for a full parametrization. 

In the lower panel we see that  $b$ increases with $\fopa$, as expected. The three thin lines approximating the  numerical results are proportional to $\log(\fopa)$, meaning that $k=10^{b}$ is linearly proportional to $\fopa$. A linear dependence of the growth timescale on the opacity has already been found by Ikoma et al. (\cite{ikomanakazawa2000}) and is found in the analytical solution of Stevenson (\cite{stevenson1982}) for purely radiative envelopes.  Rafikov (\cite{rafikov2006}) generalized this result to envelopes that are radiative at the interface to the nebula but convective below, as it is the case during phase II for the simulations shown here. 

Table \ref{tab:paramstkh} summarizes the results by giving approximate expression for $p$ and $b$ as a function of $\sigmas0$ (and therefore the core mass) and  $0.001\leq\fopa\leq 0.1$. These expression can be used together with the semi-analytical solution for rough estimates of the timescales of crossover and runaway. But we caution that full evolutionary calculations are necessary to determine them under realistic conditions, where e.g. the planetesimal accretion rate varies due to orbital migration or the competition with other cores.

\begin{table}
\caption{Parameters for $\tg=10^{b} M^{-p}$ of planets in phase II for $0.001\leq\fopa\leq 0.1$. }\label{tab:paramstkh}
\begin{center}
\begin{tabular}{lccc}
\hline\hline
$\sigmas0$ [g/cm$^{2}$]                  & $\mz$ [$\mearth$] &  $p$ & b\\\hline 
10      &     $\sim$13 &  1.84-2.46 & 10.7 + $\log(\fopa)$\\                                             
6       &  $\sim$7 & 1.44-1.95 & 10.4 + $\log(\fopa)$\\
4     &  $\sim$4 & 1.25-1.55 & 10.2 + $\log(\fopa)$ \\ \hline
\end{tabular}
\end{center}
\end{table}


\begin{thebibliography}{}

\bibitem[2008]{adamsseager2008} 
Adams, E.~R., Seager, S., \& Elkins-Tanton, L.\ 2008, \apj, 673, 1160 


\bibitem[2012]{adibekyansantos2012} 
Adibekyan, V.~Z., Santos, N.~C., Sousa, S.~G., et al.\ 2012, \aap, 543, A89 



\bibitem[2004]{alibertmordasini2004} 
 Alibert, Y., Mordasini, C., \& Benz, W.\ 2004, \aap, 417, L25 

\bibitem[2005]{alibertmordasini2005}
Alibert, Y., Mordasini, C., Benz, W., \& Winisdoerffer, C. 2005, \aap, 434, 343

\bibitem[2013]{alibertcarron2013} 
Alibert, Y., Carron, F., Fortier, A., et al.\ 2013, \aap, 558, A109 




%

\bibitem[2009]{asplundgrevesse2009} 
Asplund, M., Grevesse, N., Sauval, A.~J., \& Scott, P.\ 2009, \araa, 47, 481 





\bibitem[2008]{baraffechabrier2008}
Baraffe, I., Chabrier, G., \& Barman, T.~S. 2008, \aap, 482, 315


\bibitem[1994]{belllin1994}
Bell, K.~R. \& Lin, D. N.~C. 1994, \apj, 427,987



\bibitem[2012]{birnstielklahr2012} 
Birnstiel, T., Klahr, H., \& Ercolano, B.\ 2012, \aap, 539, A148 



\bibitem[1986]{bodenheimerpollack1986}
Bodenheimer, P.~H. \& Pollack, J.~B. 1986,  Icarus, 67, 391

\bibitem[2013]{bodenheimerdangelo2013} 
Bodenheimer, P., D'Angelo, G., Lissauer, J.~J., Fortney, J.~J., \& Saumon, D.\ 2013, \apj, 770, 120 



\bibitem[2011]{boleyhelled2011} 
Boley, A.~C., Helled, R., \& Payne, M.~J.\ 2011, \apj, 735, 30 

\bibitem[2010]{bonomosanterne2010} 
Bonomo, A.~S., Santerne, A., Alonso, R., et al.\ 2010, \aap, 520, A65 






\bibitem[2013]{broegfortier2013} 
Broeg, C., Fortier, A., Ehrenreich, D., et al.\ 2013, European Physical Journal Web of Conferences, 
47, 3005 

\bibitem[2000]{brydenlin2000} 
Bryden, G., Lin, D.~N.~C., \& Ida, S.\ 2000, \apj, 544, 481 

\bibitem[1997]{burrowsmarley1997}
Burrows, A.~S., Marley, M.~S., Hubbard, W.~B., Lunine, J.~I., Guillot, T.,  Saumon, D., Freedman, R., Sudarsky, D., \& Sharp, C. 1997, \apj, 491, 856

\bibitem[2007]{burrowshubeny2007} 
Burrows, A., Hubeny, I., Budaj, J., \& Hubbard, W.~B.\ 2007, \apj, 661, 502 

\bibitem[2011]{burrowsetal2011} 
Burrows, A., Heng, K., \& Nampaisarn, T.\ 2011, \apj, 736, 47 












\bibitem[2011]{cochranfabrycky2011} 
Cochran, W.~D., Fabrycky, D.~C., Torres, G., et al.\ 2011, \apjs, 197, 7 







\bibitem[2011]{demoryseager2011} 
Demory, B.-O., \& Seager, S.\ 2011, \apjs, 197, 12 

\bibitem[2014]{dittkristmordasini2014}
Dittkrist, K.-M., Mordasini, C., Klahr, H.,  Alibert, Y., \& Henning, T. 2014, \aap, accepted [arXiv:1402.5969]


\bibitem[2010]{fedelevandenancker2010} 
Fedele, D., van den Ancker, M.~E., Henning, T., Jayawardhana, R., \& Oliveira, J.~M.\ 2010, \aap, 510, A72 



\bibitem[2007]{fortierbenvenuto2007}
Fortier, A., Benvenuto, O.~G., and Brunini, A. 2007, \aap, 473, 311

\bibitem[2013]{fortieralibert2013} 
Fortier, A., Alibert, Y., Carron, F., Benz, W., \& Dittkrist, K.-M.\ 2013, \aap, 549, A44 



\bibitem[2007]{fortneymarley2007}
Fortney, J.~J., Marley, M.~S., \& Barnes, J.~W. 2007, \apj, 659, 1661



\bibitem[2011]{fortneyikoma2011} 
Fortney, J.~J., Ikoma, M., Nettelmann, N., Guillot, T., \& Marley, M.~S.\ 2011, \apj, 729, 32 



\bibitem[2008]{freedmanmarley2008}
Freedman, R.~S., Marley, M.~S., \& Lodders, K. 2008, \apjs, 174, 504



\bibitem[1999]{guillot1999} 
Guillot, T. 1999, Science, 286, 72



\bibitem[2006]{guillotsantos2006} 
Guillot, T., Santos, N.~C., Pont, F., et al.\ 2006, \aap, 453, L21 

\bibitem[2008]{guillot2008} 
Guillot, T.\ 2008, Physica Scripta Volume T, 130, 014023 

\bibitem[2009]{guillotgautier2009}
Guillot, T. \& Gautier, D. 2009, in Treatise on Geophysics,  ed. G. Shubert \& T. Spohn, 10, 439 


\bibitem[2001]{haischlada2001} 
Haisch, K.~E., Jr., Lada, E.~A., \& Lada, C.~J.\ 2001, \apjl, 553, L153 


\bibitem[2009]{helledschubert2009} 
Helled, R., \& Schubert, G.\ 2009, \apj, 697, 1256 


\bibitem[2011]{helledbodenheimer2011} 
Helled, R., \& Bodenheimer, P.\ 2011, \icarus, 211, 939 




\bibitem[2010]{horiikoma2010}
Hori, Y. \& Ikoma, M. 2010, \apj, 714, 1343

\bibitem[2011]{horiikoma2011} 
Hori, Y., \& Ikoma, M.\ 2011, \mnras, 416, 1419


\bibitem[2005]{hubickyjbodenheimer2005}
Hubickyj, O., Bodenheimer, P.~H., \& Lissauer, J.~J. 2005, Icarus, 179, 415

\bibitem[1993]{idamakino1993} 
Ida, S., \& Makino, J.\ 1993, \icarus, 106, 210 

\bibitem[2004]{idalin2004}
Ida, S. \& Lin, D.N.C. 2004, \apj, 604, 388

\bibitem[2000]{ikomanakazawa2000}
Ikoma, M., Nakazawa, K., \&  Emori, H. 2000, \apj, 537, 1013

\bibitem[2006]{ikomagenda2006} 
Ikoma, M., \& Genda, H.\ 2006, \apj, 648, 696 

 
%
%



\bibitem[2006]{kleydirksen2006} 
Kley, W., \& Dirksen, G.\ 2006, \aap, 447, 369 



\bibitem[2004]{kornetetal2004} 
Kornet, K., R{\'o}{\.z}yczka, M., \& Stepinski, T.~F.\ 2004, \aap, 417, 151 






\bibitem[2012]{lecontechabrier2012} 
Leconte, J., \& Chabrier, G.\ 2012, \aap, 540, A20 




\bibitem[2010]{levisonthommes2010}
Levison, H.~F., Thommes, E.~W., \& Duncan, M.~J. 2010, \aj, 139, 1297

\bibitem[1993]{lissauer1993} 
Lissauer, J.~J.\ 1993, \araa, 31, 129 

\bibitem[2009]{lissauerhubickyj2009}
Lissauer, J.~J., Hubickyj, O., D'Angelo, G., \& Bodenheimer, P.~H. 2009, Icarus, 199, 338

\bibitem[2011]{lissauerfabrycky2011} 
Lissauer, J.~J., Fabrycky, D.~C., Ford, E.~B., et al.\ 2011, \nat, 470, 53 


\bibitem[2012]{lopezfortney2012} 
Lopez, E.~D., Fortney, J.~J., \& Miller, N.\ 2012, \apj, 761, 59 




%
\bibitem[2010]{lyrapaardekooper2010}
Lyra, W., Paardekooper, S.-J., \& Low, M.-M. M.~M. 2010, \apjl, 715, L68

%


\bibitem[2014]{marcyetal2014} 
Marcy, G.~W., Isaacson, H., Howard, A.~W., et al.\ 2014, \apjs, 210, 20 




\bibitem[2011]{mayormarmier2011} 
Mayor, M., Marmier, M., Lovis, C., et al.\ 2011, \aap, submitted [arXiv:1109.2497] 


\bibitem[2008]{miguelbrunini2008} 
Miguel, Y., \& Brunini, A.\ 2008, \mnras, 387, 463 

\bibitem[2011]{miguelguilera2011} 
Miguel, Y., Guilera, O.~M., \& Brunini, A.\ 2011, \mnras, 412, 2113 

\bibitem[2008]{militzerhubbard2008} 
Militzer, B., Hubbard, W.~B., Vorberger, J., Tamblyn, I., \& Bonev, S.~A.\ 2008, \apjl, 688, L45 

\bibitem[2011]{millerfortney2011} 
Miller, N., \& Fortney, J.~J.\ 2011, \apjl, 736, L29 


\bibitem[1978]{mizunoetal1978}
Mizuno, H., Nakazawa, K., \& Hayashi, C. 1978, Prog. Th. Phys., 60, 3, 699 

\bibitem[2012]{mollieremordasini2012} 
Molli{\`e}re, P., \& Mordasini, C.\ 2012, \aap, 547, A105 

\bibitem[2006]{mordasinialibert2005}
Mordasini, C., Alibert, Y., \& Benz, W. 2005, in Tenth Anniversary of 51 Peg-b: Status of and prospects for hot Jupiter studies, ed. L. Arnold, F. Bouchy, \& C. Moutou, 84 - 86

\bibitem[2009a]{mordasinialibert2009a}
Mordasini, C., Alibert, Y., \& Benz, W. 2009a, \aap, 501, 1139

\bibitem[2009b]{mordasinialibert2009b} 
Mordasini, C., Alibert, Y., Benz, W., \& Naef, D.\ 2009b, \aap, 501, 1161 



\bibitem[2012a]{mordasinialibert2012pp} 
Mordasini, C., Alibert, Y., Benz, W., Klahr, H., \& Henning, T.\ 2012a, \aap, 541, A97 

\bibitem[2012b]{mordasinialibert2011}
Mordasini, C., Alibert, Y., Klahr, H., \& Henning, T.\ 2012b, \aap, 547, A111 

\bibitem[2012c]{mordasinialibert2012} 
Mordasini, C., Alibert, Y., Georgy, C., et al.\ 2012c, \aap, 547, A112 

\bibitem[2013]{mordasini2013} Mordasini, C.\ 2013, \aap, 558, A113 

\bibitem[2014]{mordasini2014} 
Mordasini, C. 2014, \aap, submitted (Paper II)

\bibitem[2008]{movshovitzpodolak2008} 
Movshovitz, N., \& Podolak, M.\ 2008, \icarus, 194, 368 

\bibitem[2010]{mbpl2010}
Movshovitz, N., Bodenheimer, P.~H., Podolak, M., \& Lissauer, J.~J. 2010, Icarus, 209, 616 (MBPL10)

\bibitem[2010]{nayakshin2010} 
Nayakshin, S.\ 2010, \mnras, 408, 2381 

\bibitem[2012]{nettelmannbecker2012} 
Nettelmann, N., Becker, A., Holst, B., \& Redmer, R.\ 2012, \apj, 750, 52 


\bibitem[2012]{ormelkobayashi2012} 
Ormel, C.~W., \& Kobayashi, H.\ 2012, \apj, 747, 115 

\bibitem[2013]{owenwu2013} 
Owen, J.~E., \& Wu, Y.\ 2013, \apj, 775, 105 




\bibitem[1999]{papaloizouterquem1999} 
Papaloizou, J. C. B. \& Terquem, C. 1999, \apj, 521, 823



\bibitem[1988]{podolakpollack1988} 
Podolak, M., Pollack, J.~B., \& Reynolds, R.~T.\ 1988, \icarus, 73, 163 

\bibitem[2003]{podolak2003} 
Podolak, M.\ 2003, Icarus, 165, 428 

\bibitem[1996]{pollackhubickyj1996} 
Pollack, J. B., Hubickyj, O., Bodenheimer, P., et al. 1996, Icarus, 124, 62 (P96)

\bibitem[2009]{ponthebrard2009} 
Pont, F., H{\'e}brard, G., Irwin, J.~M., et al.\ 2009, \aap, 502, 695 



\bibitem[2006]{rafikov2006} 
Rafikov, R.~R.\ 2006, \apj, 648, 666

\bibitem[2013]{rauercatala2013} 
Rauer, H., Catala, C., Aerts, C., et al.\ 2013, Experimental Astronomy, in rev. [arXiv:1310.0696] 


\bibitem[2010]{rogersseager2010} 
Rogers, L.~A., \& Seager, S.\ 2010, \apj, 712, 974 

\bibitem[2011]{rogersbodenheimer2011} 
Rogers, L.~A., Bodenheimer, P., Lissauer, J.~J., \& Seager, S.\ 2011, \apj, 738, 59 


\bibitem[1978]{rossow1978} 
Rossow, W.~B.\ 1978, \icarus, 36, 1 

\bibitem[1995]{saumonchabrier1995} 
Saumon, D., Chabrier, G. \& Van Horn, H. M. 1995, \apjs, 99, 713

\bibitem[2004]{saumonguillot2004} 
Saumon, D., \& Guillot, T.\ 2004, \apj, 609, 1170 




\bibitem[2014]{shengmordasini2014}
Sheng, S., Mordasini, C., Parmentier, V., et al. 2014, in prep.

\bibitem[2008]{shiraishiida2008}
Shiraishi, M. \& Ida, S. 2008, \apj, 684, 1416


%

\bibitem[1982]{stevenson1982} 
Stevenson, D.~J.\ 1982, \planss, 30, 755 






\bibitem[2010]{valenciaikoma2010}
Valencia, D., Ikoma, M., Guillot, T., \& Nettelmann, N. 2010, \aap,  516, A20









\bibitem[2007]{zhoulin2007}
Zhou, J.-L. \& Lin, D. N. C. 2007, \apj, 666, 447

\end{thebibliography}
\end{document}